\documentclass[conference,compsoc]{IEEEtran}

\AtBeginDocument{%
  \providecommand\BibTeX{{%
    \normalfont B\kern-0.5em{\scshape i\kern-0.25em b}\kern-0.8em\TeX}}}

\usepackage{tikz}

\usepackage[ruled,vlined, linesnumbered]{algorithm2e}
\SetNlSty{}{}{}  
\usepackage{amsmath,amssymb,amsfonts}
\usepackage{subfigure}
\usepackage{textcomp}
\usepackage{xcolor}
\usepackage{graphicx}
\usepackage{multirow}
\usepackage{textcomp}
\usepackage{tikz}
\usepackage{xcolor}
\usepackage{hyperref}
\usepackage{xspace}
\usepackage{listings}
\usepackage{bbding}
\usepackage{wasysym}
\usepackage{enumitem,kantlipsum}
\usepackage{textcomp}
\usepackage{xcolor}
\usepackage{colortbl}
\usepackage{soul}
\usepackage{interval}
\usepackage{pgfplots}
\usepackage{pifont}
\usepackage{url}
\usepackage{phonetic}
\usepackage{interval}
\usepackage{pgfplots}
\usepackage{tipa}
\usepackage{multicol}
\usepackage{multirow}
\usepackage{xspace}
\usepackage{xfrac}
\usepackage{pifont}
\usepackage{booktabs}
\usepackage{diagbox}

\newcommand*\emptycircle[1][1ex]{\raisebox{-0.5ex}{\tikz\draw (0,0) circle (#1);}}
\newcommand*\halfcircle[1][1ex]{\raisebox{-0.5ex}{%
  \begin{tikzpicture}
  \draw[fill] (0,0)-- (90:#1) arc (90:270:#1) -- cycle ;
  \draw (0,0) circle (#1);
  \end{tikzpicture}}}
\newcommand*\fullcircle[1][1ex]{\raisebox{-0.5ex}{\tikz\fill (0,0) circle (#1);}}

\newcommand{\hanqing}[1]{{\color{black} #1}}

\newcommand{\cmark}{\ding{51}}%
\newcommand{\xmark}{\ding{55}}%

\makeatletter
\long\def\@makefntext#1{\parindent 1em\noindent
\hbox to 1.8em{\hss\textsuperscript{\@thefnmark}}#1}
\makeatother

\begin{document}
\title{SoK: How Robust is Audio Watermarking in Generative AI models?
}

\author{
\IEEEauthorblockN{
Yizhu Wen\textsuperscript{1}\IEEEauthorrefmark{1}, 
Ashwin Innuganti\textsuperscript{2}\IEEEauthorrefmark{1}, 
Aaron Bien Ramos\textsuperscript{1}, 
Hanqing Guo\textsuperscript{1}, 
Qiben Yan\textsuperscript{2}
}
\IEEEauthorblockA{\textsuperscript{1}University of Hawaii at Manoa}
\IEEEauthorblockA{\textsuperscript{2}Michigan State University}
}

\maketitle
\renewcommand{\thefootnote}{\fnsymbol{footnote}} 
\footnotetext[1]{Yizhu Wen and Ashwin Innuganti contributed equally to this work.}

\begin{abstract}
Audio watermarking has been used to verify the provenance of AI-generated content from generative models. It spawns a wealth of applications for detecting AI-generated speech, protecting the music IP, and defending against voice cloning attacks. 
Generally, audio watermarking should be robust against removal attacks that distort the signal to evade detection. Many audio watermarking schemes claim robustness; however, these claims are often validated in isolation against a limited set of attacks. There is no systematic, empirical evaluation of robustness against a comprehensive set of removal attacks in the audio domain. This uncertainty complicates the practical deployment of watermarking schemes. In this paper, we survey and evaluate whether recent audio watermarking schemes claiming robustness can withstand a broad range of removal attacks. 
First, we propose a taxonomy for 22 audio watermarking schemes. Second, we summarize the audio watermark technologies and their potential vulnerabilities. Third, we conduct a large-scale, comprehensive measurement study to evaluate the robustness of existing watermarking schemes. To facilitate this analysis, we develop an evaluation framework encompassing a total of \hanqing{22} types of watermark removal attacks (109 different configurations). Our framework covers signal-level distortions, physical-level distortions, and AI-induced distortions. We reproduced 9 watermarking schemes with open-source code and identified 8 new attacks that are highly effective against all watermarks, and discovered 11 key findings that illustrate their fundamental weaknesses across 3 public datasets. Our study reveals critical insights: \emph{none of the surveyed watermarking schemes is robust enough to withstand all tested distortions in practice}. This extensive evaluation provides a comprehensive view of how well current watermarking schemes perform in the face of real-world threats. Our demo and code are available at \url{https://sokaudiowm.github.io/}.
\end{abstract} 

\section{Introduction}\label{sec-introduction}\vspace{-10pt}

Watermarking is a crucial technique for embedding secret information into media to ensure copyright protection, verify content ownership, and detect tampering~\cite{boney1996digital, cox1997secure}. 
Audio watermarking embeds imperceptible data within audio signals, such as music, voice recordings, or any digital sound content. This hidden information remains undetectable by human listeners but can be identified by software, providing a method for tracking, verifying, and protecting audio content.

With the rise of generative AI models, concerns around copyright protection and deepfake audio detection have grown significantly. AI-generated music and deepfake voices pose new challenges for content security. For instance, in 2023, fake AI-generated songs mimicking popular artists such as Drake and The Weeknd circulated on social media, stirring debates over copyright infringement and the protection of intellectual property (IP) in the music industry~\cite{harvard2023ai}.
Furthermore, voice deepfakes have been used for impersonation and financial fraud. In one high-profile case, scammers used AI to generate a deepfake of a CEO’s voice to instruct an employee to transfer \$243,000 to a fraudulent account~\cite{fortune2019deepfake}. 
\begin{figure}[t]
    \centering
    \includegraphics[width=\linewidth]{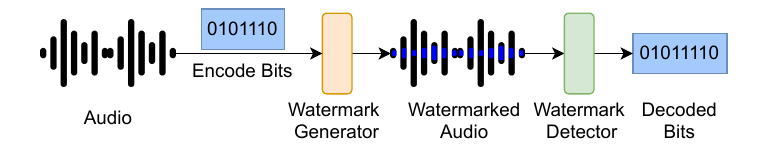}
    \vspace{-10pt}
    \caption{Audio watermarking scenario (detector $\rightarrow$ extractor)}
    \label{fig:cover}
\end{figure}
To address these concerns, researchers are turning to audio watermarking as a practical solution. When applied to AI-generated content, watermarking can help detect and defend against synthetic attacks, providing a means to verify the authenticity of the audio and distinguish it from human-created content. For human-created audio, watermarking serves to protect IP, ensuring that ownership is preserved and that unauthorized use or tampering can be detected. 


Fig.~\ref{fig:cover} demonstrates the process of audio watermarking.
First, a watermark generator encodes human-provided bits (e.g., ``0101110") into the audio signal, producing a watermarked audio. A watermark detector then extracts and decodes these bits from the audio, verifying the content’s authenticity. 
Although audio watermarking shows promise, its robustness in real-world applications hinges on overcoming core challenges. \emph{Robustness} represents the watermark's resilience to various modifications, such as compression, noise, format conversion, and other common manipulations, which allows it to be accurately extracted even under these conditions. If the watermark can be easily removed or degraded, its utility as a tool for IP protection and authentication diminishes.
Besides robustness, prior work~\cite{LI2021171, Lukas2021SoKHR} emphasized that any watermarking system also needs to keep a high \emph{fidelity}. In the case of AI-generated content, generative models may produce variations of audio that could inadvertently weaken or obscure the embedded watermark, making detection more difficult. Therefore, designing watermarks that are not only imperceptible but also durable under adversarial conditions is crucial for ensuring long-term effectiveness.

Although there are related works, \cite{hua2016twenty} surveyed audio watermarking schemes only up to 2016, analyzing 15 digital‐level attacks but providing no experimental evaluations. A more recent study \cite{liu2024audiomarkbench} surveyed and reproduced 3 watermarking schemes under 15 attack scenarios, each with 5 different settings, using 2 datasets limited to the speech domain. In this work, we conduct a systematic evaluation and propose a taxonomy for audio watermarking schemes and attacks. To comprehensively assess current methods, we reproduce \hanqing{9} recently proposed audio watermarking schemes (all with publicly available source code) and test them against \hanqing{22} distinct black‐box watermark‐removal attacks, each configured with over five parameter settings. We evaluate these schemes based on robustness and fidelity and categorize attacks into three groups: \emph{Signal‐Level Distortions}, \emph{Physical‐Level Distortions}, and \emph{AI‐Induced Distortions}. Our 3 test sets cover both speech and music.

\begin{itemize}
    \item In \emph{Signal-Level Distortions}, the attacker directly modifies the audio signal (e.g., through noise addition, compression, or format shifting) to erase or degrade the watermark.
    \item In \emph{Physical-Level Distortions}, the attacker manipulates the environmental sound during audio playback or re-recording to disrupt the watermark’s integrity under real-world device and environmental conditions.
    \item In \emph{AI-Induced Distortions}, the attacker employs generative models for style transfer or other transformations to extract the unwatermarked version of the audio.
\end{itemize}

As shown in Table~\ref{tab:cover}, while all of the investigated watermarks claimed robustness, they mainly focus on \emph{Signal-Level Distortion} (e.g., applying noise, quantization, format shifting, or compression). One scheme~\cite{liu2023detecting} evaluates the robustness in terms of \emph{AI-induced Distortion} but is restricted by limited distortion types. None of the evaluated watermarks demonstrate robustness against \emph{Physical-Level Distortion}; this raises the question of whether the existing watermark could guarantee applicability in real-life adversarial scenarios. For example, an attacker could play and re-record music with a watermark, facilitating music piracy. 

\begin{table}[t!]
\centering
\begin{tabular}{|l|c|c|c|}
\hline & \multicolumn{3}{c|}{Robustness Claims} \\ \hline
Watermark & Signal-Level & Physical-Level & AI-Induced \\ \hline
AudioSeal~\cite{san2024proactive} & 
\fullcircle    &  \emptycircle  & \emptycircle\\ \hline
Timbre~\cite{liu2023detecting} &
\fullcircle    &  \emptycircle  & \halfcircle \\ \hline
Wavmark~\cite{chen2023wavmark}  &
\fullcircle    &  \emptycircle  & \emptycircle\\ \hline
 SilentCipher~\cite{singh2024silentcipher} &
 \fullcircle    &  \emptycircle  & \emptycircle\\ \hline
 audiowmark~\cite{audiowmark}&
\halfcircle    &  \emptycircle  & \emptycircle\\ \hline
 RobustDNN~\cite{pavlovic2022robust} &
\halfcircle    &  \emptycircle  & \emptycircle    \\ \hline
Patchwork~\cite{natgunanathan2017patchwork}  &
\halfcircle    &  \emptycircle  & \emptycircle  \\ \hline
Norm-space~\cite{saadi2019novel}&
\fullcircle    &  \emptycircle  & \emptycircle \\ \hline
FSVC~\cite{zhao2021desynchronization} &
\fullcircle    &  \emptycircle  & \emptycircle \\ \hline
\end{tabular}
\vspace{5pt}
\caption{Robustness of Existing Audio Watermarking Techniques as Claimed in Literature, \halfcircle means the watermark claims partial robustness against such distortions; \fullcircle indicates that the watermark claims full robustness, and \emptycircle represents that the watermark does not claim robustness.}
\label{tab:cover}
\end{table}

\noindent\textbf{Our Contributions.} In this paper, we systematize the state-of-the-art research on audio watermarking and categorize them based on different technical principles. We focus on the potential limitations of the current audio watermark deployed in practical applications. Considering influential factors in real-world usage, we benchmark existing techniques and provide a summary of key observations, unresolved challenges, and future research directions. Specifically, we make the following contributions.
\begin{itemize}
    \item \textbf{We are the first to measure audio watermarks on a large scale}; Our work surveys 22 watermarks and reproduces 9 watermark approaches for 2 different watermarking tasks: watermarking speech and watermarking music. We test 22 attacks (including very time-consuming attacks such as training 35 different Voice Conversion (VC) and text-to-speech (TTS) models), and 109 configurations, involving attacks of more than 200,000 watermarked audios.
   
    \item \textbf{We systematically analyze state-of-the-art audio watermarking schemes and identify several vulnerabilities.}
    We divide existing audio watermarking methods into 2 main categories and 6 subcategories, summarizing fundamental techniques to craft audio watermarks. We identify the potential vulnerabilities associated with each approach based on the watermarking scheme. For instance, if a watermark is crafted using frequency-domain patterns, it may be vulnerable to the types of attacks that involve frequency-level manipulations.
   
    \item  \textbf{We propose and discover 8 new black-box attacks that effectively remove the watermark. We are the first to incorporate AI-induced distortions into watermark evaluations.} Our analysis reveals common vulnerabilities in existing audio watermarks, particularly in defending against using generative AI strategies to remove watermarks.

    \item \textbf{We focus on the quality of the modified audio during the attack process.} Unlike previous watermark removal techniques that focus on attack success without fully guaranteeing audio quality, our study presents a more practical scenario where attackers aim to remove the watermark while minimizing the loss of audio quality. 
    \item \textbf{We open-source our code and provide a demo at \url{https://sokaudiowm.github.io/}}

\end{itemize}

\noindent\textbf{Key Finding.} This work reveals that \emph{none of the existing audio watermarking techniques are fully reliable against all forms of attack.} While certain watermarks may withstand specific signal-level distortions, they fall short in providing robust IP protection against more complex, real-world scenarios, especially when facing physical distortions and evolving attacks driven by generative AI. This suggests that watermarking may not be a viable long-term IP protection strategy, as it cannot be modified or ``patched" once deployed. Once a watermark is publicly accessible, adversaries might exploit numerous pathways to compromise it, ultimately limiting its effectiveness in safeguarding IP.


\section{Background}\label{sec-background}\vspace{-10pt}
\subsection{Generative Audio Models}
Generative audio models are neural networks crafted to produce high-quality audio data that closely mimics human-generated sounds based on a provided training dataset. Formally, a generative model \(G\) maps low-dimensional latent vectors \(z\in \mathbb{Z}\) to the space of audio waveform signals \(\mathcal{X}\subseteq \mathbb{R}^T \), where \(T\) denotes the number of time samples in the waveform:
\[
G: \mathbb{Z} \mapsto \mathcal{X}.
\]
These models capture the underlying distribution of the training audio data, enabling the generation of synthetic audio samples that are statistically similar to real audio. Popular architectures for generative audio models include Variational Autoencoders (VAEs)~\cite{kingma2022autoencodingvariationalbayes}, Generative Adversarial Networks (GANs)~\cite{goodfellow2014generativeadversarialnetworks}, and diffusion models~\cite{ho2020denoising, rombach2022highresolutionimagesynthesislatent}. Notable implementations include RAVE~\cite{caillon2021ravevariationalautoencoderfast}, HiFi-GAN~\cite{kong2020hifigangenerativeadversarialnetworks}, and  AudioLDM~\cite{liu2023audioldmtexttoaudiogenerationlatent}. 
These models have a wide range of applications, including speech synthesis~\cite{shen2018natural}, music generation~\cite{dhariwal2020jukebox}, and voice cloning~\cite{arik2018neural}.
Training a generative model involves optimizing a loss function \( \mathcal{L} \) that quantifies the difference between the generated audio \( G(z) \) and the real audio data \(x\in\mathbb{X}\) it is trying to map to. This optimization is commonly performed using gradient descent methods~\cite{ruder2017overviewgradientdescentoptimization}.

\subsection{Audio Watermarking}
The primary purposes of audio watermarking are copyright protection, authentication, and provenance tracking of audio content~\cite{cox1997secure, Ren2024CopyrightPI}. Audio watermarking works by embedding a unique signal, or ``watermark", into an audio file in a way that is imperceptible to human listeners but detectable by specialized algorithms. Unlike metadata tagging, which can be easily stripped or lost during audio processing, watermarks are embedded directly within the audio waveform or spectral properties. This enhances their resilience to transformations. Typically, the watermark contains information that links the audio back to its source, such as a unique identifier, copyright details, or other provenance data~\cite{cox2002digital}.

Traditional watermarking techniques rely on deterministic signal processing to embed and retrieve watermarks in audio through time-domain, frequency-domain, or hybrid-domain methods. Time-domain methods~\cite{ERFANI2009809, Kanhe_2015, Hua_2015}, such as least significant bit (LSB) embedding~\cite{Chetan2021, Cvejic2004} and echo hiding~\cite{Wei2009}, directly inject watermarks based on audio amplitude or timing. While generally simpler and less complex, time-domain watermarking approaches often lack robustness~\cite{HUA2016222}. In contrast, frequency-domain watermarking techniques~\cite{zhao2021desynchronization, saadi2019novel, audiowmark} embed or modulate watermark information into selected frequency coefficients, providing enhanced resilience. Examples include Discrete Wavelet Transform (DWT)~\cite{Karajeh_2019}, Discrete Fourier Transform (DFT)~\cite{Singh2014}, and spread spectrum techniques~\cite{CVEJIC2004207}. Although more complex to implement, frequency-domain approaches typically offer greater robustness than time-domain techniques. Hybrid-domain watermarking schemes leverage the strengths of both time-domain and frequency-domain techniques for improved robustness, imperceptibility, and security. Saadi \textit{et al.}~\cite{saadi2019novel} performed a watermark embedding in the DWT-transformed signal, while Discrete Cosine Transform (DCT) was used for additional processing, ensuring both time-localized and frequency-specific manipulation. 

A newer approach, AI-based watermarking, offers a more adaptable and flexible solution to watermark design~\cite{san2024proactive, chen2023wavmark, liu2023detecting, LI2021171}. These watermarks leverage learned patterns and complex audio features, providing higher resilience against advanced AI-based detection and removal techniques than traditional methods. AudioSeal~\cite{san2024proactive} is a novel time-domain watermarking scheme and WavMark~\cite{chen2023wavmark} transforms the audio signal into a frequency domain by applying the Short-Time Fourier Transform (STFT). Additionally, hybrid-domain AI-based watermarking techniques such as Timbre Watermarking~\cite{liu2023detecting} and AudioQR~\cite{QU2023} utilize both the time and frequency features to generate an audio watermark.









\section{Audio Watermark Formalization}\label{sec-tax}
\begin{figure}[t]
\includegraphics[width=0.5\textwidth]{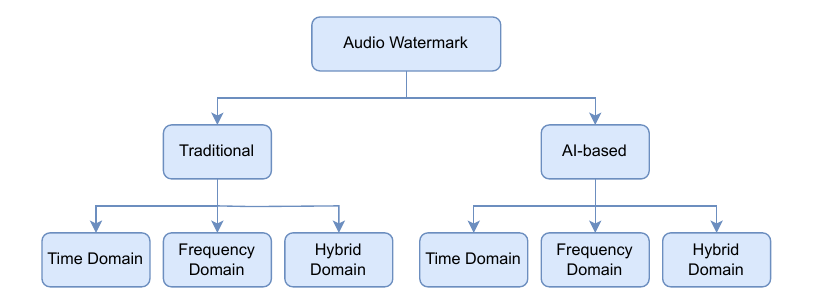}
\vspace{-10pt}
\caption{Watermark taxonomy.}
\label{fig:hierarchy}
\end{figure}

\vspace{-10pt}
To systematically evaluate the robustness of audio watermarking schemes in generative AI models, we propose a taxonomy in Fig.~\ref{fig:hierarchy} that categorizes these schemes based on how the watermark is embedded into the audio signal.


\subsection{Audio Watermark Definition}\label{subsec:3.1}
Audio watermarking involves embedding hidden information, typically an \( n \)-bit string, into audio signals in a way that remains imperceptible to listeners but can be algorithmically detected or extracted using a secret watermarking key.


An audio watermarking scheme consists of two primary functions:
1.~\textbf{Embedding Function (\( E \))}: Integrates a $N$ bit-string message \( w \in \{0, 1\}^N \) into an original audio signal \( x \in \mathcal{X} \), resulting in a watermarked audio signal \( x_w \in \mathcal{X} \):

\begin{equation}
    E(x, w) = x_w
\end{equation}

2.~\textbf{Extraction Function (\( \text{EXT} \))}: Attempts to detect or extract the watermark \( w \in \{0, 1\}^N \) from a unknown audio signal \( x' \in \mathcal{X} \):

\begin{equation}
    \begin{array}{l}
\text{For zero-bit:} \quad \textbf{EXT}(x') = 
\begin{cases} 
1 & \text{if watermark detected}, \\
0 & \text{otherwise}.
\end{cases} \\[10pt]
\text{For multi-bit:} \quad \textbf{EXT}(x') = w'
\end{array}
\end{equation}

where \( w' \in \{0, 1\}^N \) is the extracted watermark bit-string.
The embedding process aims to keep the watermark imperceptible, minimizing perceptual changes to the original audio. Meanwhile, the detection process must be robust enough to recover the watermark reliably, even after potential distortions or attacks.
The extraction accuracy, \( \textbf{Acc}(w,w') \), measures the proportion of matching bits between the original and extracted bit-strings. 
A watermark is considered successfully detected in audio if the extraction accuracy exceeds a predefined decision threshold, and otherwise it fails. A random guess bits will result in \textbf{Acc} around 0.5.

\subsection{Audio Watermark Purposes}
\begin{figure}[t]
\centering     
\subfigure[Model-level watermark]{\label{fig:mlw}\includegraphics[width=0.5\textwidth]{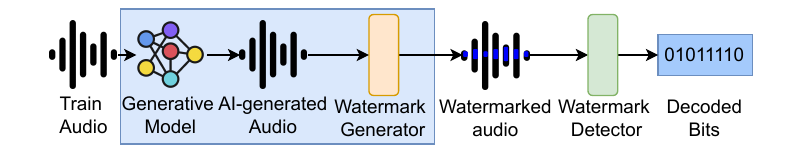}}
\subfigure[User-level watermark]{\label{fig:ulw}\includegraphics[width=0.5\textwidth]{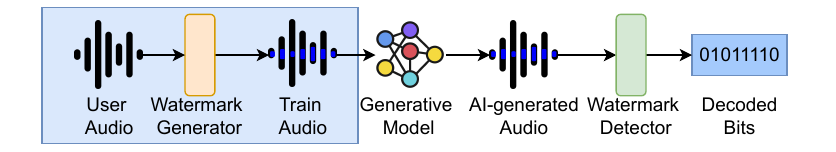}}
\vspace{-10pt}
\caption{Different watermark usage}
\end{figure}

When adopting watermarking into a generative AI scenario, there are two types of watermarking based on the pipeline. The first is \textbf{model-level watermarking} (Fig.~\ref{fig:mlw}), where a watermark generator is added right after the generative model. This ensures that AI-generated content directly contains the watermark, which can later be detected by the watermark detector. While this approach is robust, it requires that all generative models integrate an additional watermarking generator layer, which poses challenges in aligning with the tech giants in the field and may not be practical on a large scale.
The second approach is \textbf{user-level} watermarking (Fig.~\ref{fig:ulw}), where users embed watermarks into their audio files before sharing them online. If adversaries scrape this watermarked audio and use it as training data for a generative model, the generated content will inherit the watermark from the training data. This allows the watermark to be detected even in AI-generated outputs, providing a practical way for users to protect their IP without relying on modifications to generative models.

We classify audio watermarking schemes into two broad categories: traditional methods, which rely on classical signal processing techniques to embed watermarks, and AI-based methods, which utilize machine learning and generative models to create robust, adaptive watermarking techniques. More detailed information can be found in Section~\ref{sec-system}.

\subsection{Formalizing Watermark Requirements}\label{subsec:3.3}

\subsubsection{Robustness}
Robustness describes the ability of a watermark to endure and remain detectable even after the audio undergoes various distortions, transformations, and adversarial attacks. A robust watermark should withstand attacks such as signal processing (e.g., compression, filtering), physical distortions (e.g., re-recording), and, importantly, AI-driven transformations like VC and TTS models.

Formally, let \( x \in \mathcal{X} \) denote the original audio sample and \( x_w \in \mathcal{X} \) represent the watermarked version of \( x \), generated by an embedding function \( E \). For a given distortion function \( D \) (e.g., pitch shifting, time stretching, or VC), let \( x_w' = D(x_w) \) be the distorted watermarked audio sample. The extracted watermark \( E(x_w') = w' \) should match the original watermark \( w \) with high detection/extraction accuracy \( \textbf{Acc} \). Mathematically,
  \begin{equation}
      \textbf{Acc}(\textbf{EXT}(D(x_w), w) \geq \tau
  \end{equation}

where \(\tau\) is a predefined threshold indicating acceptable robustness; typically, this is close to 1.

\subsubsection{Fidelity}
Imperceptibility demands that the watermark remains inaudible to human listeners and does not noticeably degrade audio quality. A watermarked audio file should be indistinguishable from the original in terms of sound quality. To assess this, a perceptual quality metric function \( M \) measures perceived audio quality, ensuring that the watermark does not impact the listening experience or reveal the presence of the watermark.
Given an original audio sample \( x \) and its watermarked counterpart \( x_w \), the perceptual difference \(M(x) - M(x_w) = S(x, x_w) \) should not exceed a minimum threshold, where \( M \) represents a perceptual quality metric (e.g., PESQ – Perceptual Evaluation of Speech Quality, and ViSQOL – Virtual Speech Quality Objective Listener), and \( S \) evaluates perceived quality differences between \( x_w \) and \( x \). 


\subsection{Watermark Attacks}




Let \( x_w \) be the watermarked audio, and \( D \) represent a distortion function. The distorted audio \( D(x_w) \) is generated by applying \( D \) to the watermarked signal. \(P\) represents the probability of probability detection. The goal is to minimize the detectability of the watermark while maintaining perceptual similarity to the original audio. Mathematically:
  \begin{equation}
\min_{D}P(\textbf{EXT}(D(x_w)) = w),~\text{s.t}~S(D_{\text{signal}}(x_w), x)
  \end{equation}
To comprehensively evaluate watermarking robustness, attacks are tested across a range of distortion functions \( D \), each representing different types of adversarial methods (signal-level distortions, physical-level distortions, and AI-induced distortions).

\section{Threat model} \vspace{-10pt}

\subsection{Attacker Goals}
The attacker’s main objective is to remove the watermark from the audio while maintaining high audio quality (fidelity). A watermark removal attack is considered effective if the resulting audio shows low accuracy in recovering the original bits.

\subsection{Attacker Capabilities}
We assume the attacker has \textbf{zero knowledge of the watermarking schemes} and does not have access to the internal watermark generator or extractor. \emph{Specifically, the attacker is unaware of the watermarking scheme’s architecture, gradients, or output labels.} The attacker is assumed to possess a set of watermarked samples and, in some cases, a collection of clean, unwatermarked samples. The attacker cannot directly infer the watermarking model or train a substitute model to replicate the watermarking process. While the adversary is restricted in accessing or replicating the watermarking algorithms, we assume they possess basic knowledge of signal processing techniques and can apply standard transformations.




\subsection{Attack Scenario}
There are two distinct attack scenarios to consider. In the first scenario, the attacker possesses only watermarked samples. They apply various distortions to these samples, aiming to disrupt the watermark. Additionally, the attacker may use the watermarked samples to train AI models to simulate the target voice, expecting the models to generate new audio that inherently lacks the watermark. In the second scenario, the attacker has access to both watermarked and a few unwatermarked samples. By using the unwatermarked samples as a reference, the attacker can employ modification techniques, such as identifying characteristics that differentiate watermarked from unwatermarked samples, or leveraging VC models to mimic the style of unwatermarked audio, thereby attempting to remove the watermark. Throughout this process, the attacker aims not only to remove the watermark but also to ensure that the audio’s fidelity is preserved, maintaining a high-quality output.



\section{Audio Watermarking Taxonomy}\label{sec-system}\vspace{-10pt}
\subsection{Traditional Watermarking Methods}
Traditional watermarking methods use signal processing techniques to embed watermarks by directly altering the properties of the audio signal, typically leveraging human auditory perception to keep the watermark imperceptible. These approaches are well-established and widely adopted in the audio domain. Depending on the domain in which the watermark message is embedded, we categorize these methods into three types: \emph{time-domain}, \emph{frequency-domain}, and \emph{hybrid-domain} techniques.

\noindent{\textbf{Time-Domain Watermarking:}}
Time-domain audio watermarking schemes embed watermarks by directly modifying the amplitude or timing of the audio waveform in a way that is imperceptible to human listeners. Techniques like amplitude modulation subtly alter sample amplitudes to encode information, as demonstrated by Akira~\cite{Nishimura2014AudioWB}, who combined modulation masking with amplitude changes for secure and high-quality embedding. Echo hiding introduces imperceptible echoes with precise delays for decoding, while the time-spread echo method~\cite{Ko2005TimespreadEM, Yong_dual-channel} uses pseudo-noise sequences to minimize perceptual distortion and maintain decoding performance. Improvements by Natgunanathan and Xiang~\cite{Natgunanathan2009ANP} enhanced robustness and imperceptibility by introducing novel sequences and advanced decoding techniques.

\begin{figure*}[t]
\centering
\includegraphics[width=0.9\textwidth]{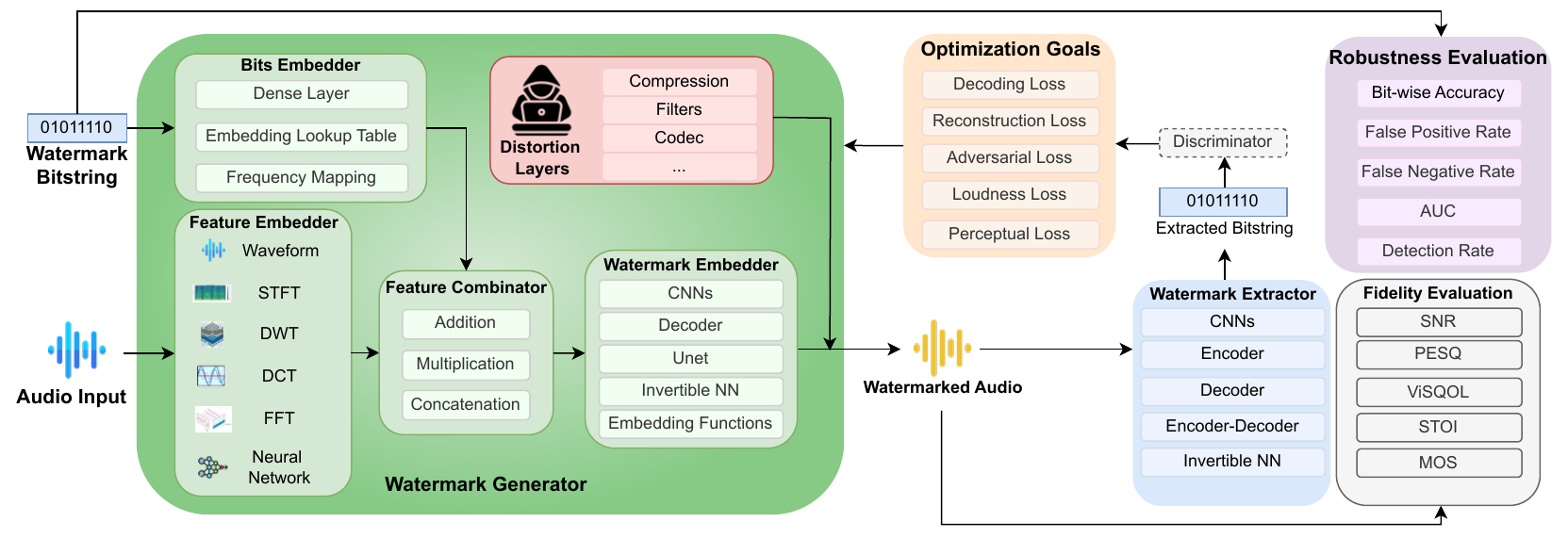}
\vspace{-10pt}
\caption{Watermark System Design}
\label{fig:systemdesign}
\end{figure*}

\noindent \textbf{Frequency-Domain Watermarking:} 
Frequency-domain methods embed watermarks by modifying the audio signal in the frequency spectrum using techniques such as the Fast Fourier Transform (FFT), Discrete Wavelet Transform (DWT)~\cite{Guofu+M-sequency}, spectral shaping, or Discrete Cosine Transform (DCT)~\cite{HU2014115, Lalitha_DCT, qiuling_adaptive}. Spectral shaping alters specific frequency bands, typically higher frequencies, to ensure imperceptibility, as demonstrated by Hu~\textit{et al.}~\cite{Hu2015TheUO}, who combined DWT with spectral shaping to enhance robustness and audio quality. Phase coding techniques embed information by modifying the phase spectrum, with Ngo and Unoki~\cite{Ngo2015} employing quantization index modulation, and Wang~\textit{et al.}~\cite{Wang2021} using phase shifting to exploit the uniform distribution of the phase spectrum. Yang~\cite{yang2024improvedphasecodingaudio} further refined phase coding by embedding data into dynamically segmented mid-frequency components, enhancing resistance to steganalysis, simplifying computation, and improving security.

\noindent \textbf{Hybrid Domain Watermarking:}
Hybrid domain techniques combine time and frequency information to enhance robustness and fidelity in audio watermarking. By leveraging multi-resolution analysis methods such as STFT and Discrete Wavelet Packet Transform (DWPT), these approaches enable precise and imperceptible watermark embedding. Techniques using STFT facilitate robust embedding through psychoacoustic guidance~\cite{Esmaili_2003}, while time-frequency plane methods optimize embedding in low-energy regions to improve resilience~\cite{zhang2020timefrequencyperspectiveaudiowatermarking}. Combining STFT with Singular Value Decomposition (SVD) further enhances robustness by embedding watermarks in adaptive key coefficients~\cite{Hamza_2005}. Hybrid techniques demonstrate strong resistance to attacks like pitch-shifting, time-scaling, resampling, and lossy compression, making them a promising---albeit computationally intensive---approach for imperceptible and robust watermarking.

\subsection{AI-Based Watermarking Methods}
AI-based watermarking methods utilize machine learning models, particularly deep neural networks (DNNs), to embed and extract watermarks with enhanced adaptability, robustness, and imperceptibility. These methods leverage the power of AI to learn and optimize watermarking schemes, making them resilient to advanced adversarial attacks and distortions. Depending on the domain of application, AI-based watermarking can be classified into \emph{time-domain}, \emph{frequency-domain}, and \emph{hybrid domain} approaches.

\noindent{\textbf{Time-Domain Watermarking:}}
In time-domain AI-based watermarking, neural networks are trained to embed and extract watermarks directly within the audio waveform. For example, the AudioSeal~\cite{san2024proactive} method introduces localized audio watermarking by embedding watermarks in the time domain, enabling robust detection after audio editing operations like time modification or audio effects. This technique ensures that the watermark remains resilient to common audio processing operations.

\noindent{\textbf{Frequency-Domain Watermarking:}}
In the realm of audio watermarking, frequency-domain approaches have been significantly enhanced by deep learning techniques, leading to improved robustness and imperceptibility. For example, DeAR~\cite{liu2023dear} not only transforms audios into the frequency domain by a differential DWT but incorporates frequency domain distortions into its training process. This makes it resilient to the challenges posed by real-world re-recording scenarios.

\noindent{\textbf{Hybrid Domain Watermarking:}}
Hybrid-domain AI-based watermarking combines time and frequency-domain information to exploit the strengths of both domains, enabling robust and imperceptible embedding. These methods leverage multi-resolution analysis, such as the STFT~\cite{chen2023wavmark, singh2024silentcipher, Moritz_2024, Chuxuan_enhance} or DCT~\cite{Maha_non-security}, to optimize the balance between watermark robustness and audio quality. Advanced models like IDEAW \cite{li2024ideaw} utilize invertible neural networks for dual-domain embedding, enhancing resilience to attacks such as compression, pitch-shifting, and resampling. By integrating deep learning techniques, hybrid-domain approaches achieve superior performance in robustness, capacity, and imperceptibility, representing a significant advancement over traditional methods for safeguarding audio content.

\subsection{Generalized Watermarking Pipeline}
Understanding the different watermark schemes, we generalize a comprehensive watermarking pipeline (Fig.~\ref{fig:systemdesign}). From the left, we list the watermark generator, detailed into five components. The \emph{Feature Embedder} extracts the feature from audio input, the \emph{Bits Embedder} encodes the bits into high dimension feature, the \emph{Feature Combinator} fuse the bits feature and input feature, and fed to \emph{Watermark Embedder}, which optimized by different \emph{Distortion Layers}, and generate watermarked Audio.

The watermark decoder employs neural network layers (e.g., convolutional layers) and inverse transformations (e.g., ISTFT, IDWT, IDCT) to reconstruct the watermarked audio and extract the embedded watermark. To ensure robustness, distortion layers simulate real-world audio distortions such as compression, noise addition, and filtering. As discussed in Section~\ref{subsec:3.1}, a watermark is considered successfully detected only if it can be accurately extracted, which is why most state-of-the-art AI-based watermarking algorithms include an additional neural network dedicated to this task. These generator and extractor networks are typically trained together (optionally with a discriminator) to minimize losses such as decoding, reconstruction, loudness, perceptual, and adversarial losses, all of which aim to improve watermark robustness and fidelity, as highlighted in Section~\ref{subsec:3.3}.

The robustness evaluation system assesses the watermark's ability to withstand distortions based on metrics such as bit-wise accuracy, FPR, and FNR. Additionally, the fidelity evaluation module measures the quality of the watermark using common audio quality indicators like Signal-to-Noise Ratio (SNR), PESQ, and ViSQOL. Together, these modules ensure that the watermarking scheme achieves a balance between robustness and audio fidelity.

\section{Removal attacks}\vspace{-10pt}
\subsection{Watermark removal attacks} 



\subsubsection{Signal-Level Distortions}
With the use of the term signal-level distortions, we refer to digital modifications, which were applied directly either to the waveform, or frequency spectrum properties of an audio file. Besides the conventional transformation, we introduce the listing of new distortions for evaluating watermarks.


\noindent\textbf{Pitch Shift:}
Changing the pitch of the audio without affecting its duration. This alters the frequency components of the signal, therefore making it hard for watermarking algorithms relying on certain features in frequency to detect this watermarked audio. Slight changes in pitch evade the detection of a watermark but may still keep the speech recognizable, hence it finds wide applications in attacks where intelligibility is to be preserved.

\noindent\textbf{Time Stretch:}
A process that modifies the audio duration without altering its pitch. Time-stretching destroys watermarks based on temporal patterns since it stretches or compresses the signal. Speeding up or slowing down the audio may render embedded numeric patterns unreadable to watermark detectors.

\noindent{\textbf{Cutting Audio:}} 
The removal of sections from the audio. This destroys temporal patterns relied on for watermark detection. Audio cropping is often applied in an attempt to cut out parts of the audio that may contain the watermark. In audio splicing, this is generally effective for most watermarks embedded in repeated patterns throughout the audio duration.

\noindent{\textbf{Sample Suppression:}} 
Randomly suppress or remove samples within the audio to create dropouts. This attack disrupts continuous watermark signals without affecting overall intelligibility. Sample suppression is effective in watermarking schemes relying on continuous signal properties because it introduces gaps in the watermark pattern.
For the complete signal distortion, refer to Appendix A.



\begin{figure*}[t]
    \centering
    \includegraphics[width=\linewidth]{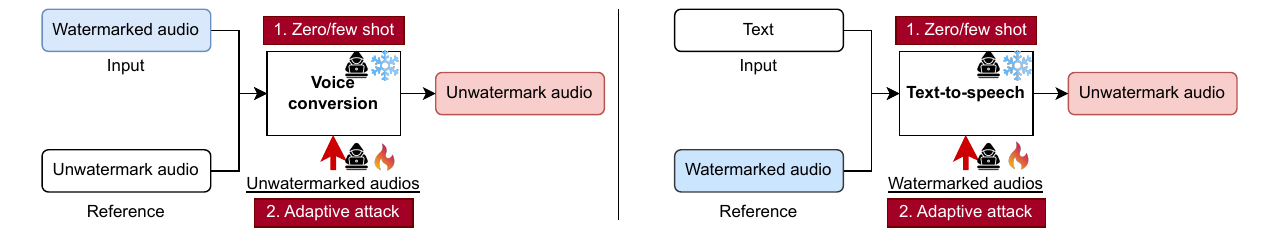}
    \vspace{-20pt}
    \caption{Zero/few shot attack and adaptive attack to introduce AI distortion}
    \label{fig:ai_sys}
\end{figure*}

\subsubsection{Physical-Level Distortions}
Physical-level distortions occur when audio is played back and re-recorded in real-world environments using varying distances and devices. These distortions introduce environmental artifacts, such as noise, reflections, and acoustic variations, that compromise the integrity of the watermark. By replicating practical audio usage scenarios, these attacks pose a significant challenge to the resilience of watermarking schemes.

\noindent{\textbf{Re-recording at Different Distances:}}
This attack involves playing watermarked audio through speakers and re-recording it using microphones placed at varying distances. Increased distance between the speaker and the microphone amplifies the impact of ambient noise, room acoustics, and signal degradation. Longer distances introduce more reverberation and weaken the fidelity of the signal, making it harder for the watermark to survive. Such attacks mimic real-world environments where audio is recorded from diverse setups, such as public spaces or large rooms, making them particularly effective in compromising watermark integrity.

\noindent{\textbf{Re-recording with Different Devices:}}
In this attack, watermarked audio is played and re-recorded using a variety of speaker and microphone configurations. Different devices have unique audio characteristics, such as frequency response, sensitivity, and distortion levels, which alter the signal in distinct ways. For example, re-recording with low-quality or mismatched devices can significantly degrade the watermark while preserving the overall perceptibility of the audio. This variability makes it challenging for watermarking schemes to retain embedded data consistently across diverse hardware environments.

\subsubsection{AI Distortions} AI-induced distortions use advanced VC and TTS models to transform or re-synthesize audio, effectively removing watermark features while preserving the original content and quality. Our experiments utilized VC and TTS models to evaluate the robustness of watermarking schemes against removal attacks. Each of them will have two attack settings: zero/few-shot setting and adaptive attack setting.

\noindent{\textbf{VC Models:}}
For VC model attacks, the input is watermarked audio, and it is assumed that the attacker has access to an unwatermarked reference sample. The goal of the VC model is to convert the watermarked audio into unwatermarked audio while preserving linguistic content. Various VC models are used to achieve this, including AdaIn-VC~\cite{chou2019oneshotvoiceconversionseparating}, FragmentVC~\cite{lin2021fragmentvcanytoanyvoiceconversion}, MediumVC~\cite{gu2021mediumvcanytoanyvoiceconversion}, YourTTS VC~\cite{casanova2023yourttszeroshotmultispeakertts}, and Retrieval-based Voice Conversion (RVC)~\cite{rvc_project}. AdaIn-VC uses adaptive instance normalization to adjust style and prosody, effectively removing original voice features. FragmentVC processes small fragments independently, disrupting watermark patterns. MediumVC provides lightweight conversion with minimal artifacts, modifying pitch, timbre, and pacing to evade watermark detection. YourTTS VC supports multilingual and multi-speaker conversions, enabling speaker features to be adjusted to bypass watermarks. RVC specializes in core feature manipulation, generating converted audio that leaves no traceable watermark remnants.

\noindent\textbf{TTS Models:}
In addition to VC models, we consider attackers using TTS models to remove watermarks. Unlike VC attacks, where attackers may have access to unwatermarked reference samples, TTS attacks assume the attacker has only watermarked audio samples. The attacker uses the watermarked audio along with arbitrary text content as input, aiming for the TTS model to generate unwatermarked audio output matching the arbitrary text. Alternatively, under the adaptive setting, the attacker uses watermarked audio as the label and its corresponding transcription as input to fine-tune the TTS model. By learning the audio characteristics of the watermarked samples during fine-tuning, the TTS model may remove watermark features while preserving only the acoustic characteristics. TTS models such as Tacotron2~\cite{shen2018natural}, FastSpeech2~\cite{ren2022fastspeech2fasthighquality}, and YourTTS~\cite{casanova2023yourttszeroshotmultispeakertts} generate entirely new audio directly from text, leaving no residual data from the original watermarked file. Tacotron2 synthesizes high-fidelity speech with natural prosody, ensuring watermark-free outputs. FastSpeech2 achieves similarly watermark-free results with high intelligibility and rapid synthesis speed, while YourTTS allows advanced voice attribute control and cross-lingual synthesis, producing audio that retains vocal characteristics while eliminating watermarks.

\begin{table*}[t]
\scalebox{0.65}{
\begin{tabular}{|l|c|ccccc|c|c|c|c|c|c|}
\hline
\multicolumn{1}{|c|}{} &  & \multicolumn{5}{c|}{\cellcolor{blue!20}\textbf{Watermark Generator}} &  &  &  &  &  & \cellcolor[HTML]{FFCCC9}\\ \cline{3-7}
\multicolumn{1}{|c|}{\multirow{-2}{*}{\textbf{\begin{tabular}[c]{@{}c@{}}Water\\ mark\end{tabular}}}} & \multirow{-2}{*}{\textbf{\begin{tabular}[c]{@{}c@{}}Model \\ Type\end{tabular}}} & \multicolumn{1}{c|}{\cellcolor{blue!20}\textbf{\begin{tabular}[c]{@{}c@{}}Feature \\ Embedder\end{tabular}}} & \multicolumn{1}{c|}{\cellcolor{blue!20}\textbf{\begin{tabular}[c]{@{}c@{}}Bits \\ Embedder\end{tabular}}} & \multicolumn{1}{c|}{\cellcolor{blue!20}\textbf{\begin{tabular}[c]{@{}c@{}}Feature \\ Combinator\end{tabular}}} & \multicolumn{1}{c|}{\cellcolor{blue!20}\textbf{\begin{tabular}[c]{@{}c@{}}Watermark \\ Embedder\end{tabular}}} & \cellcolor{blue!20}\textbf{\begin{tabular}[c]{@{}c@{}}Training \\ Distortion\end{tabular}} & \multirow{-2}{*}{\textbf{\begin{tabular}[c]{@{}c@{}}Watermark \\ Extractor\end{tabular}}} & \multirow{-2}{*}{\textbf{Loss Function}} & \multirow{-2}{*}{\textbf{Bits}} & \multirow{-2}{*}{\textbf{\begin{tabular}[c]{@{}c@{}}Robustness\\  Measure\end{tabular}}} & \multirow{-2}{*}{\textbf{Fidelity}} & \multirow{-2}{*}{\cellcolor[HTML]{FFCCC9}\textbf{Vulnerbility}} \\ \hline
~\cite{zhao2021desynchronization} & \begin{tabular}[c]{@{}c@{}}Trad\\ (Freq)\end{tabular} & \multicolumn{1}{c|}{DCT} & \multicolumn{1}{c|}{(N/A)} & \multicolumn{2}{c|}{\begin{tabular}[c]{@{}c@{}}Embed to \\ FSVC\end{tabular}} & (N/A) & Extract FSVC & (N/A) & Any & BER & DWR, ODG & \begin{tabular}[c]{@{}c@{}}Susceptible to \\ cropping and other \\ desync \\ attacks\end{tabular} \\ \hline
~\cite{natgunanathan2017patchwork} & \begin{tabular}[c]{@{}c@{}}Trad\\ (Freq)\end{tabular} & \multicolumn{1}{c|}{DCT} & \multicolumn{1}{c|}{(N/A)} & \multicolumn{2}{c|}{\begin{tabular}[c]{@{}c@{}}Embed to \\ DCT coefficient\end{tabular}} & (N/A) & Extract DCT & (N/A) & Any & DR & ODG & \begin{tabular}[c]{@{}c@{}}Susceptible to collusion \\ attacks\end{tabular} \\ \hline
~\cite{saadi2019novel} & \begin{tabular}[c]{@{}c@{}}Trad\\ (Freq)\end{tabular} & \multicolumn{1}{c|}{DWT} & \multicolumn{1}{c|}{(N/A)} & \multicolumn{2}{c|}{\begin{tabular}[c]{@{}c@{}}Embed to \\ DCT coefficient\end{tabular}} & (N/A) & Extract DCT & (N/A) & Any & BER,  NC & \begin{tabular}[c]{@{}c@{}}SNR, SSNR,\\  MOS\end{tabular} & \begin{tabular}[c]{@{}c@{}}Susceptible to \\ cropping and \\ other desync attacks\end{tabular} \\ \hline
~\cite{audiowmark} & \begin{tabular}[c]{@{}c@{}}Trad\\ (Freq)\end{tabular} & \multicolumn{1}{c|}{FFT} & \multicolumn{1}{c|}{(N/A)} & \multicolumn{2}{c|}{\begin{tabular}[c]{@{}c@{}}Embed to \\ frequency\end{tabular}} & (N/A) & \cellcolor[HTML]{FFFFFF}Extract Frequency & (N/A) & Up to 128 & BER & \begin{tabular}[c]{@{}c@{}}Informal \\ listening tests\end{tabular} & \begin{tabular}[c]{@{}c@{}}Not robust on\\ pitch-invariant \\ time-stretching\end{tabular} \\ \hline
~\cite{Liu_2019} & \begin{tabular}[c]{@{}c@{}}Trad\\ (Freq)\end{tabular} & \multicolumn{1}{c|}{DCT} & \multicolumn{1}{c|}{(N/A)} & \multicolumn{2}{c|}{\begin{tabular}[c]{@{}c@{}}Embed to \\ FDLM\end{tabular}} & (N/A) & Extract FDLM & (N/A) & Any & BER & \begin{tabular}[c]{@{}c@{}}SNR, ODG, \\ SDG\end{tabular} & \begin{tabular}[c]{@{}c@{}}Susceptible to desync \\ attacks that \\ affect lengths of \\ different frames\end{tabular} \\ \hline
~\cite{Chetan2021} & \begin{tabular}[c]{@{}c@{}}Trad\\ (Time)\end{tabular} & \multicolumn{1}{c|}{(N/A)} & \multicolumn{1}{c|}{(N/A)} & \multicolumn{2}{c|}{\begin{tabular}[c]{@{}c@{}}Embed to \\ LSB plane\end{tabular}} & (N/A) & Extract LSB & (N/A) & Any &\begin{tabular}[c]{@{}c@{}} (N/A) \end{tabular} & \begin{tabular}[c]{@{}c@{}}PESQ, PEAQ, \\ SNR\end{tabular} & \begin{tabular}[c]{@{}c@{}}LSB watermarks \\ can be easily removed\end{tabular} \\ \hline
~\cite{Wang2021} & \begin{tabular}[c]{@{}c@{}}Trad\\ (Freq)\end{tabular} & \multicolumn{1}{c|}{FFT} & \multicolumn{1}{c|}{(N/A)} & \multicolumn{2}{c|}{Shift phase} & (N/A) & Extract phase shift & (N/A) & Any & BDR & LSD, ODG & \begin{tabular}[c]{@{}c@{}}Less robust with longer \\ watermark bitstrings\end{tabular} \\ \hline
~\cite{yang2024improvedphasecodingaudio} & \begin{tabular}[c]{@{}c@{}}Trad\\ (Freq)\end{tabular} & \multicolumn{1}{c|}{FFT} & \multicolumn{1}{c|}{(N/A)} & \multicolumn{2}{c|}{\begin{tabular}[c]{@{}c@{}}Embed to \\ frequency\end{tabular}} & (N/A) & \cellcolor[HTML]{FFFFFF}Extract Frequency & (N/A) & Any & BER & \begin{tabular}[c]{@{}c@{}}Phase\\  Comparison\end{tabular} & \begin{tabular}[c]{@{}c@{}}Not tested against any\\  watermark attacks\end{tabular} \\ \hline
~\cite{Kanhe_2015} & \begin{tabular}[c]{@{}c@{}}Trad\\ (Time)\end{tabular} & \multicolumn{1}{c|}{(N/A)} & \multicolumn{1}{c|}{\begin{tabular}[c]{@{}c@{}}AES \\ encryption\end{tabular}} & \multicolumn{2}{c|}{\begin{tabular}[c]{@{}c@{}}Embed to \\ LSB plane\end{tabular}} & (N/A) & Extract LSB & (N/A) & 16 & \begin{tabular}[c]{@{}c@{}}Correlation\\ Coefficient \end{tabular} & \begin{tabular}[c]{@{}c@{}}SNR,\\ Listener Test\end{tabular} & \begin{tabular}[c]{@{}c@{}}LSB watermarks can \\ be easily removed\end{tabular} \\ \hline
~\cite{Karajeh_2019} & \begin{tabular}[c]{@{}c@{}}Trad\\ (Freq)\end{tabular} & \multicolumn{1}{c|}{DWT} & \multicolumn{1}{c|}{(N/A)} & \multicolumn{2}{c|}{\begin{tabular}[c]{@{}c@{}}Embed to Schur \\ decomposition\end{tabular}} & (N/A) & \begin{tabular}[c]{@{}c@{}}Extract \\ decomposition\end{tabular} & (N/A) & Any & BER,  NC & \begin{tabular}[c]{@{}c@{}}SNR, ODG, \\ SDG\end{tabular} & \begin{tabular}[c]{@{}c@{}}Not tested against \\ advanced \\ desync attacks\end{tabular} \\ \hline

~\cite{qiuling_adaptive} & \begin{tabular}[c]{@{}c@{}}Trad\\ (Freq)\end{tabular} & \multicolumn{1}{c|}{DWT} & \multicolumn{1}{c|}{(N/A)} & \multicolumn{2}{c|}{\begin{tabular}[c]{@{}c@{}}Embed to DCT \\ coefficient\end{tabular}} & (N/A) & \begin{tabular}[c]{@{}c@{}}Extract \\ SVD\end{tabular} & (N/A) & Any & \begin{tabular}[c]{@{}c@{}}BER, Correlation\\ Coefficient\end{tabular} & \begin{tabular}[c]{@{}c@{}}SNR, ODG\end{tabular} & \begin{tabular}[c]{@{}c@{}}Susceptible to \\ desync attacks \\(pitch/time-scale \\ modification) \end{tabular} \\ \hline

~\cite{Yong_dual-channel} & \begin{tabular}[c]{@{}c@{}}Trad\\ (Freq)\end{tabular} & \multicolumn{1}{c|}{DWT} & \multicolumn{1}{c|}{(N/A)} & \multicolumn{2}{c|}{\begin{tabular}[c]{@{}c@{}}Embed to SVD \\ coefficient\end{tabular}} & (N/A) & \begin{tabular}[c]{@{}c@{}}Extract \\ DCT\end{tabular} & (N/A) & Any & BER, NC & \begin{tabular}[c]{@{}c@{}}SNR\end{tabular} & \begin{tabular}[c]{@{}c@{}}Susceptible to \\ high-pass filter\end{tabular} \\ \hline

~\cite{Guofu+M-sequency} & \begin{tabular}[c]{@{}c@{}}Trad\\ (Time)\end{tabular} & \multicolumn{1}{c|}{(N/A)} & \multicolumn{1}{c|}{Add Echo} & \multicolumn{2}{c|}{\begin{tabular}[c]{@{}c@{}}Embed to Subsignals \end{tabular}} & (N/A) & \begin{tabular}[c]{@{}c@{}}Extract \\ FFT\end{tabular} & (N/A) & Any & Detection Rate & \begin{tabular}[c]{@{}c@{}}SNR\end{tabular} & \begin{tabular}[c]{@{}c@{}}Susceptible to desync and\\ cropping attacks\end{tabular} \\ \hline

~\cite{liu2023detecting} & \begin{tabular}[c]{@{}c@{}}AI\\ (Hybrid)\end{tabular} & \multicolumn{1}{c|}{STFT} & \multicolumn{1}{c|}{\cellcolor[HTML]{FFFFFF}\begin{tabular}[c]{@{}c@{}}Linear \\ Layer +\\ Frequency\end{tabular}} & \multicolumn{1}{c|}{\begin{tabular}[c]{@{}c@{}}Concate-\\ -nation\end{tabular}} & \multicolumn{1}{c|}{2D CNN} & \begin{tabular}[c]{@{}c@{}}ISTFT, \\ Normalization, \\ Mel-transform, \\ GriffinLim\end{tabular} & \cellcolor[HTML]{FFFFFF}\begin{tabular}[c]{@{}c@{}}Decoder\\ (CNNs)\end{tabular} & \begin{tabular}[c]{@{}c@{}}Decoding Loss, \\ Reconstruction\\  Loss, \\ Adversarial Loss\end{tabular} & 10 & BRA & \begin{tabular}[c]{@{}c@{}}SNR, PESQ,\\  SECS, MOS\end{tabular} & \begin{tabular}[c]{@{}c@{}}Not trained on \\ non-English\\  or singing data\end{tabular} \\ \hline
~\cite{san2024proactive} & \begin{tabular}[c]{@{}c@{}}AI\\ (Time)\end{tabular} & \multicolumn{1}{c|}{CNN} & \multicolumn{1}{c|}{\begin{tabular}[c]{@{}c@{}}Embedding \\ Lookup \\ Table\end{tabular}} & \multicolumn{1}{c|}{Addition} & \multicolumn{1}{c|}{\begin{tabular}[c]{@{}c@{}}Decoder \\ (CNNs + \\ LSTM)\end{tabular}} & \begin{tabular}[c]{@{}c@{}}BF, HF, LF, \\ Speed, Resample, \\ BA, DA, \\ Echo, PN, WN, \\ Smooth, \\ ACC, MP3, \\ EnCodec\end{tabular} & \cellcolor[HTML]{FFFFFF}\begin{tabular}[c]{@{}c@{}}Encoder-Decoder\\ (CNNs + LSTM)\end{tabular} & \begin{tabular}[c]{@{}c@{}}Decoding Loss, \\ Loudness Loss, \\ Perceptual Loss, \\ Detection Loss\end{tabular} & 16 & \begin{tabular}[c]{@{}c@{}}Accuracy, \\ TPR, \\ FPR, AUC\end{tabular} & \begin{tabular}[c]{@{}c@{}}PESQ ViSQOL, \\ STOI, SO-SNR,\\  MUSHRA\end{tabular} & \begin{tabular}[c]{@{}c@{}}Susceptible to \\ cropping attack\end{tabular} \\ \hline
~\cite{chen2023wavmark} & \begin{tabular}[c]{@{}c@{}}AI\\ (Hybrid)\end{tabular} & \multicolumn{1}{c|}{STFT} & \multicolumn{1}{c|}{\begin{tabular}[c]{@{}c@{}}Linear \\ Layer +\\ STFT\end{tabular}} & \multicolumn{1}{c|}{\begin{tabular}[c]{@{}c@{}}Concate-\\ -nation \\ and Addition\end{tabular}} & \multicolumn{1}{c|}{\begin{tabular}[c]{@{}c@{}}Invertible \\ NN\end{tabular}} & \begin{tabular}[c]{@{}c@{}}UN, SS, LF, \\ Resample, AS, \\ Compression, \\ Quantization, \\ Echo, TS\end{tabular} & \cellcolor[HTML]{FFFFFF}Invertible NN & \begin{tabular}[c]{@{}c@{}}Decoding Loss, \\ Perceptual Loss, \\ Adversarial Loss\end{tabular} & 16 & BER & PESQ, SNR & \begin{tabular}[c]{@{}c@{}}Only works with\\  audio \textgreater{}= 1\\  second in length, \\ only trained against \\ subtle attacks, \\ slow brute-force detection\end{tabular} \\ \hline
~\cite{singh2024silentcipher} & \begin{tabular}[c]{@{}c@{}}AI\\ (Hybrid)\end{tabular} & \multicolumn{1}{c|}{STFT} & \multicolumn{1}{c|}{\begin{tabular}[c]{@{}c@{}}Linear \\ Layer +\\  STFT\end{tabular}} & \multicolumn{1}{c|}{\begin{tabular}[c]{@{}c@{}}Concate-\\ -nation\end{tabular}} & \multicolumn{1}{c|}{\begin{tabular}[c]{@{}c@{}}Decoder \\ (CNNs + \\ FC)\end{tabular}} & \begin{tabular}[c]{@{}c@{}}WN, TJ, RE, \\ MP3, OGG, \\ AAC\end{tabular} & \cellcolor[HTML]{FFFFFF}\begin{tabular}[c]{@{}c@{}}Decoder \\ (CNNs + FC)\end{tabular} & Decoding Loss & 32/40 & Accuracy & \begin{tabular}[c]{@{}c@{}}Mean \\ Inaudibility\end{tabular} & \begin{tabular}[c]{@{}c@{}}Trained against only a \\ few minor distortion attacks, \\ robustness against \\ cropping requires \\ slow brute-force detection\end{tabular} \\ \hline
~\cite{pavlovic2022robust} & \begin{tabular}[c]{@{}c@{}}AI\\ (Hybrid)\end{tabular} & \multicolumn{1}{c|}{STFT} & \multicolumn{1}{c|}{\begin{tabular}[c]{@{}c@{}}Frequency \\ Mapping\end{tabular}} & \multicolumn{1}{c|}{\begin{tabular}[c]{@{}c@{}}Concate-\\ -nation\end{tabular}} & \multicolumn{1}{c|}{Unet} & UN, LF, SS & \cellcolor[HTML]{FFFFFF}\begin{tabular}[c]{@{}c@{}}Encoder\\ (CNNs + FC)\end{tabular} & \begin{tabular}[c]{@{}c@{}}Decoding Loss, \\ Perceptual Loss\end{tabular} & 512 & BRA & PESQ, SNR & \begin{tabular}[c]{@{}c@{}}Susceptible to \\ false positives, \\ vulnerable against \\ desync attacks\end{tabular} \\ \hline
~\cite{liu2023dear} & \begin{tabular}[c]{@{}c@{}}AI\\ (Freq)\end{tabular} & \multicolumn{1}{c|}{DWT} & \multicolumn{1}{c|}{CNN} & \multicolumn{1}{c|}{\begin{tabular}[c]{@{}c@{}}Embed to \\ DWT \\ coefficient\end{tabular}} & \multicolumn{1}{c|}{\begin{tabular}[c]{@{}c@{}}Encoder \\ (CNNs + \\ ResNet)\end{tabular}} & \begin{tabular}[c]{@{}c@{}}IR augmentation, \\ band-pass \\ filtering, \\ Gaussian noise\end{tabular} & \cellcolor[HTML]{FFFFFF}\begin{tabular}[c]{@{}c@{}}Decoder\\  (CNNs + ResNet)\end{tabular} & \begin{tabular}[c]{@{}c@{}}Encoding Loss, \\ Adversarial Loss,\\  Decoding Loss\end{tabular} & 100 & BRA & SNR & \begin{tabular}[c]{@{}c@{}}Not trained against \\ desync attacks\end{tabular} \\ \hline
~\cite{QU2023} & \begin{tabular}[c]{@{}c@{}}AI\\ (Hybrid)\end{tabular} & \multicolumn{1}{c|}{CNN} & \multicolumn{1}{c|}{\begin{tabular}[c]{@{}c@{}}Linear \\ Layers\end{tabular}} & \multicolumn{1}{c|}{\begin{tabular}[c]{@{}c@{}}Concate-\\ -nation\end{tabular}} & \multicolumn{1}{c|}{\begin{tabular}[c]{@{}c@{}}Encoder\\  (1D CNNs + \\ FC)\end{tabular}} & \begin{tabular}[c]{@{}c@{}}BG noise, \\ IR augmentation, \\ BG music, \\ Gaussian noise\end{tabular} & \cellcolor[HTML]{FFFFFF}\begin{tabular}[c]{@{}c@{}}Decoder\\  (CNNs)\end{tabular} & \begin{tabular}[c]{@{}c@{}}Decoding Losses, \\ Perceptual Losses\end{tabular} & 50 & BER/BRA & SNR & \begin{tabular}[c]{@{}c@{}}Trained against very\\  few basic attacks\end{tabular} \\ \hline

~\cite{Maha_non-security} & \begin{tabular}[c]{@{}c@{}}AI\\ (Hybrid)\end{tabular} & \multicolumn{1}{c|}{DCT} & \multicolumn{1}{c|}{\begin{tabular}[c]{@{}c@{}}Linear \\ Layers\end{tabular}} & \multicolumn{2}{c|}{\begin{tabular}[c]{@{}c@{}}Embed to \\ DCT \\ coefficient\end{tabular}} & \begin{tabular}[c]{@{}c@{}}(N/A)\end{tabular} & \cellcolor[HTML]{FFFFFF}\begin{tabular}[c]{@{}c@{}}Decoder\\  (FC)\end{tabular} & \begin{tabular}[c]{@{}c@{}}MSE\end{tabular} & Any & BER, NC & SNR, ODG & \begin{tabular}[c]{@{}c@{}}Depend on \\Network prediction,\\ Use single coefficient \\ per frame for each bit, \\ Susceptible to \\ desync attacks  \end{tabular} \\ \hline

~\cite{Chuxuan_enhance} & \begin{tabular}[c]{@{}c@{}}AI\\ (Hybrid)\end{tabular} & \multicolumn{1}{c|}{STFT} & \multicolumn{1}{c|}{\begin{tabular}[c]{@{}c@{}}CNN \end{tabular}} & \multicolumn{1}{c|}{\begin{tabular}[c]{@{}c@{}}Concate-\\ -nation\end{tabular}} & \multicolumn{1}{c|}{\begin{tabular}[c]{@{}c@{}}Encoder\\  (2D CNNs + \\ Transformer)\end{tabular}} & \begin{tabular}[c]{@{}c@{}}Perturbation \\ generator\end{tabular} & \cellcolor[HTML]{FFFFFF}\begin{tabular}[c]{@{}c@{}}Decoder\\  (CNNs)\end{tabular} & \begin{tabular}[c]{@{}c@{}}Decoding Loss, \\ Perturbation Loss, \\ Spectrogram Loss, \\ Watermark Loss\end{tabular} & Any & BER, MSE & \begin{tabular}[c]{@{}c@{}}Document-\\to-\\Watermark \\ Ratio, PESQ\end{tabular} & \begin{tabular}[c]{@{}c@{}}Trained against perturbation \\without attacks, \\ Distortions not compared \\with SOTA methods\end{tabular} \\ \hline

\end{tabular}
}
\caption{Summary of Existing Audio Watermarks (The abbreviations are explained in Appendix F)}
\label{tab:sysdesign}
\end{table*}

\subsection{Watermark Vulnerability} 
Table~\ref{tab:sysdesign} summarizes the components, processes, and vulnerabilities of traditional and AI-based watermarking schemes. Traditional methods typically combine feature and watermark embedding into coefficients like LSB, DCT, or DWT, while AI-based systems separate these processes, using neural networks and advanced architectures like ResNet and INN. Traditional methods are vulnerable to basic attacks like noise addition and desynchronization, while frequency-domain methods face synchronization and collusion issues. AI-based schemes improve resistance to noise, compression, and filtering but remain prone to desynchronization, false positives, and overfitting to domain-specific training data, limiting generalization. Both approaches struggle with multi-attack resilience, with AI methods further challenged by fidelity loss when embedding large data. Desynchronization and poor generalization emerge as the most critical vulnerabilities, underscoring the need for adaptive strategies and robust evaluation metrics. 

\vspace{-10pt}
\section{Experiment}\label{sec-exp}\vspace{-10pt}
\subsection{Metric Settings}
\vspace{-10pt}

\noindent\textbf{Bit Recovery Accuracy (ACC):}
ACC measures the proportion of correctly recovered watermark bits compared to the original. Higher ACC indicates greater robustness in retaining encoded information under various attacks.

\noindent\textbf{Watermark Probability:}
This metric indicates the likelihood that an audio sample is watermarked, with a probability of 1 assigned to detected watermarks and 0 otherwise. It reflects watermark presence even if bit accuracy is low, ensuring consistency across scenarios.

\noindent\textbf{Watermark Fidelity:}
Watermark fidelity assesses the quality of watermarked audio compared to the original, ensuring minimal perceptual impact. High fidelity preserves the natural audio quality and intelligibility.

\noindent\textbf{Attack Fidelity:}
Attack fidelity evaluates the quality of audio after watermark removal attacks, with higher fidelity indicating minimal degradation. Metrics like SNR, PESQ, and ViSQOL measure the usability of attacked audio in practical scenarios.

\noindent\textbf{Fidelity-Tolerant Robustness:}
This metric measures watermark robustness under attacks that maintain a specified level of audio quality. It ensures that watermark removal efforts preserve usable audio fidelity in real-world conditions.

\subsection{Experiment setting}
\vspace{-10pt}
\noindent\textbf{Setup:}
We reproduced 9 audio watermarking schemes, including Wavmark~\cite{chen2023wavmark}, AudioSeal~\cite{san2024proactive}, Timbre~\cite{liu2023detecting}, RobustDNN~\cite{pavlovic2022robust}, audiowmark~\cite{audiowmark}, Norm-space~\cite{saadi2019novel}, Patchwork~\cite{natgunanathan2017patchwork}, FSVC~\cite{zhao2021desynchronization}, SilentCipher~\cite{singh2024silentcipher}. We introduce 22 removal attacks (109 configurations) using our custom-built Watermark-Robustness-Framework, developed with PyTorch as the backend. 
All experiment is conducted using a server equipped with NVIDIA RTX A6000 GPUs.

\noindent\textbf{Dataset:} We evaluate the robustness of watermarking schemes using 3 datasets: LJSpeech~\cite{ljspeech17},  LibriSpeech~\cite{panayotov2015librispeech} and, M4Singer~\cite{zhang2022msinger}. LJSpeech is a publicly available speech dataset containing 13,100 short audio clips from a single speaker reading passages from books; we randomly sampled 2,000 clips for our experiments. LibriSpeech is a large-scale open-source corpus of read English speech derived from audiobooks, from which we randomly selected 200 audio clips. In contrast, M4Singer is a large-scale dataset of singing voice recordings designed for singing voice synthesis, featuring diverse vocal styles and genres; we sampled 2,000 music clips to evaluate robustness in musical contexts. Together, these datasets enable a comprehensive assessment of watermark resilience across both speech and singing domains.

\noindent\textbf{Robustness Evaluation Pipeline:} 
Our evaluation pipeline begins by getting clean samples from each dataset. We then embed watermarks into these samples using various watermarking schemes. Following this, we apply different types of watermark removal attacks, including signal-level, physical distortion, and AI-induced attacks. At each step of the pipeline, we measure and report metrics such as fidelity, which indicates audio quality, and accuracy, which reflects the success of watermark encoding and decoding. 


\subsection{Watermark Reproduction}
Before evaluating the robustness of selected 9 watermark schemes, we reproduce them and report the benign performance of using them to protect audio. Specifically, we measure the \textbf{Watermark Fidelity}, \textbf{Bit Recovery Accuracy}, and \textbf{Watermark Probability}. 
Due to the length limit,
we report the result in Appendix B. For quick results, all evaluated watermark performs well in terms of fidelity, accuracy, and probability. The only concern is some watermarks(e.g., Timbre Watermark) do not work well on the Music dataset.

\begin{figure*}[t]
\centering
\subfigure[Pitch Shift]{\label{fig:Pitch_Shift}\includegraphics[width=42mm]{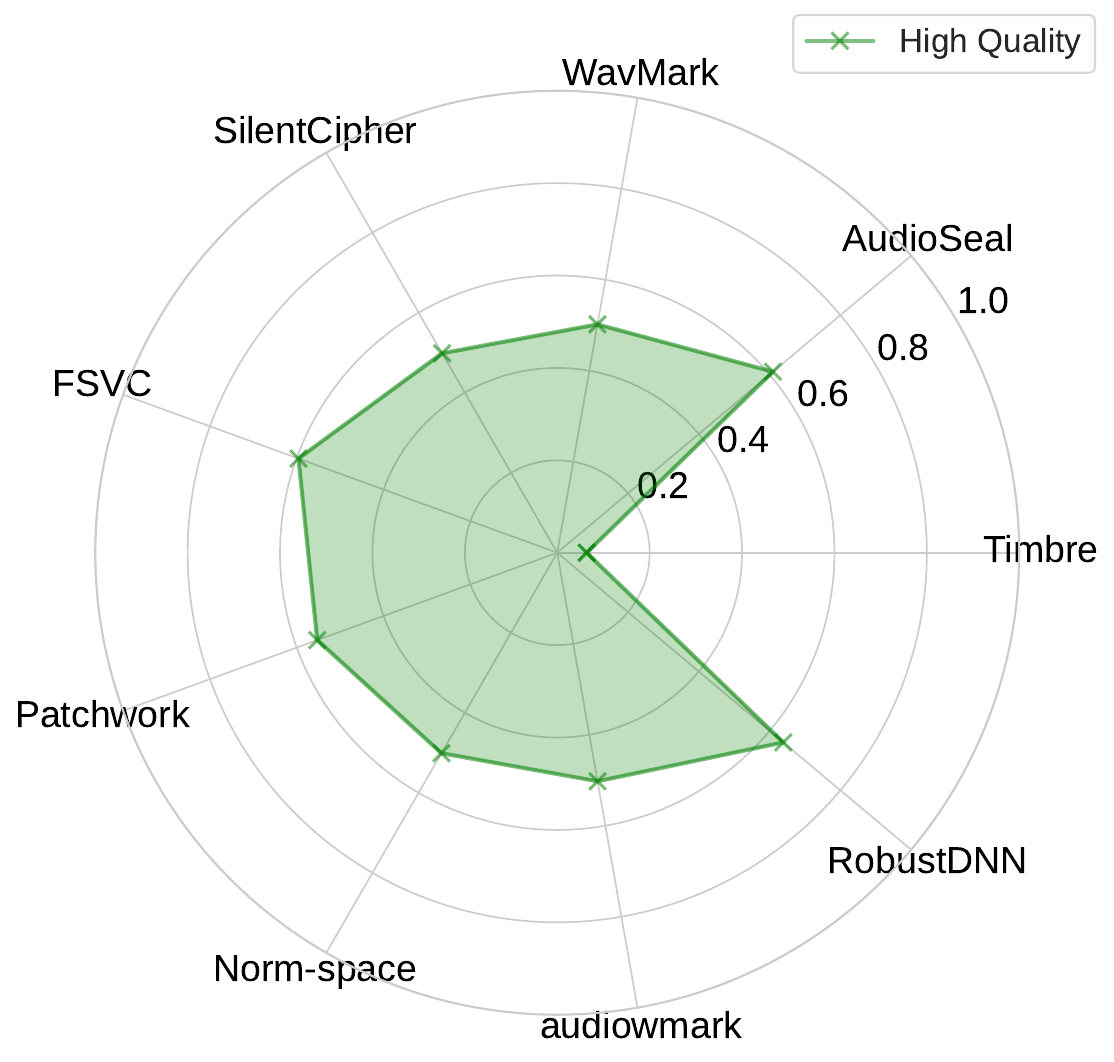}}
\subfigure[Time Stretch]{\label{fig:Time_Stretch}\includegraphics[width=42mm]{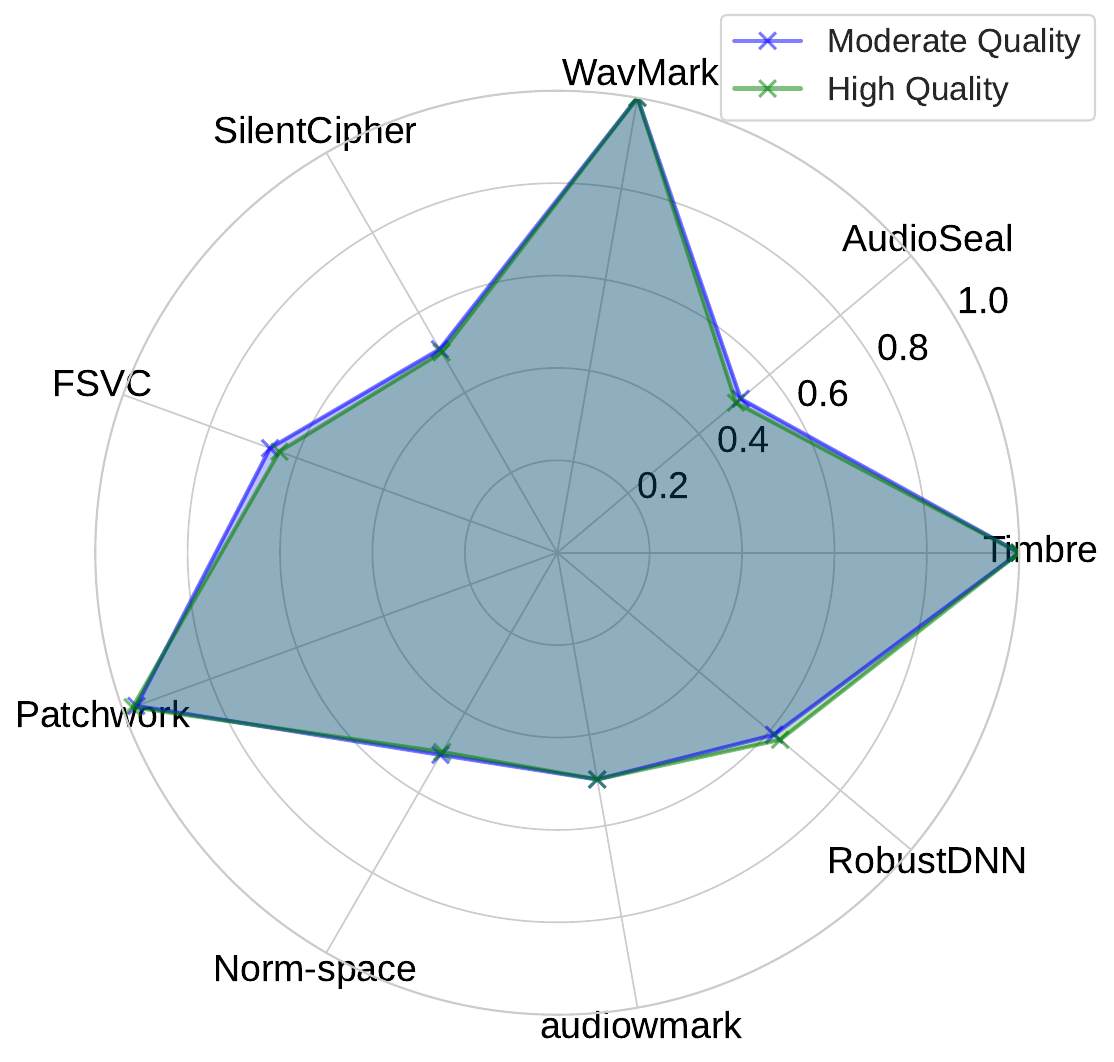}}
\subfigure[Gaussian Noise]{\label{fig:Gaussian_Noise}\includegraphics[width=42mm]{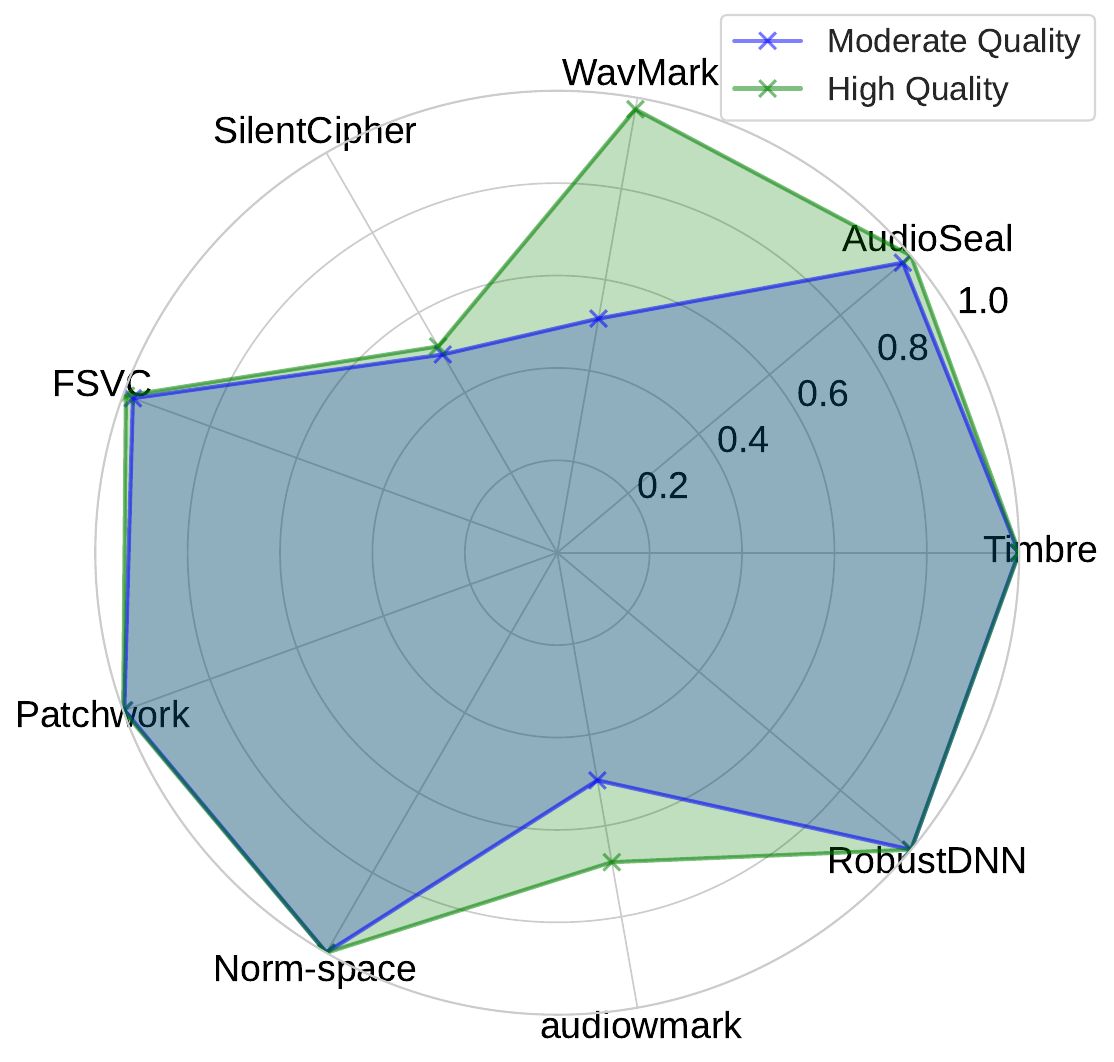}}
\subfigure[Bitcrush]{\label{fig:Bitcrush}\includegraphics[width=42mm]{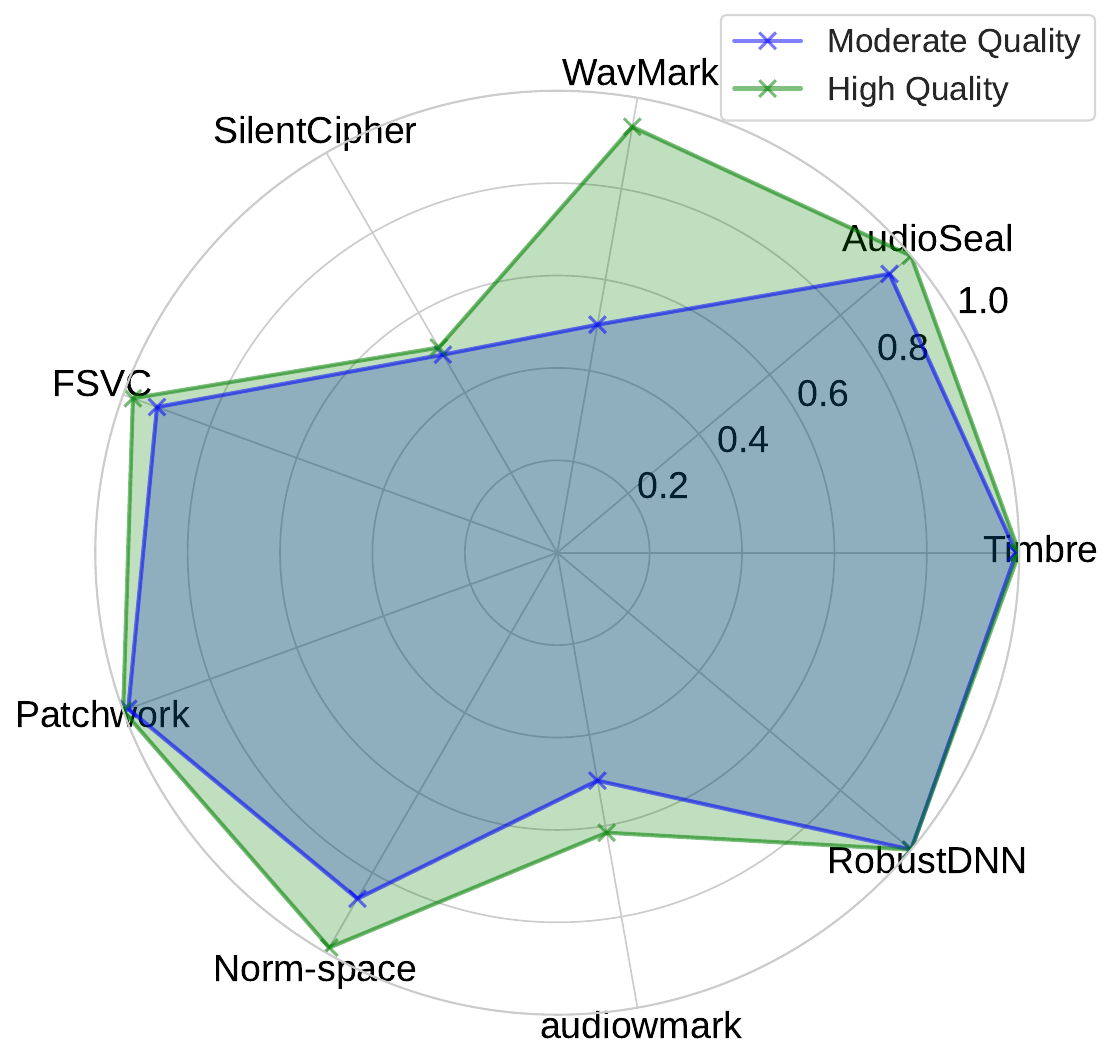}}
\subfigure[MP3 Compression]{\label{fig:MP3_Compression}\includegraphics[width=42mm]{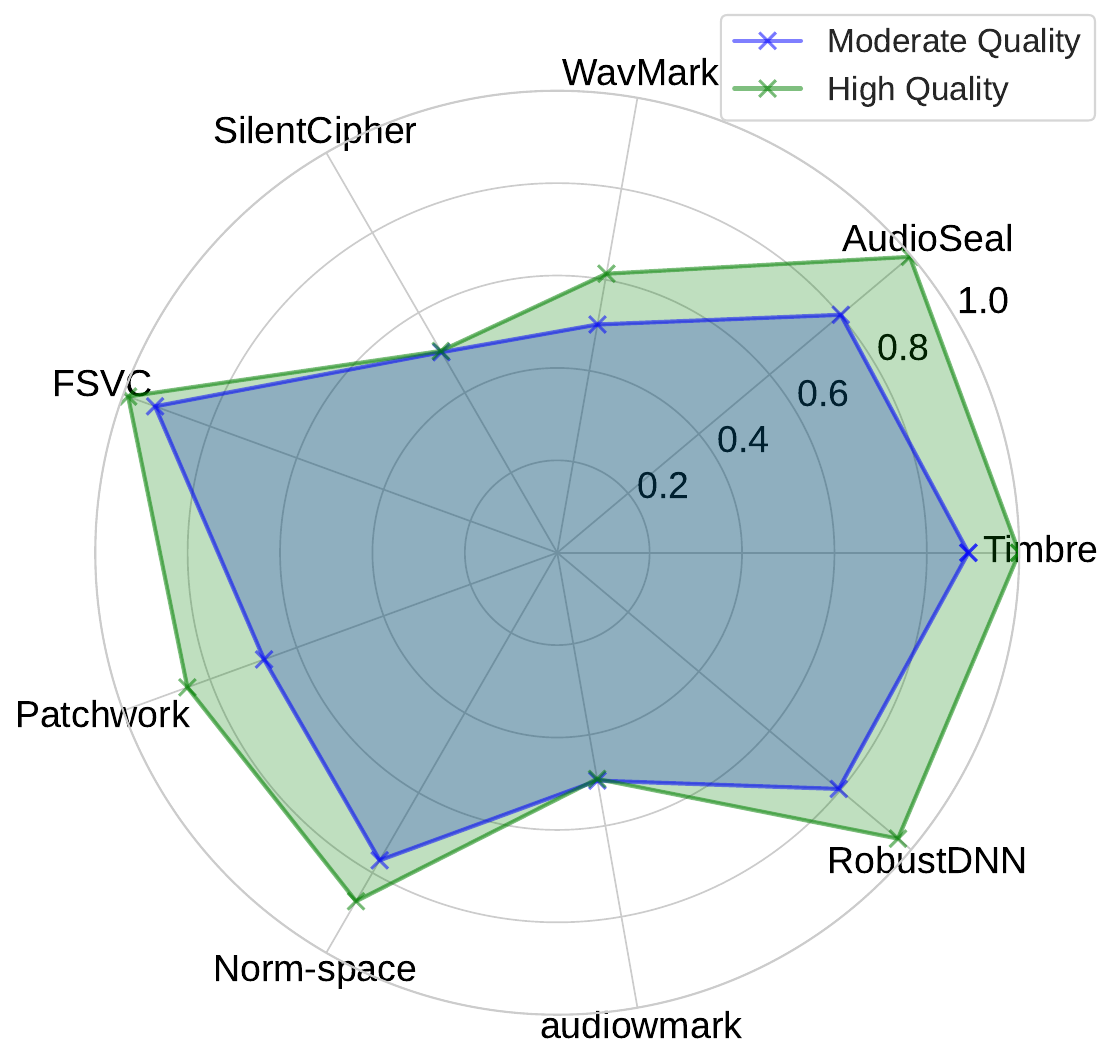}}
\subfigure[Background Noise]{\label{fig:Background_Noise}\includegraphics[width=42mm]{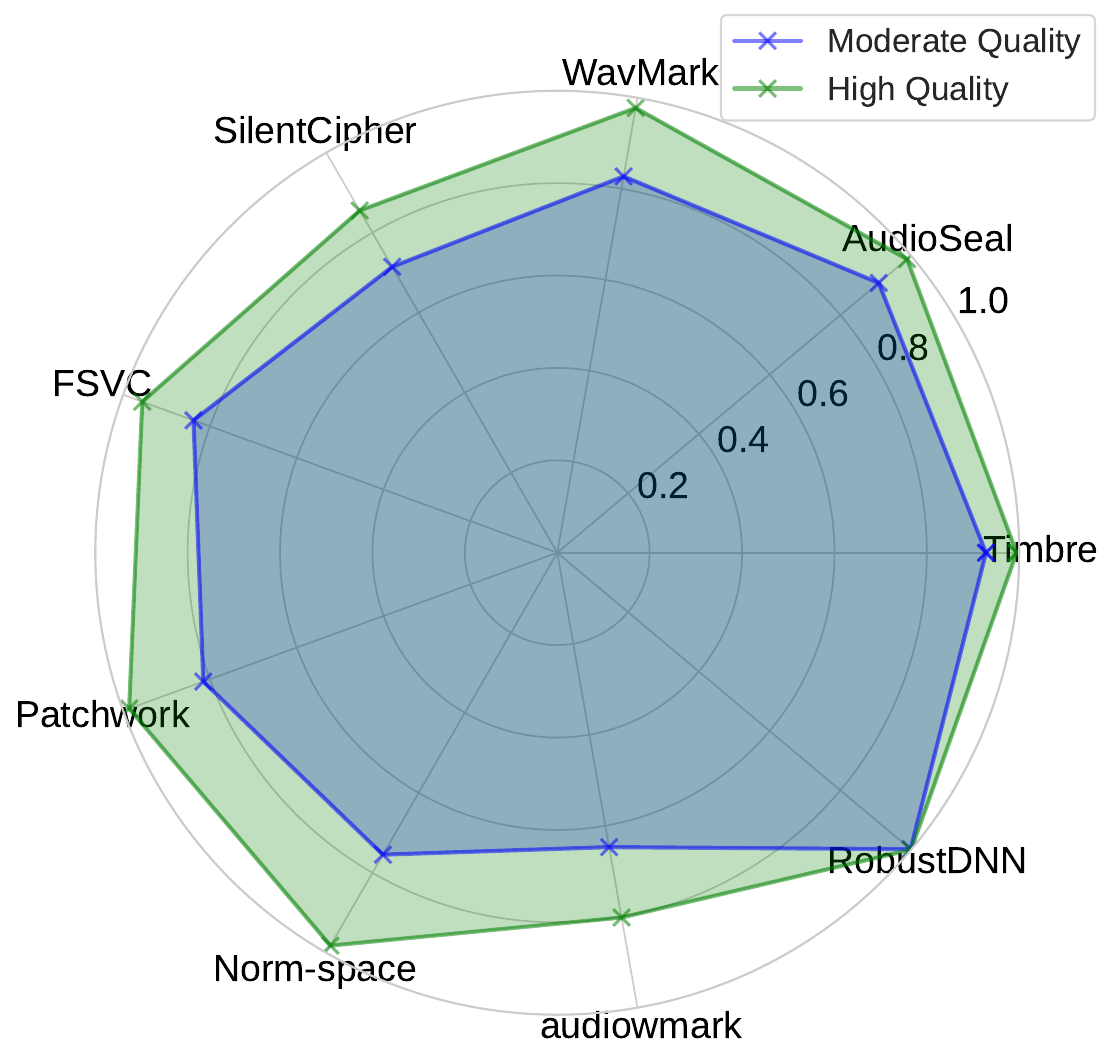}}
\subfigure[High-Pass Filter]{\label{fig:High-Pass_Filter}\includegraphics[width=42mm]{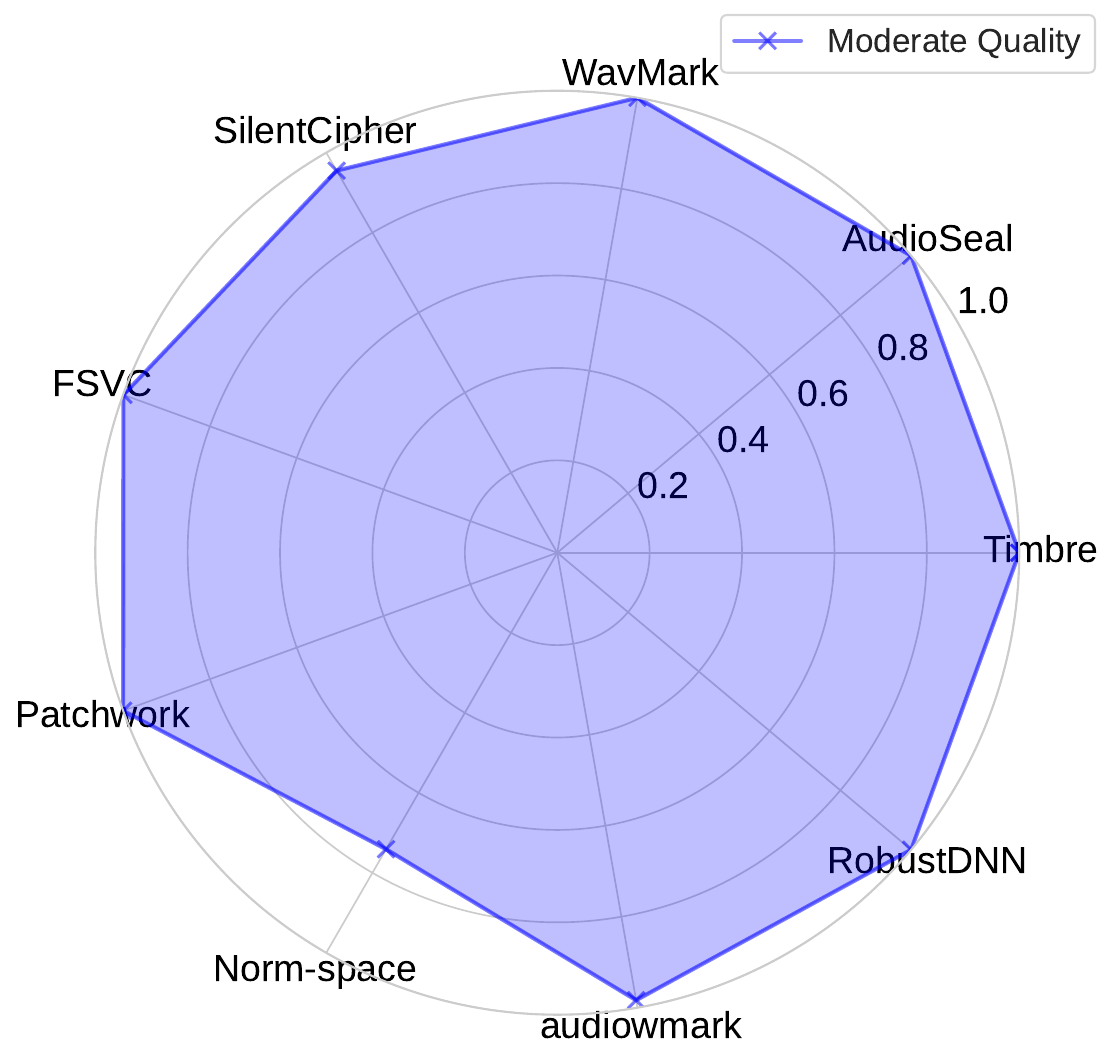}}
\subfigure[Low-Pass Filter]{\label{fig:Low-Pass_Filter}\includegraphics[width=42mm]{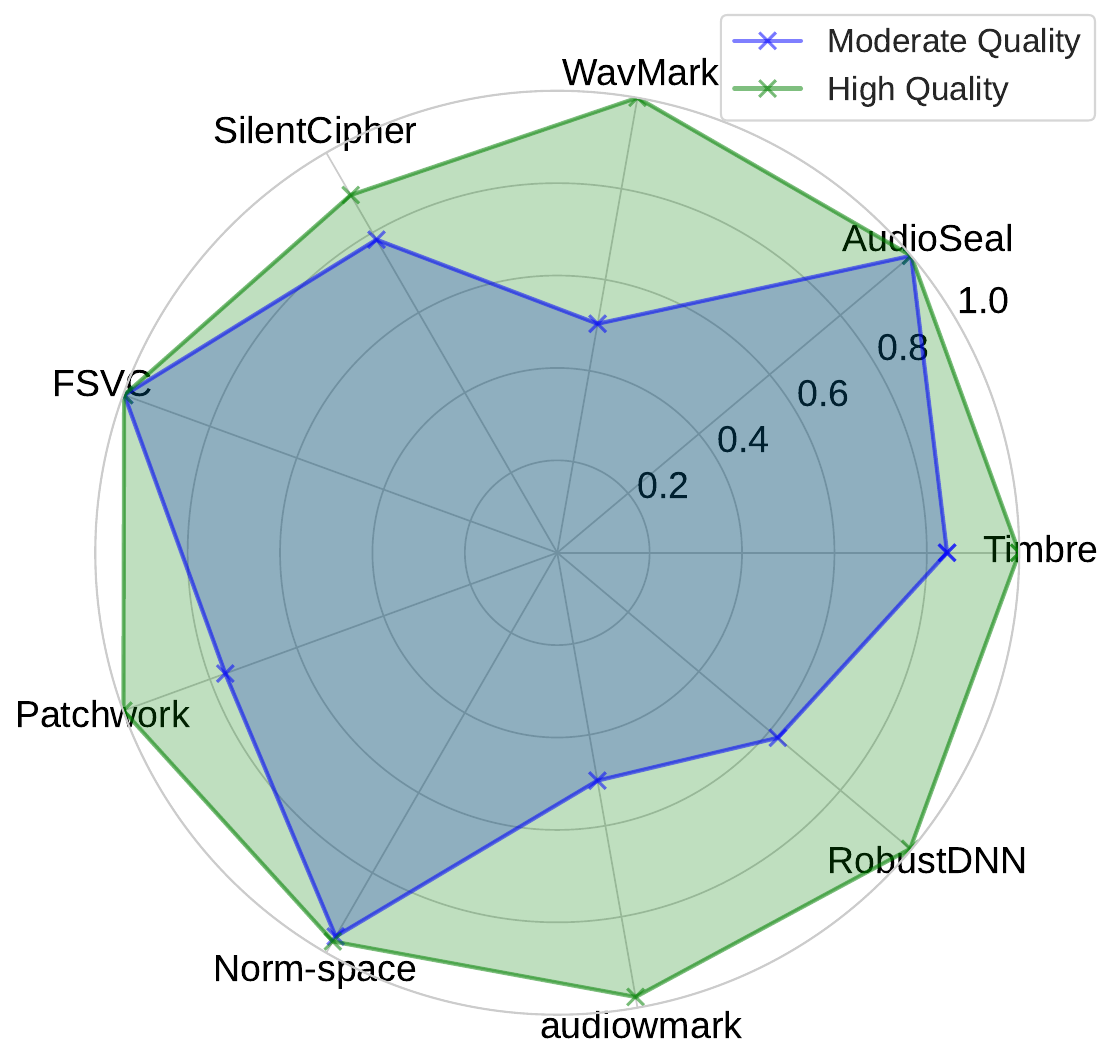}}
\subfigure[Sample Suppression]{\label{fig:Sample_Suppression}\includegraphics[width=42mm]{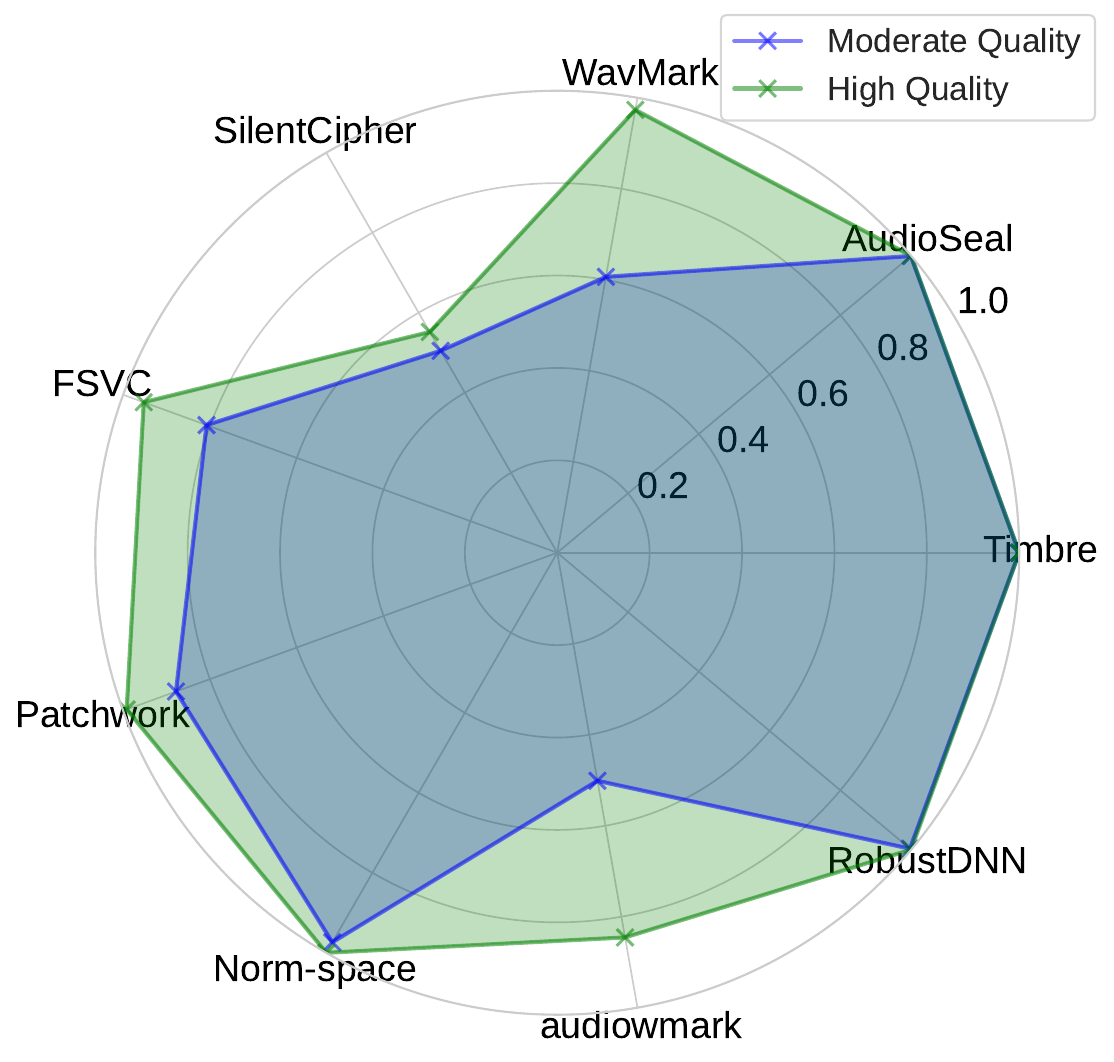}}
\subfigure[Resampling]{\label{fig:Resampling}\includegraphics[width=42mm]{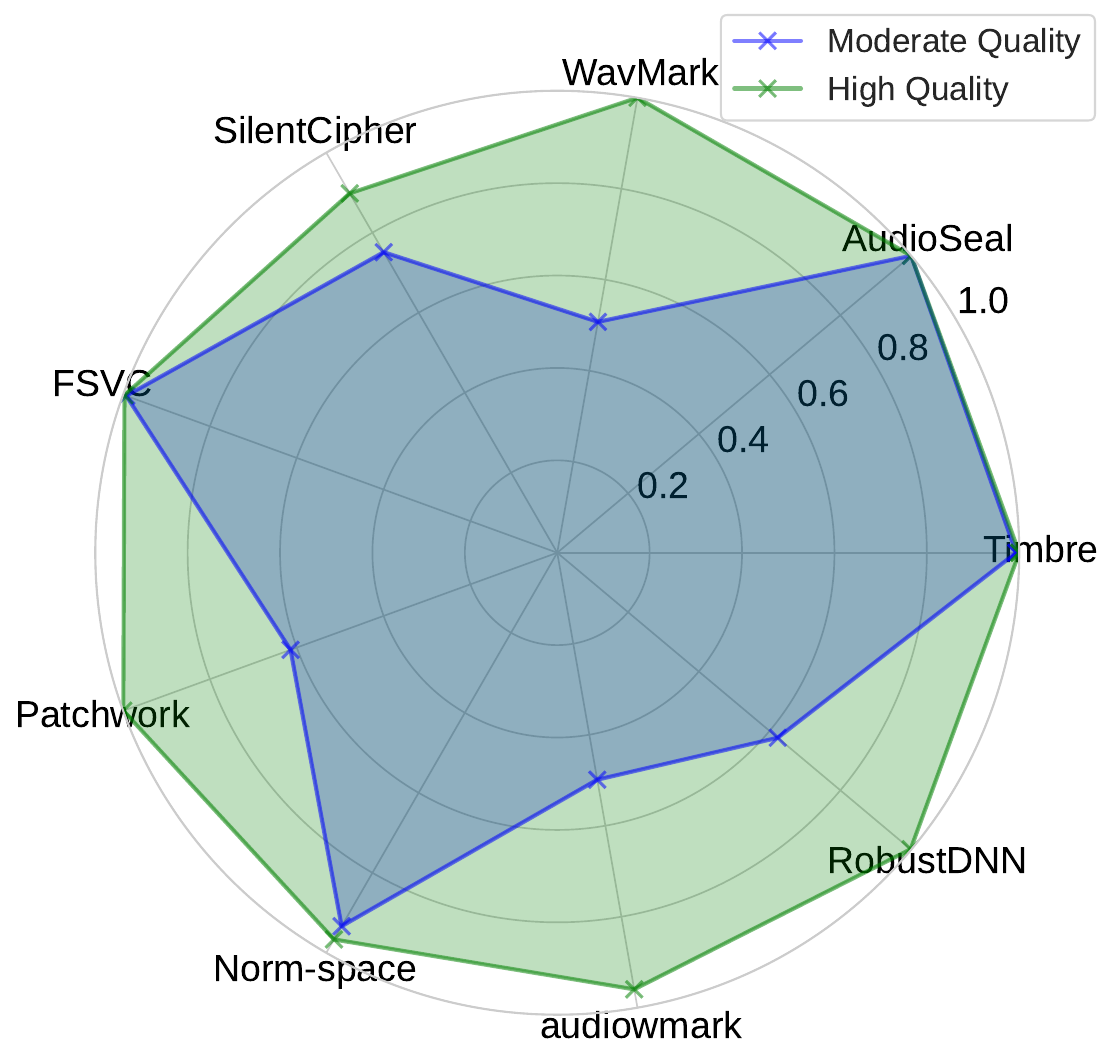}}
\subfigure[Cutting Audio]{\label{fig:Cutting_Audio}\includegraphics[width=42mm]{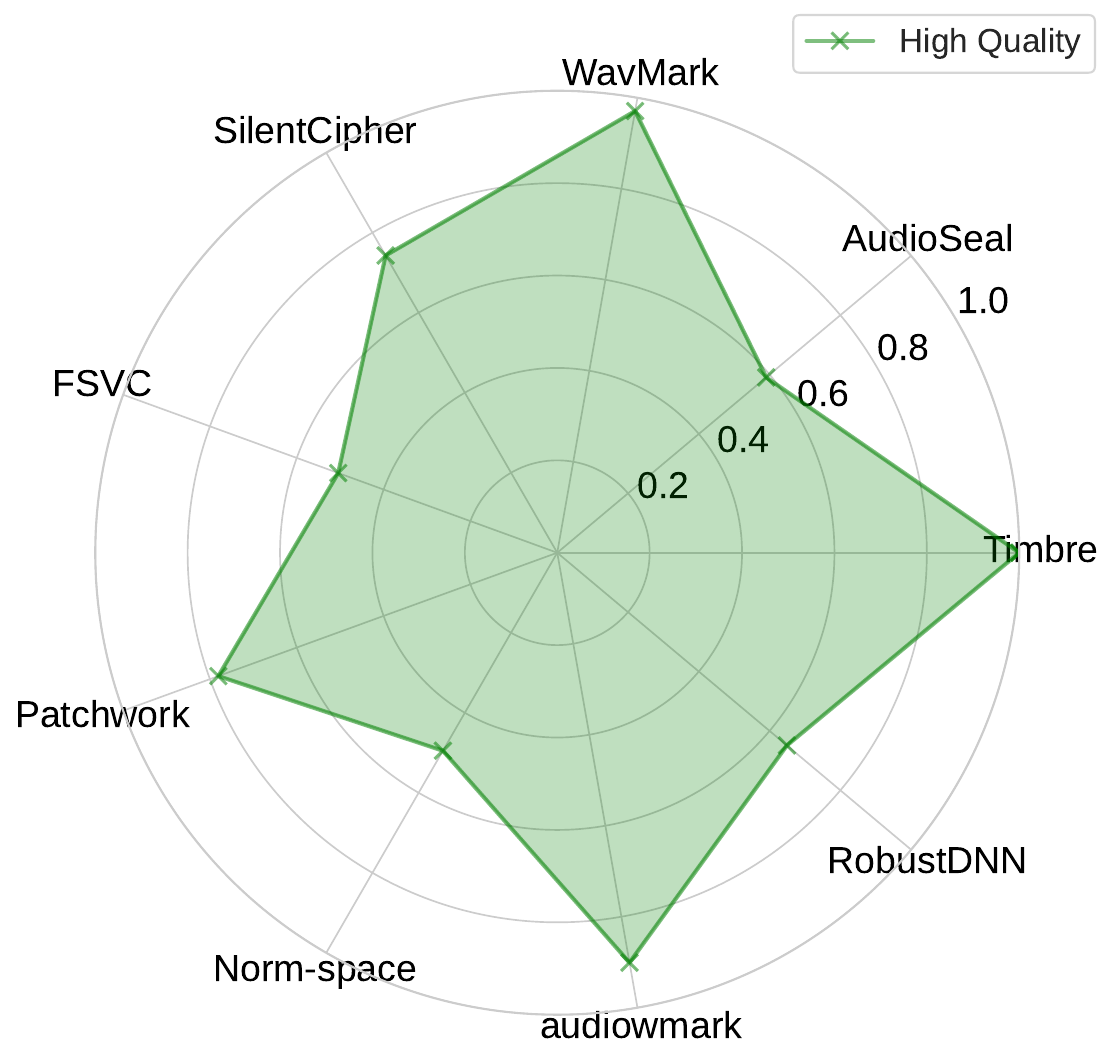}}
\subfigure[Impulse Response]
{\label{fig:Impulse_Response}\includegraphics[width=42mm]{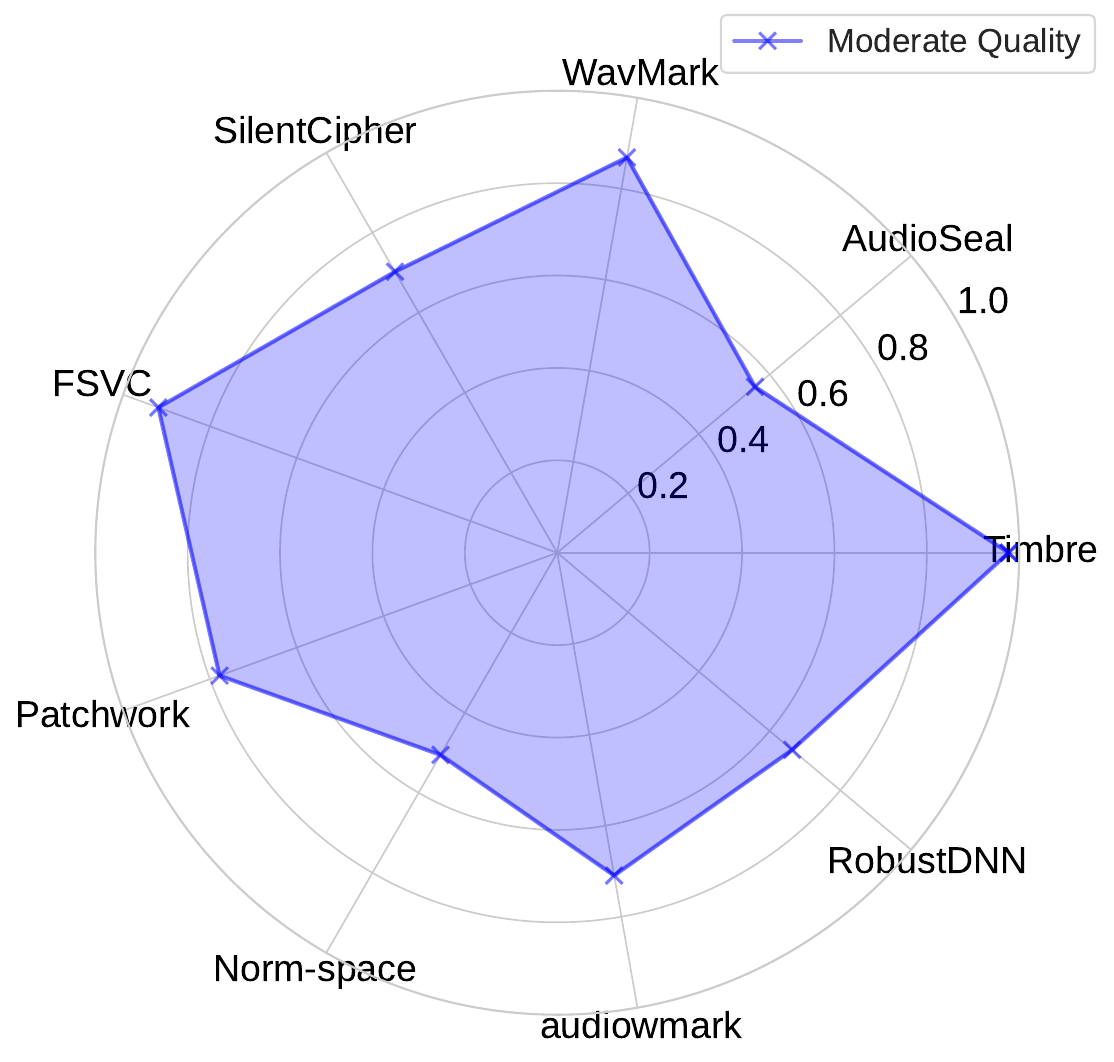}}
\vspace{-10pt}
\caption{Signal Distortion vs. Watermark}
\label{fig:radar_signal}
\end{figure*}

\subsection{Signal-Level Distortion Attacks}

In this evaluation, we assess the robustness of various watermarking schemes against different signal-level distortion attacks, with adversarial constraints to maintain either \emph{moderate} or \emph{high} audio quality. The quality constraints are defined as follows: under the high-quality constraint, attackers are limited to minimal modifications, resulting in high perceptual quality (ViSQOL $\geq$ 4.0); under the moderate-quality constraint, attackers are allowed slightly greater distortion, retaining a perceptually moderate quality (ViSQOL $\geq$ 3) while maintaining intelligibility and usability.

We present our result in Fig.~\ref{fig:radar_signal}.
Each radar plot represents a different type of signal-level distortion attack, including pitch shift, time stretch, Gaussian noise, bitcrush, MP3 compression, background noise, high-pass and low-pass filtering, cutting audio, and impulse response. The axes of each radar chart show the watermarking schemes (e.g., WavMark, AudioSeal, Timbre, RobustDNN, etc.), and the plotted values represent Bit Recovery Accuracy. Higher ACC values indicate greater robustness of the watermarking scheme under the given attack. The two shaded regions—red for moderate quality and blue for high quality—represent the robustness achieved under each respective quality constraint. A larger shaded area in either region reflects better watermark resilience, as it implies higher bit recovery accuracy across the schemes.

\noindent\textbf{Key Finding 1: All watermark schemes are vulnerable to pitch shift attacks}, which can effectively remove the watermark while maintaining high audio quality (e.g., ViSQOL score above 4.0 under high-quality constraints). Under pitch shift, all watermarks show a Bit Recovery Accuracy below 0.6, indicating near-complete removal. This vulnerability arises from the distortion of frequency-based patterns, reducing ACC to near-random levels.

\begin{figure}[t]
\centering
\subfigure[Pitch Shift]{\label{fig:libri_pitch_shift}\includegraphics[width=0.23\textwidth]{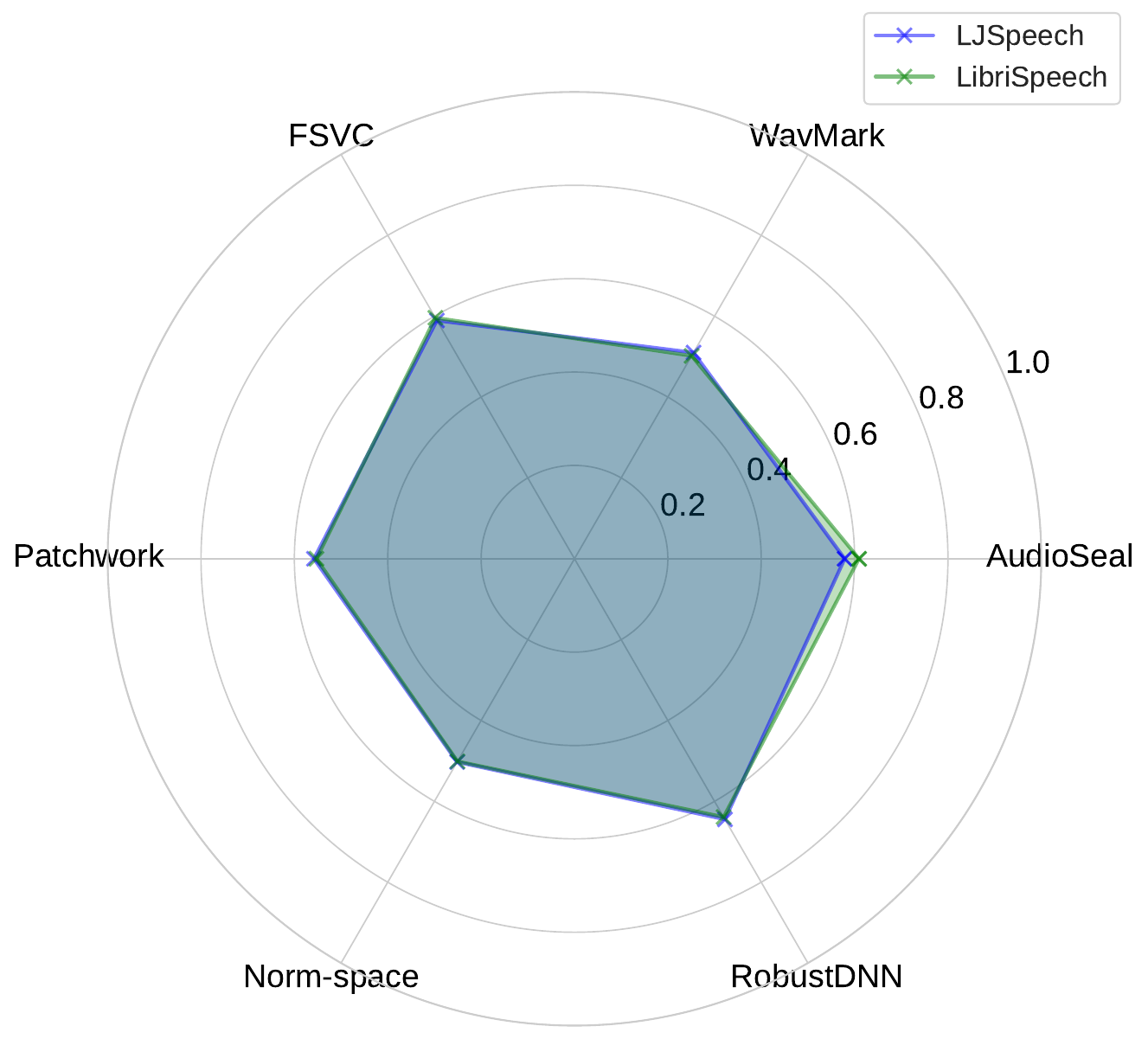}}
\subfigure[Time Stretch]{\label{fig:libri_time_stretch}\includegraphics[width=0.23\textwidth]{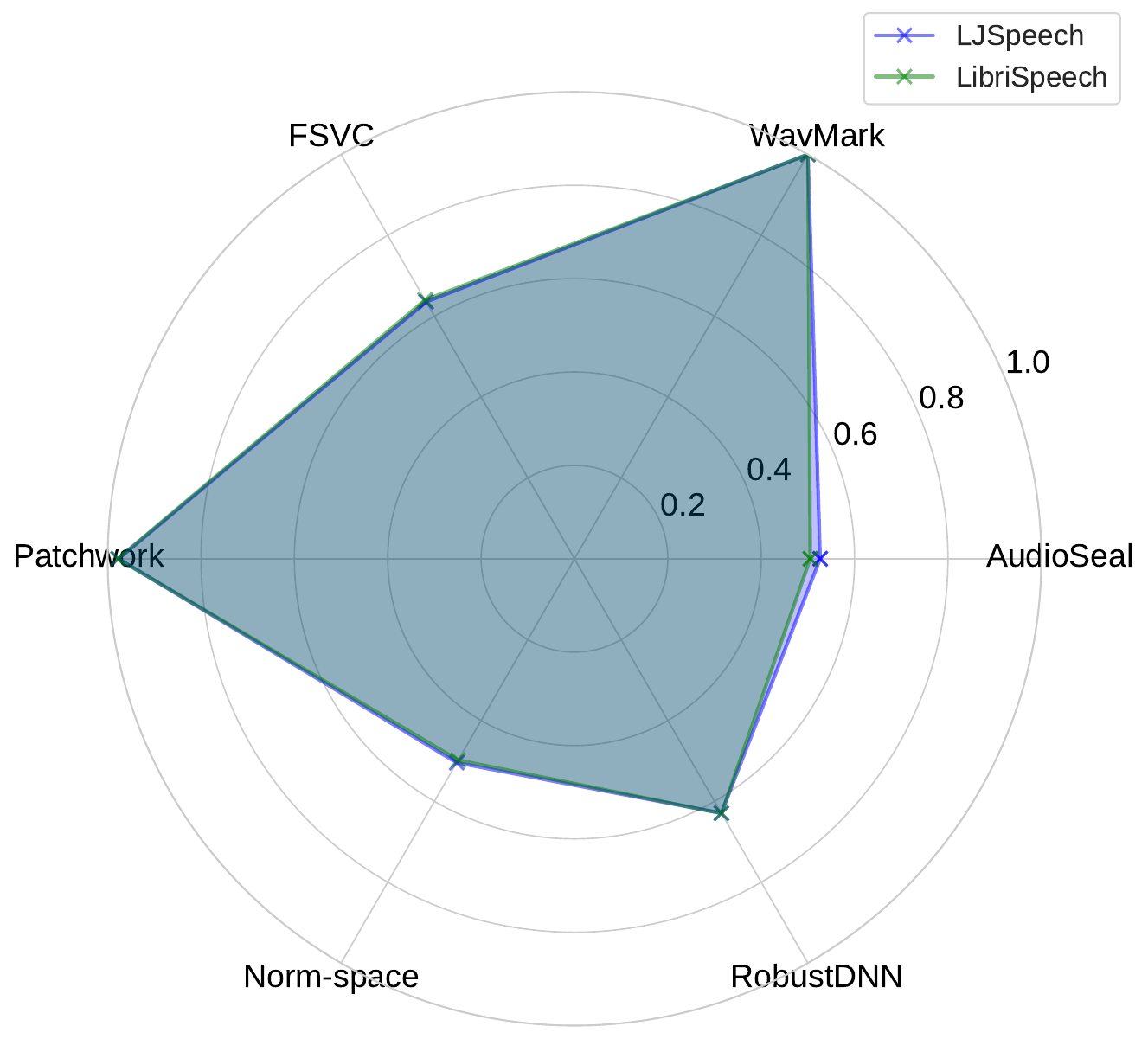}}
\subfigure[Cutting Audio]{\label{fig:libri_cutting_audio}\includegraphics[width=0.23\textwidth]{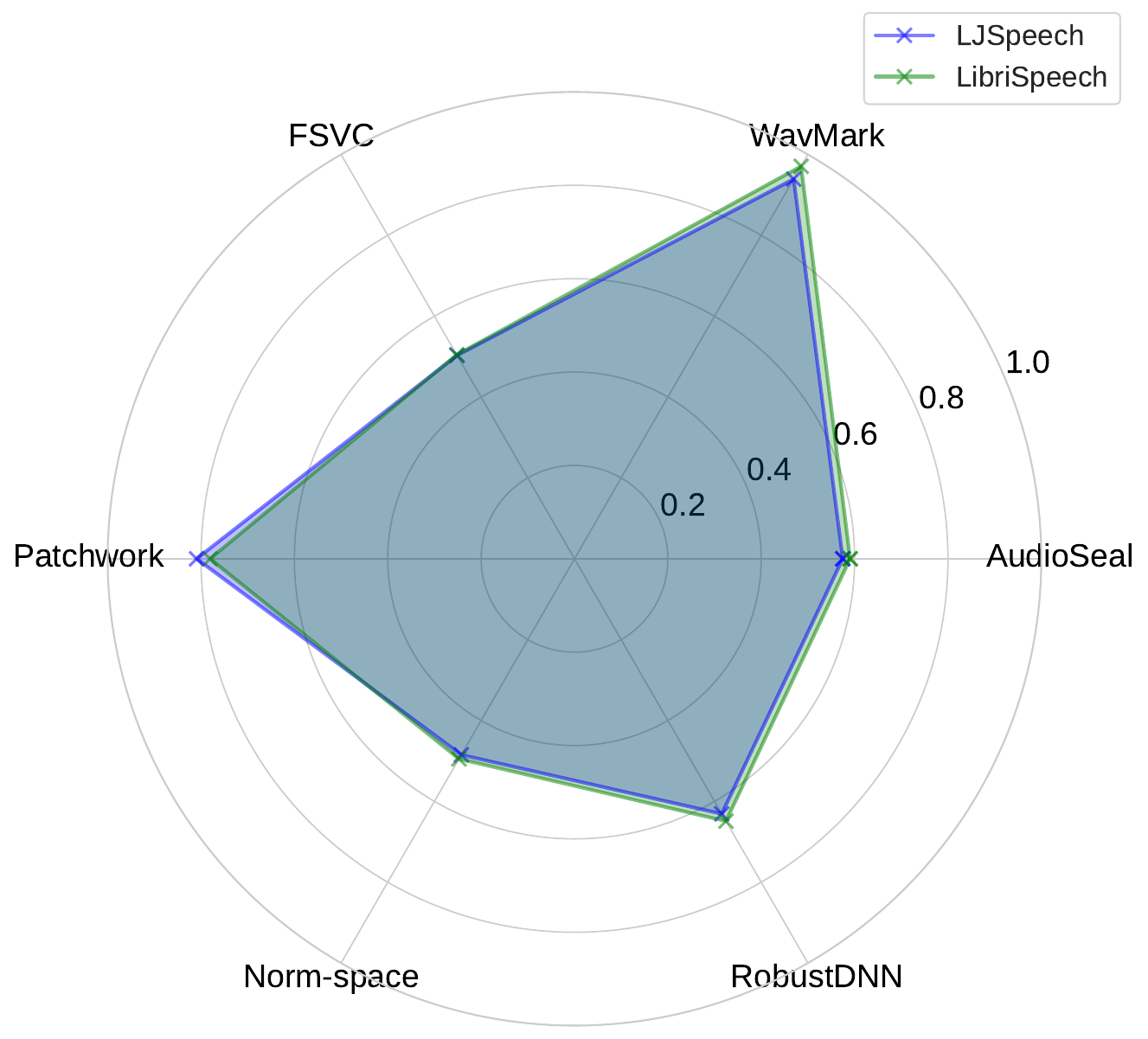}}
\vspace{-10pt}
\caption{LibriSpeech Signal Distortion Accuracy vs. Watermark}
\label{fig:libri_acc}
\end{figure}

\noindent\textbf{Key Finding 2: Most watermark schemes, except Timbre~\cite{liu2023detecting}, WavMark~\cite{chen2023wavmark}, and Patchwork~\cite{natgunanathan2017patchwork}, are sensitive to time-domain modifications like time stretch and cutting audio.} For instance, time stretch attacks with moderate quality constraints result in ACC values below 0.5 for most schemes. In contrast, Timbre~\cite{liu2023detecting}, WavMark~\cite{chen2023wavmark}, and Patchwork~\cite{natgunanathan2017patchwork} maintain ACC values above 0.8 due to repetitive embedding of watermark patterns, which improves resilience to time-based distortions. To ensure unbiased results, we evaluate 6 watermarking schemes on 200 audio samples from the LibriSpeech dataset under 3 distortion types: pitch shift, time stretch, and audio cutting. As shown in Fig.~\ref{fig:libri_acc}, the performance of the watermarking schemes is similar across both the LibriSpeech and LJSpeech datasets.

\noindent\textbf{Key Finding 3: AudioSeal~\cite{san2024proactive}, Timbre~\cite{liu2023detecting}, RobustDNN~\cite{pavlovic2022robust}, and FSVC~\cite{zhao2021desynchronization} demonstrate the highest robustness among all watermark schemes}, performing well against Gaussian noise, MP3 compression, sample suppression, and resampling. These schemes achieve ACC values of 0.7 or higher, with AudioSeal~\cite{san2024proactive} and Timbre~\cite{liu2023detecting} often exceeding 0.8. Their robustness stems from incorporating these distortions during training and the inherent design of their models, making them suitable for real-world scenarios involving common audio modifications.

\noindent\textbf{Key Finding 4: Most watermark schemes withstand high-pass filters but are vulnerable to low-pass filters.} High-pass filtering results in ACC values above 0.8 for most schemes. However, low-pass filtering with moderate quality constraints reduces the ACC of many schemes below 0.6, except for AudioSeal~\cite{san2024proactive}, FSVC~\cite{zhao2021desynchronization}, Norm-space~\cite{saadi2019novel}, and Timbre~\cite{liu2023detecting}, which maintain ACC values above 0.8. This weakness is due to watermark patterns heavily relying on high-frequency components, which are removed by low-pass filters.

\noindent\textbf{Key Finding 5: Timbre~\cite{liu2023detecting} is the most resilient watermarking scheme among those tested}, successfully defending against all attacks except pitch shift. Against distortions like time stretch, Gaussian noise, and MP3 compression, Timbre~\cite{liu2023detecting} achieves ACC values of 0.75 or higher, with some reaching 0.85. This adaptability to diverse distortions highlights Timbre~\cite{liu2023detecting} as the most robust watermarking scheme for applications facing broad audio modifications.

To analyze watermark robustness against the most effective distortions, we identify \emph{pitch shift, time stretch, cutting audio, and sample suppression} as the four most impactful signal-level attacks. Detailed results are available in Appendix C.

\subsection{Physical Distortion Attacks}
\begin{figure}[t]
\centering
\subfigure[Default setting]{\label{fig:physical_real_1}\includegraphics[width=0.23\textwidth]{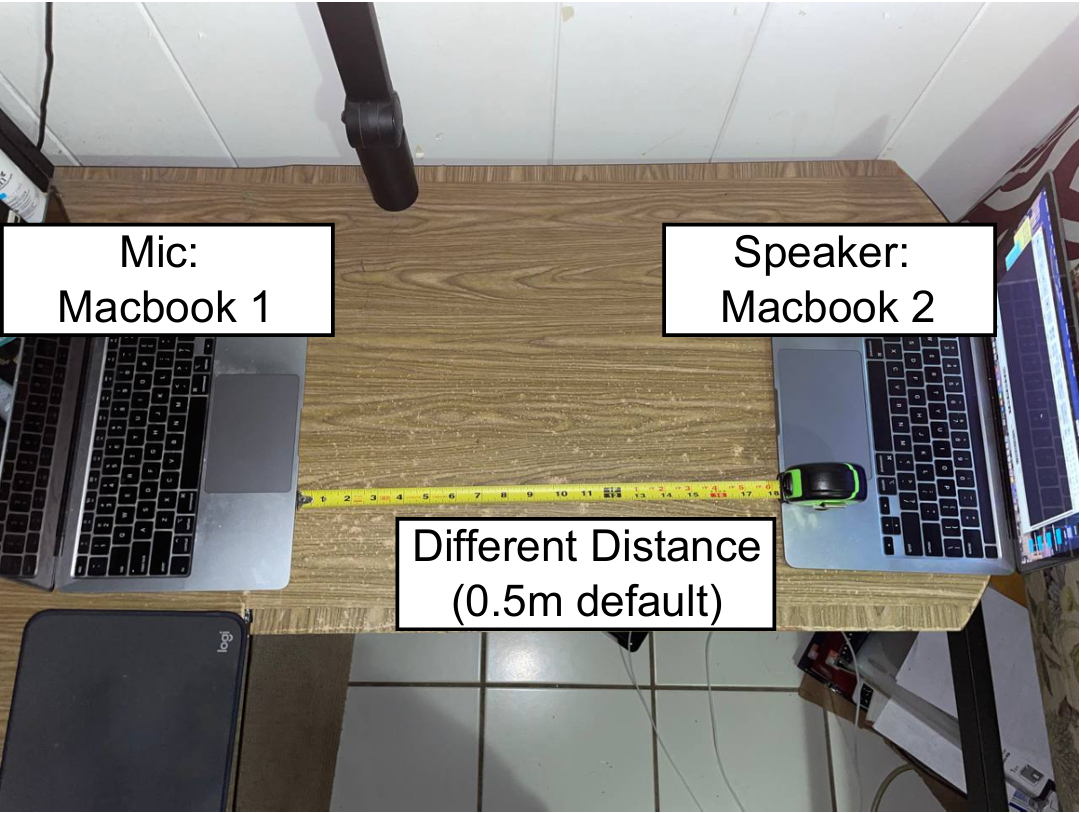}}
\subfigure[Different devices]{\label{fig:physical_real_2}\includegraphics[width=0.23\textwidth]{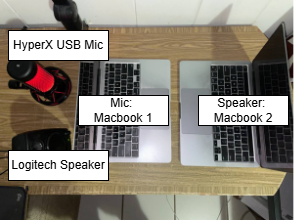}}
\vspace{-10pt}
\caption{Physical distortion experiment setup}
\label{fig:realworld}
\end{figure}

\noindent\textbf{Setup:} We generate 5 watermarked samples for each of our 9 watermarking schemes. We then play back and record them with different devices and at different distances. The detailed experiment setting is listed in Table~\ref{tab:setup}.
\begin{table}[h]
\scalebox{0.7}{
\begin{tabular}{|c|ccc|cc|}
\hline
           & \multicolumn{3}{c|}{Distance Distortion}                                                                                                                                                                                                      & \multicolumn{2}{c|}{Device Distortion}                                                                                                                   \\ \hline
Distance   & \multicolumn{1}{c|}{0.5m}                                                             & \multicolumn{1}{c|}{2.5m}                                                             & 5m                                                               & \multicolumn{1}{c|}{0.5m}                                                              & 0.5m                                                              \\ \hline
Microphone & \multicolumn{1}{c|}{\begin{tabular}[c]{@{}c@{}}Macbook Pro \\ Built-in\end{tabular}} & \multicolumn{1}{c|}{\begin{tabular}[c]{@{}c@{}}Macbook Pro \\ Built-in\end{tabular}} & \begin{tabular}[c]{@{}c@{}}Macbook Pro \\ Built-in\end{tabular} & \multicolumn{1}{c|}{\begin{tabular}[c]{@{}c@{}}HyperX USB \\ Microphone\end{tabular}} & \begin{tabular}[c]{@{}c@{}}Macbook Pro \\ Built-in\end{tabular}  \\ \hline
Speaker    & \multicolumn{1}{c|}{\begin{tabular}[c]{@{}c@{}}Macbook Pro \\ Built-in\end{tabular}} & \multicolumn{1}{c|}{\begin{tabular}[c]{@{}c@{}}Macbook Pro \\ Built-in\end{tabular}} & \begin{tabular}[c]{@{}c@{}}Macbook Pro \\ Built-in\end{tabular} & \multicolumn{1}{c|}{\begin{tabular}[c]{@{}c@{}}Macbook Pro \\ Built-in\end{tabular}}  & \begin{tabular}[c]{@{}c@{}}Logitech Z130 \\ Speaker\end{tabular} \\ \hline

\end{tabular}
}
\caption{Physical setting list}
\label{tab:setup}
\end{table}
The setup experiment is conducted in Fig.~\ref{fig:realworld}. 





\begin{figure}[t]
\centering
\subfigure[Distance Accuracy]{\label{fig:physical_distance_accuracy}\includegraphics[width=0.23\textwidth]{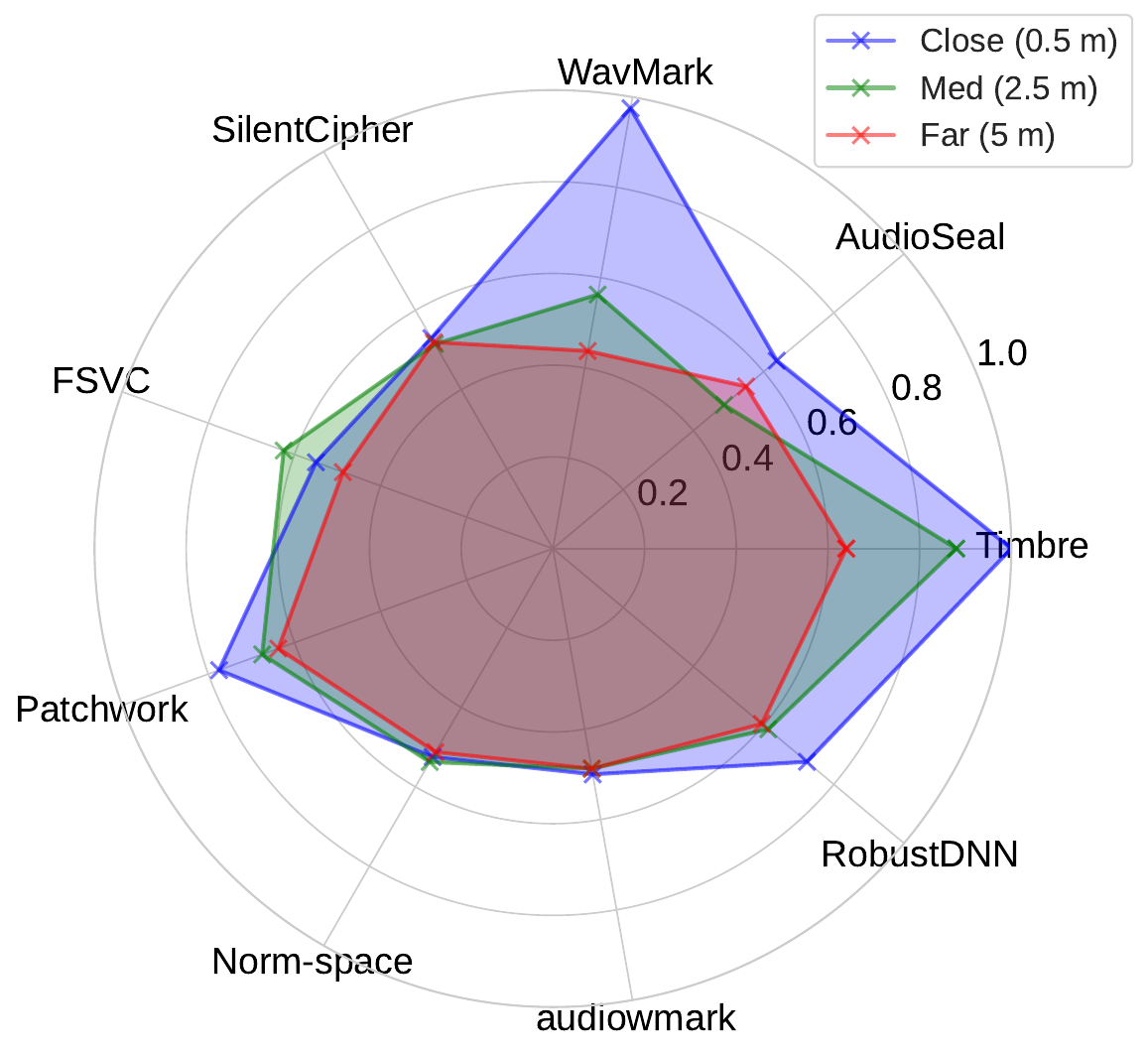}}
\subfigure[Device Accuracy]{\label{fig:physical_device_accuracy}\includegraphics[width=0.23\textwidth]{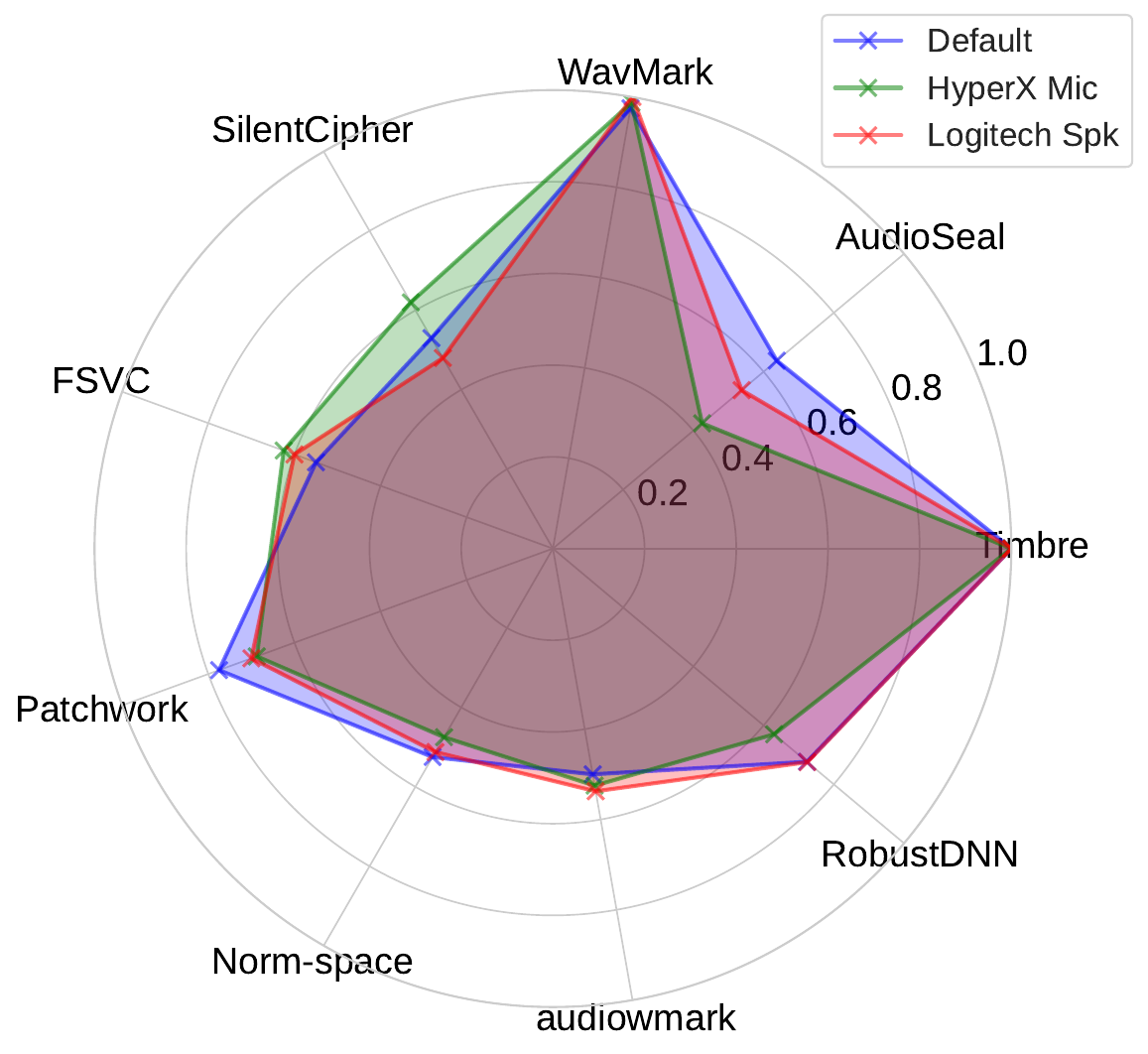}}
\vspace{-10pt}
\caption{Physical Accuracy}
\label{fig:phy_acc}
\end{figure}

\noindent\textbf{Physical Distortion on Bit Recovery Accuracy:} According to Fig.~\ref{fig:phy_acc}, we have the following findings.

\noindent\textbf{Key Finding 6: Most watermarks are vulnerable to physical re-recording.}  
While initial bit recovery accuracy for watermarks is close to 1.0 in the digital domain, physical re-recording significantly reduces their performance. As shown in Fig.~\ref{fig:physical_distance_accuracy}, all watermarks, except WavMark~\cite{chen2023wavmark} and Timbre~\cite{liu2023detecting}, experience a substantial drop in accuracy during re-recording, producing nearly random bits (ACC ~0.5) that fail to match the original watermark bits. This highlights the vulnerability of most schemes to real-world physical attacks.

\noindent\textbf{Key Finding 7: All watermarks are vulnerable to far-distance re-recording.}  
Re-recording at farther distances causes a sharp decline in watermark accuracy. According to Fig.~\ref{fig:physical_distance_accuracy}, schemes like WavMark~\cite{chen2023wavmark} and Timbre~\cite{liu2023detecting} perform well at close distances (0.5 m), maintaining high accuracy. However, as the recording distance increases to 2.5 m and 5 m, the accuracy of most schemes drops below 0.6, with some reaching values close to 0.4 or lower. This demonstrates that increased distance significantly degrades watermark performance, reducing the robustness of most schemes in practical scenarios.

\noindent\textbf{Key Finding 8: Different devices do not significantly impact watermark performance.}  
As shown in Fig.~\ref{fig:physical_device_accuracy}, watermark accuracy remains relatively stable across various recording and playback devices, including the MacBook Pro built-in microphone and speaker, HyperX USB microphone, and Logitech external speaker. This stability suggests that watermarking schemes are robust to device changes, maintaining consistent performance regardless of the hardware used. \emph{However, this observation may not be entirely reliable, as many watermarks already exhibit significant performance degradation, rendering the default setting a low accuracy benchmark.} For more physical distortion results, refer to Appendix D.

\subsection{AI Distortion Attacks}


\begin{table*}[t]
\scalebox{0.75}{
\centering
\begin{tabular}{|l|l|lllll|lllllll|}
\hline
\multirow{4}{*}{\textbf{Fidelity}} & \multirow{4}{*}{\textbf{No Attack}} & \multicolumn{5}{c|}{\multirow{2}{*}{\textbf{Voice Conversion}}}                                                                                                                                            & \multicolumn{7}{c|}{\textbf{Text-to-speech}}                                                                                                                                                                                                                                                       \\ \cline{8-14} 
                          &                            & \multicolumn{5}{c|}{}                                                                                                                                                                             & \multicolumn{1}{c|}{\multirow{2}{*}{\textbf{YourTTS}}} & \multicolumn{3}{c|}{\textbf{Tacotron2}}                                                                                & \multicolumn{3}{c|}{\textbf{Fastspeech2}}                                                                        \\ \cline{3-7} \cline{9-14} 
                          &                            & \multicolumn{1}{l|}{\textbf{AdaIn-VC}}  & \multicolumn{1}{c|}{\textbf{FragmentVC}} & \multicolumn{1}{c|}{\textbf{YourTTS VC}} & \multicolumn{1}{c|}{\textbf{MediumVC}} & \multicolumn{1}{c|}{\textbf{RVC}} & \multicolumn{1}{c|}{}                                  & \multicolumn{1}{c|}{\textbf{Griffin-Lim}} & \multicolumn{1}{c|}{\textbf{H-GAN}} & \multicolumn{1}{c|}{\textbf{H-GAN*}} & \multicolumn{1}{c|}{\textbf{G-Lim}} & \multicolumn{1}{c|}{\textbf{H-GAN}} & \multicolumn{1}{c|}{\textbf{H-GAN*}} \\ \cline{3-14} 
                          &                            & \multicolumn{1}{l|}{\cellcolor{blue!20}zero-shot} & \multicolumn{1}{l|}{\cellcolor{blue!20}zero-shot}           & \multicolumn{1}{l|}{\cellcolor{blue!20}zero-shot}           & \multicolumn{1}{l|}{\cellcolor{blue!20}few-shot}          & \cellcolor{red!20}adaptive                          & \multicolumn{1}{l|}{\cellcolor{blue!20}zero-shot}                         & \multicolumn{1}{l|}{\cellcolor{red!20}adaptive}             & \multicolumn{1}{l|}{\cellcolor{red!20}adaptive}       & \multicolumn{1}{l|}{\cellcolor{red!50}adaptive*}       & \multicolumn{1}{l|}{\cellcolor{red!20}adaptive}       & \multicolumn{1}{l|}{\cellcolor{red!20}adaptive}       & \cellcolor{red!50}adaptive*                            \\ \hline
PESQ                      & 4.17                       & \multicolumn{1}{l|}{1.71}      & \multicolumn{1}{l|}{1.14}                & \multicolumn{1}{l|}{1.87}                & \multicolumn{1}{l|}{1.07}              & \cellcolor{green!20}2.33                              & \multicolumn{1}{l|}{1.13}                              & \multicolumn{1}{l|}{1.17}                 & \multicolumn{1}{l|}{1.13}           & \multicolumn{1}{l|}{1.13}            & \multicolumn{1}{l|}{1.17}           & \multicolumn{1}{l|}{1.13}           & 1.14                                 \\ \hline
ViSQOL                    & 4.87                       & \multicolumn{1}{l|}{\cellcolor{green!20}3.82}      & \multicolumn{1}{l|}{\cellcolor{green!20}3.02}                & \multicolumn{1}{l|}{\cellcolor{green!20}3.96}                & \multicolumn{1}{l|}{\cellcolor{green!20}2.54}              & \cellcolor{green!20}4.04                              & \multicolumn{1}{l|}{2.35}                              & \multicolumn{1}{l|}{1.81}                 & \multicolumn{1}{l|}{\cellcolor{green!20}2.69}           & \multicolumn{1}{l|}{\cellcolor{green!20}2.72}            & \multicolumn{1}{l|}{1.82}           & \multicolumn{1}{l|}{\cellcolor{green!20}2.71}           & \cellcolor{green!20}2.75                                 \\ \hline
SECS                      & 0.99                       & \multicolumn{1}{l|}{\cellcolor{green!20}0.84}      & \multicolumn{1}{l|}{\cellcolor{green!20}0.75}                & \multicolumn{1}{l|}{\cellcolor{green!20}0.90}                & \multicolumn{1}{l|}{\cellcolor{green!20}0.67}              & \cellcolor{green!20}0.93                              & \multicolumn{1}{l|}{\cellcolor{green!20}0.83}                              & \multicolumn{1}{l|}{\cellcolor{green!20}0.69}                 & \multicolumn{1}{l|}{\cellcolor{green!20}0.87}           & \multicolumn{1}{l|}{\cellcolor{green!20}1.10}            & \multicolumn{1}{l|}{\cellcolor{green!20}0.71}           & \multicolumn{1}{l|}{\cellcolor{green!20}0.89}           & \cellcolor{green!20}1.13                                 \\ \hline

\end{tabular}}
\caption{Generative Models used for AI distortion}
\label{tab:ai_distort}
\end{table*}

\noindent\textbf{Setup:} We evaluate the robustness of 
7 watermark schemes with two types of AI distortions---VC and TTS. We present the used AI distortion models in Table~\ref{tab:ai_distort}. There are five VC models (AdaIn-VC~\cite{chou2019oneshotvoiceconversionseparating}, FragmentVC~\cite{lin2021fragmentvcanytoanyvoiceconversion}, YourTTS VC~\cite{casanova2023yourttszeroshotmultispeakertts}, MediumVC~\cite{gu2021mediumvcanytoanyvoiceconversion}, and RVC~\cite{rvc_project}) and two TTS models (Tacotron2~\cite{shen2018natural} and Fastspeech2~\cite{ren2022fastspeech2fasthighquality}), each tested with various vocoders (Griffin-Lim~\cite{Griffin_1984}, Universal HiFi-GAN, and fine-tuned HiFi-GAN~\cite{oprea2021hganpowerganshands}). \textbf{Zero-shot} means the attacker directly feeds watermarked samples to a pre-trained, frozen generative model without modification. \textbf{Adaptive} under the VC Model category indicates the adversary fine-tuned the VC model with \emph{unwatermarked samples} to enhance watermark removal success, while 
\textbf{adaptive} in the context of TTS indicates that the adversary trained the TTS models with \emph{watermarked samples} to simulate the same style of victims. The \textbf{adaptive*} label means that both the generative model and its vocoder are trained or fine-tuned. Note that in evaluating seven different watermark schemes, each adaptive setting requires training the VC and TTS models with different watermarks. 



\noindent\textbf{AI Distortion Fidelity}:
To evaluate the impact of generative AI-based watermark removal on audio quality, we conducted a fidelity analysis using three key metrics: Perceptual Evaluation of Speech Quality (PESQ), Virtual Speech Quality Objective Listener (ViSQOL), and Speaker Encoder Cosine Similarity (SECS). Table~\ref{tab:ai_distort} provides a comparative view of these metrics across various generative models. The ``No Attack" column represents the baseline fidelity of watermarked audio without any generative model applied, averaging the metric scores across seven different watermarking schemes. For the ``No Attack" condition, the average PESQ, ViSQOL, and SECS scores are 4.17, 4.87, and 0.99 respectively, indicating high audio quality before any interference.
Following this baseline, we applied both VC and TTS models in various configurations (zero-shot, few-shot, adaptive, and adaptive*) to attempt watermark removal and measured the resulting audio quality. The table uses color coding to visually highlight the fidelity results: light green cells denote scores that are over 50\% of the corresponding "No Attack" baseline, suggesting that audio quality remains relatively high despite the application of these generative AI models, indicating minimal quality degradation relative to the original watermarked audio.

\noindent\textbf{Key Finding 9: AI distortion does not significantly reduce watermarked audio quality across generative models}, with most configurations maintaining substantial fidelity; this is indicated by ViSQOL scores generally above 2.43 and SECS scores above 0.5. Across all AI distortions, VC models consistently deliver better audio quality than TTS models in watermark removal tasks. For instance, the RVC model in the adaptive setting achieves a PESQ score of 2.33 and a ViSQOL score of 4.04, both closer to the "No Attack" baseline than any TTS model. Additionally, adaptive models show improved audio quality, with ViSQOL scores reaching up to 2.72 for Tacotron2 H-GAN* and 2.75 for Fastspeech2 H-GAN*, suggesting that fine-tuning generative models enhances fidelity.

\subsubsection{AI Distortion---VC}
\begin{figure}[t]
\centering
\subfigure[Zero-shot VC attack]{\label{fig:zeroshot_accuracy_radar}\includegraphics[width=0.23\textwidth]{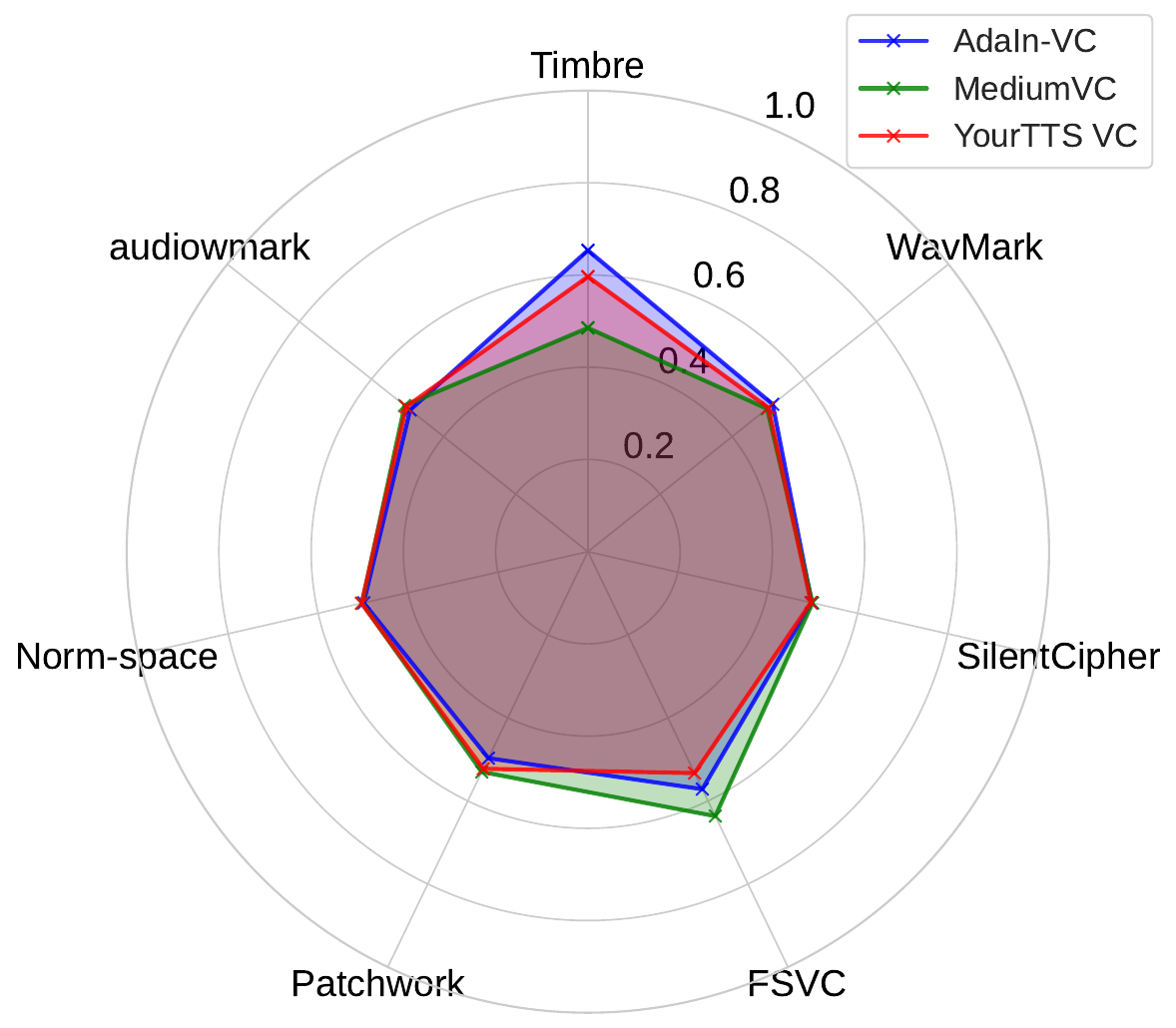}}
\subfigure[Few-shot/adaptive VC attack]{\label{fig:fewshot_accuracy_radar}\includegraphics[width=0.23\textwidth]{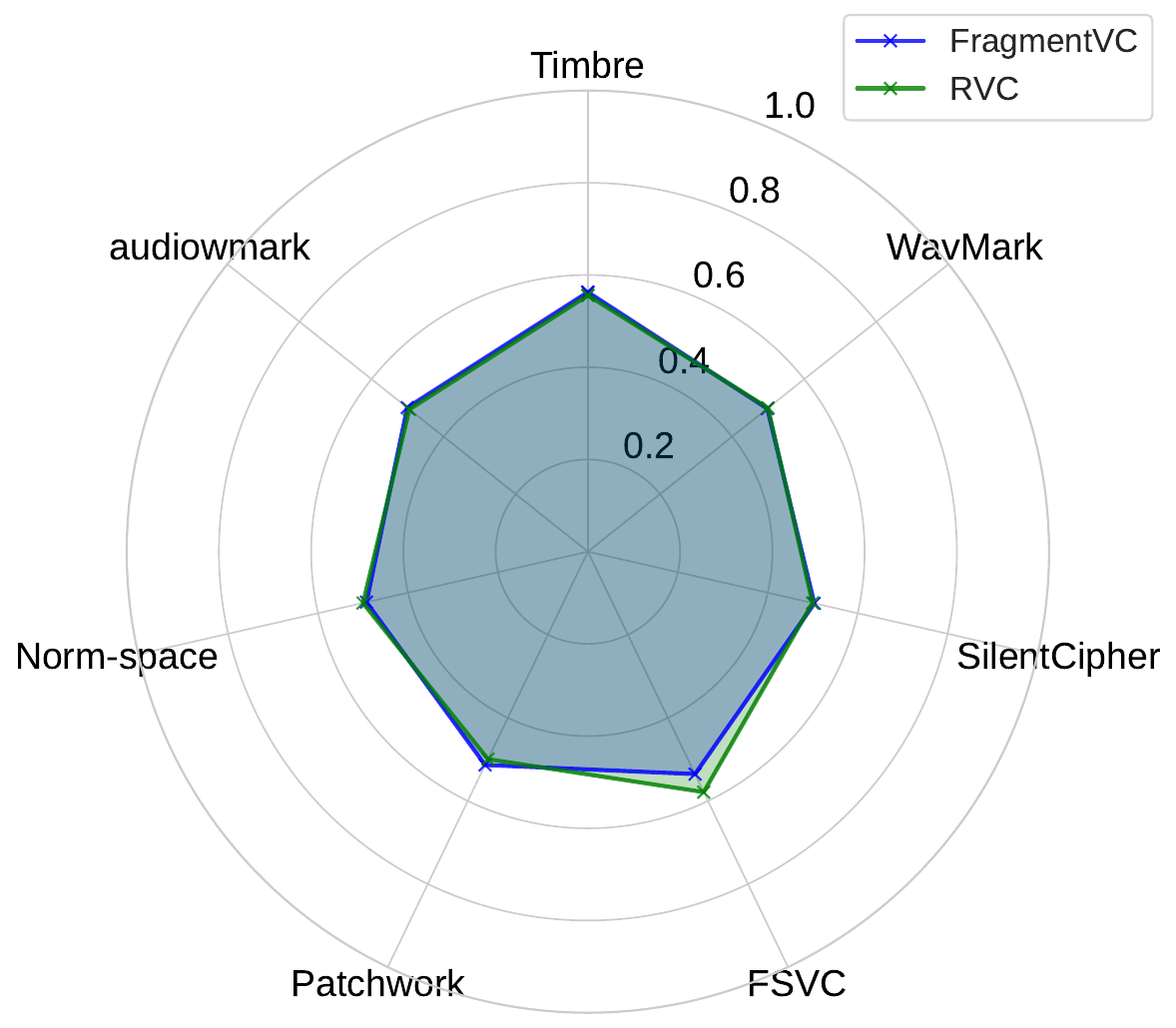}}
\vspace{-10pt}
\caption{Robustness under VC attacks}
\label{fig:vc_robust}
\end{figure}

\begin{figure*}[t]
\centering
\subfigure[Zero-shot TTS attack]{\label{fig:yourtts_accuracy_radar}\includegraphics[width=42mm]{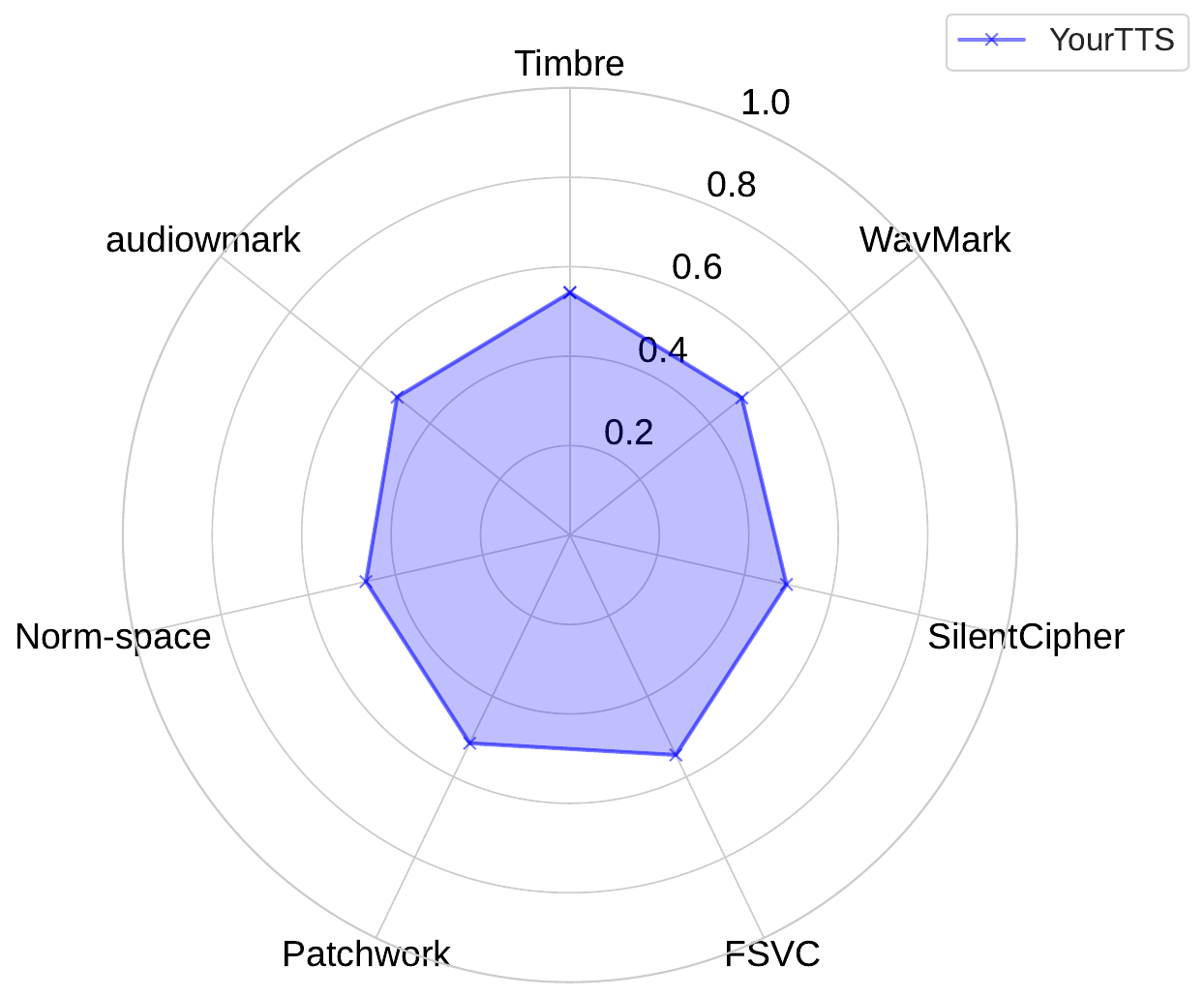}}
\subfigure[Adaptive TTS + G-Lim vocoder attack]{\label{fig:glim_accuracy_radar}\includegraphics[width=042mm]{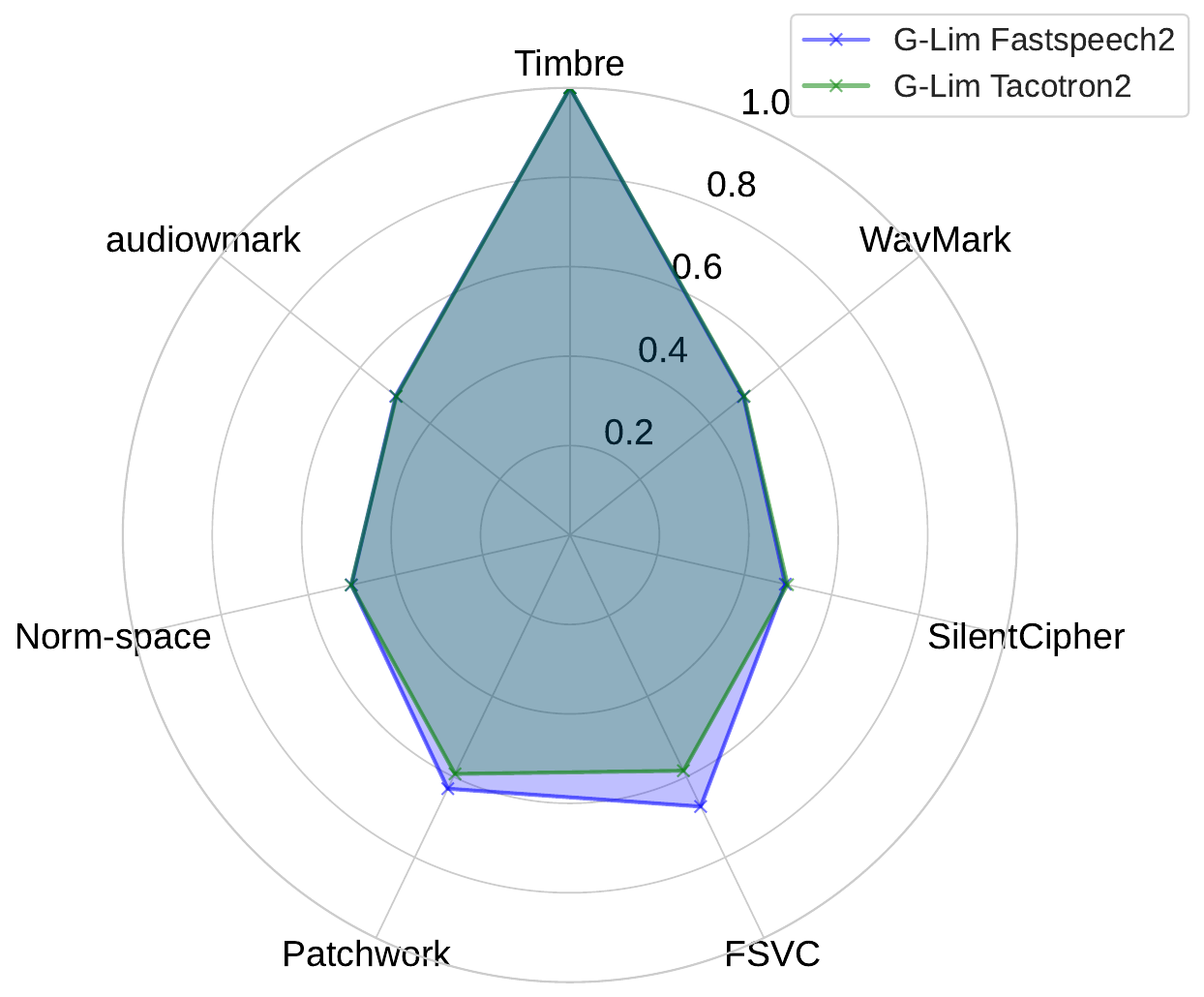}}
\subfigure[Adaptive TTS + universal H-GAN vocoder attack]{\label{fig:hgan_accuracy_radar}\includegraphics[width=042mm]{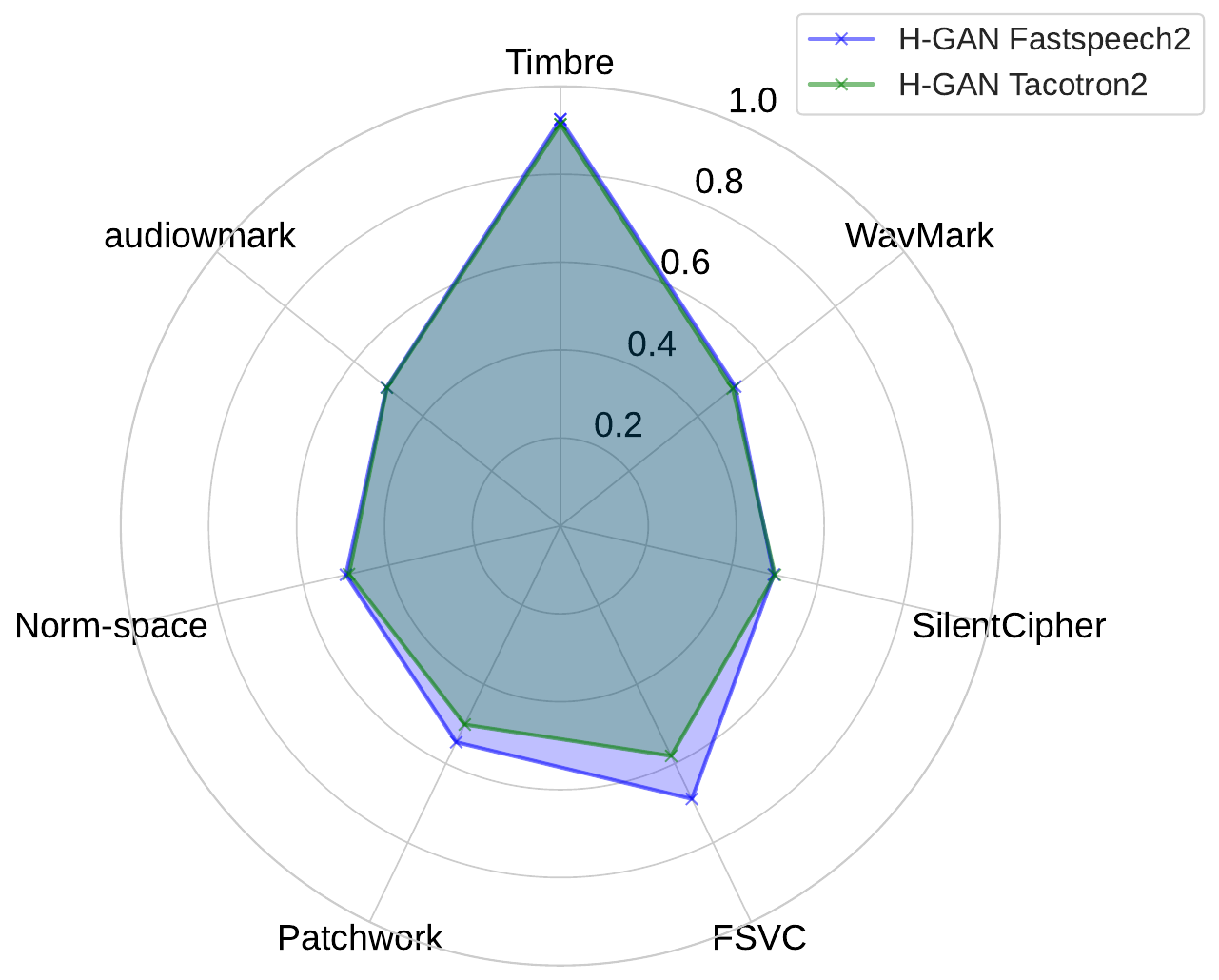}}
\subfigure[Adaptive TTS + adaptive H-GAN* vocoder attack]{\label{fig:hganfinetune_accuracy_radar}\includegraphics[width=42mm]{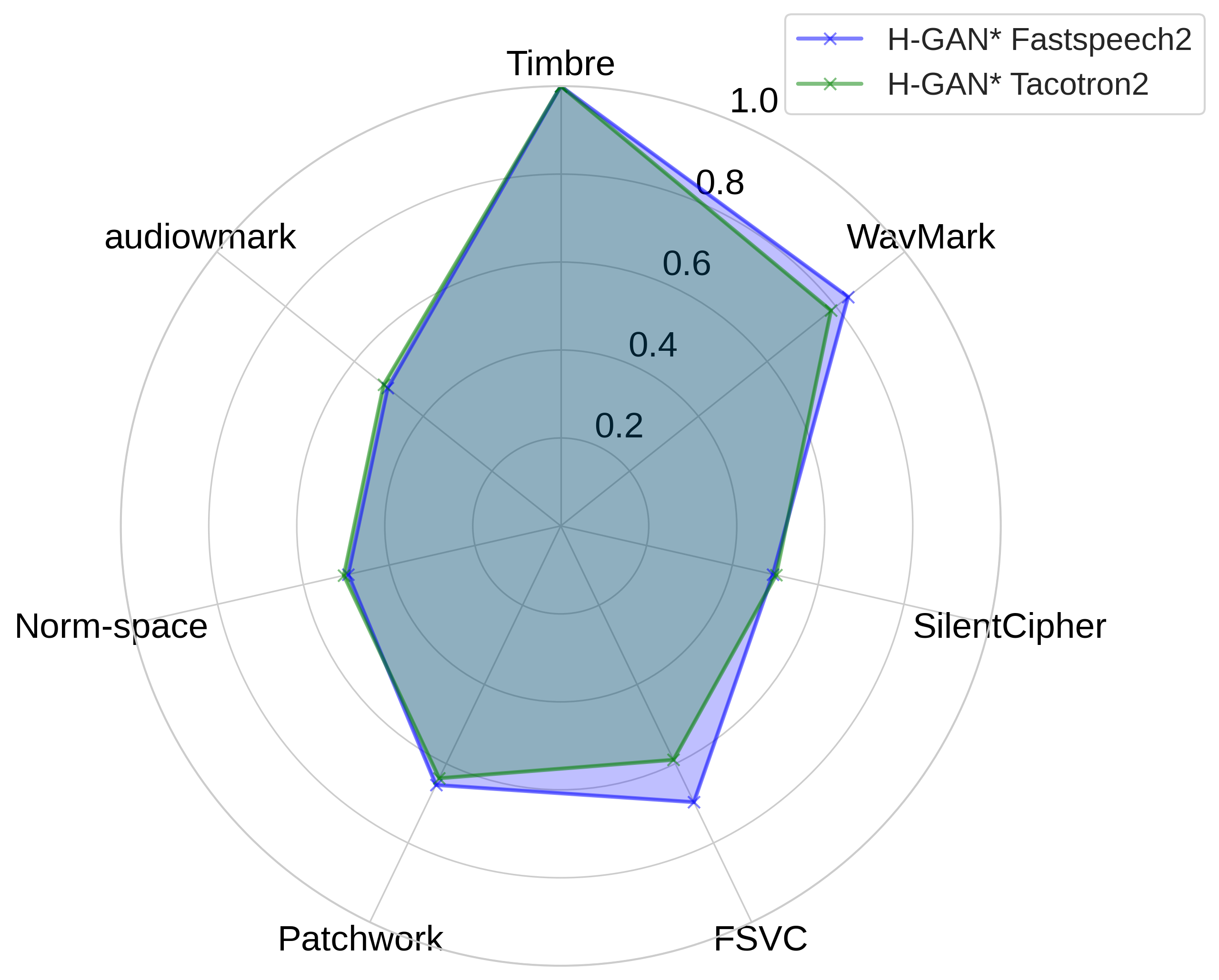}}
\vspace{-10pt}
\caption{Robustness under TTS}
\label{fig:tts_robust}
\end{figure*}
We launched attacks on seven different watermark schemes using a variety of VC models under zero/few-shot and adaptive attack settings. In the zero/few-shot setting, the model parameters are frozen, whereas in the adaptive attack setting, the model is fine-tuned to optimize its conversion effectiveness. The watermarked robustness to AI distortions is evaluated using the bit recovery accuracy. We present the results in Fig.~\ref{fig:vc_robust}. On the left, we report the ACC when applying a zero-shot VC model to the watermarked samples. On the right, we show the results for a few-shot and a fine-tuned VC model.

\noindent\textbf{Key Finding 10: Both zero-shot and adaptive/few-shot VC models effectively reduce the watermark bits recovery accuracy across various watermarking schemes}, bringing the recovery rate down to approximately 50\%, comparable to a random guess. This result indicates that, regardless of configuration, these VC models can substantially impair the reliability of watermark detection, demonstrating their effectiveness in watermark removal. 
\vspace{-5pt}
\subsubsection{AI Distortion---TTS}
For the zero-shot configuration, the attacker feeds the text that aligns with the watermarked audio, and also provides the watermarked audio as a reference to the pre-trained TTS model, to generate a sound that seems like the watermarked audio source, without including the watermark. For the adaptive case, the attacker constructs a watermarked dataset to fine-tune the TTS model to get a better audio similarity of the victim's sound. Then inference of the fine-tuned TTS model with different text. Similar to the previous VC model evaluation, we also check the bits recovery accuracy to identify the robustness of watermarks. The result is present in Fig.~\ref{fig:tts_robust}. We observe that all evaluated watermarks are susceptible to zero-shot TTS attacks, resulting in a bit recovery accuracy of approximately 50\%, similar to a random guess. However, in cases where the attacker fine-tunes the TTS model with watermarked samples, only Timbre demonstrates robustness against TTS regeneration. Specifically, when Timbre watermarked samples are used to train the TTS model, the regenerated audio consistently retains the watermark with about 100\% accuracy, regardless of whether the vocoder is fine-tuned. This finding highlights the resilience of Timbre Watermarking in maintaining its integrity even under adversarial TTS modifications.

\noindent\textbf{Key Finding 11: All watermarks exhibit vulnerability to zero-shot TTS attacks, while only Timbre Watermarking can survive on the condition that the TTS model is fine-tuned with its watermarked samples.} 
\vspace{-10pt}
\subsection{Overall Performance}
\vspace{-10pt}
\begin{table*}[h]
\centering

\scalebox{0.9}{
\begin{tabular}{lccccccccc}
\toprule
 \textbf{Attack} &  \textbf{Wavmark} & \textbf{AudioSeal} & \textbf{Timbre} & \textbf{RobustDNN} & \textbf{audiowmark} & \textbf{Norm-space} & \textbf{Patchwork} & \textbf{FSVC} & \textbf{SilentCipher} \\
 & \cite{chen2023wavmark} & \cite{san2024proactive} & \cite{liu2023detecting} & \cite{pavlovic2022robust} & \cite{audiowmark} & \cite{saadi2019novel} & \cite{natgunanathan2017patchwork} & \cite{zhao2021desynchronization} & \cite{singh2024silentcipher} \\
\midrule

\multicolumn{10}{l}{\textbf{Signal Distortions}} \\
Pitch Shift  & \cellcolor{red!50}\xmark/\xmark &\cellcolor{red!50}\xmark/\xmark &\cellcolor{red!50}\xmark/\xmark &\cellcolor{red!50}\xmark/\xmark &\cellcolor{red!50}\xmark/\xmark &\cellcolor{red!50}\xmark/\xmark &\cellcolor{red!50}\xmark/\xmark &\cellcolor{red!50}\xmark/\xmark &\cellcolor{red!50}\xmark/\xmark \\
Time Stretch  &\cellcolor{red!20}\cmark/\xmark &\cellcolor{red!50}\xmark/\xmark &\cellcolor{green!20}\cmark/\cmark &\cellcolor{red!50}\xmark/\xmark &\cellcolor{red!50}\xmark/\xmark &\cellcolor{red!50}\xmark/\xmark &\cellcolor{green!20}\cmark/\cmark &\cellcolor{red!50}\xmark/\xmark &\cellcolor{red!50}\xmark/\xmark \\
Gaussian Noise & \cellcolor{red!20}\cmark/\xmark &\cellcolor{green!20}\cmark/\cmark &\cellcolor{red!20}\cmark/\xmark &\cellcolor{red!20}\cmark/\xmark &\cellcolor{red!50}\xmark/\xmark &\cellcolor{green!20}\cmark/\cmark &\cellcolor{red!20}\cmark/\xmark &\cellcolor{red!20}\cmark/\xmark &\cellcolor{red!50}\xmark/\xmark \\
Bitcrush  & \cellcolor{red!20}\cmark/\xmark &\cellcolor{green!20}\cmark/\cmark &\cellcolor{green!20}\cmark/\cmark &\cellcolor{red!20}\cmark/\xmark &\cellcolor{red!50}\xmark/\xmark &\cellcolor{red!20}\cmark/\xmark &\cellcolor{red!20}\cmark/\xmark &\cellcolor{red!20}\cmark/\xmark &\cellcolor{red!50}\xmark/\xmark \\
MP3 Compression & \cellcolor{red!50}\xmark/\xmark &\cellcolor{green!20}\cmark/\cmark &\cellcolor{green!20}\cmark/\cmark &\cellcolor{red!20}\cmark/\xmark &\cellcolor{red!50}\xmark/\xmark &\cellcolor{red!20}\cmark/\xmark &\cellcolor{red!20}\cmark/\xmark &\cellcolor{green!20}\cmark/\cmark &\cellcolor{red!50}\xmark/\xmark \\
Background Noise  & \cellcolor{green!20}\cmark/\cmark &\cellcolor{green!20}\cmark/\cmark &\cellcolor{green!20}\cmark/\cmark &\cellcolor{green!20}\cmark/\cmark &\cellcolor{red!20}\cmark/\xmark &\cellcolor{red!20}\cmark/\xmark &\cellcolor{red!20}\cmark/\xmark &\cellcolor{green!20}\cmark/\cmark &\cellcolor{red!20}\cmark/\xmark \\

High-Pass Filter  & \cellcolor{green!20}\cmark/\cmark &\cellcolor{green!20}\cmark/\cmark &\cellcolor{green!20}\cmark/\cmark &\cellcolor{green!20}\cmark/\cmark &\cellcolor{green!20}\cmark/\cmark &\cellcolor{red!50}\xmark/\xmark &\cellcolor{green!20}\cmark/\cmark &\cellcolor{green!20}\cmark/\cmark &\cellcolor{green!20}\cmark/\cmark \\
Low-Pass Filter  & \cellcolor{red!20}\cmark/\xmark &\cellcolor{green!20}\cmark/\cmark &\cellcolor{green!20}\cmark/\cmark &\cellcolor{red!20}\cmark/\xmark &\cellcolor{red!20}\cmark/\xmark &\cellcolor{green!20}\cmark/\cmark &\cellcolor{red!20}\cmark/\xmark &\cellcolor{green!20}\cmark/\cmark &\cellcolor{red!20}\cmark/\xmark \\

Sample Suppression   & \cellcolor{red!20}\cmark/\xmark &\cellcolor{green!20}\cmark/\cmark &\cellcolor{green!20}\cmark/\cmark &\cellcolor{green!20}\cmark/\cmark &\cellcolor{red!20}\cmark/\xmark &\cellcolor{green!20}\cmark/\cmark &\cellcolor{red!20}\cmark/\xmark &\cellcolor{red!20}\cmark/\xmark &\cellcolor{red!50}\xmark/\xmark \\

Resampling   & \cellcolor{red!20}\cmark/\xmark &\cellcolor{green!20}\cmark/\cmark &\cellcolor{green!20}\cmark/\cmark &\cellcolor{red!20}\cmark/\xmark &\cellcolor{red!20}\cmark/\xmark &\cellcolor{green!20}\cmark/\cmark &\cellcolor{red!20}\cmark/\xmark &\cellcolor{green!20}\cmark/\cmark &\cellcolor{red!20}\cmark/\xmark \\

Cutting Audio  & \cellcolor{green!20}\cmark/\cmark &\cellcolor{red!50}\xmark/\xmark &\cellcolor{green!20}\cmark/\cmark &\cellcolor{red!50}\xmark/\xmark &\cellcolor{green!20}\cmark/\cmark &\cellcolor{red!50}\xmark/\xmark &\cellcolor{red!50}\xmark/\xmark &\cellcolor{red!50}\xmark/\xmark &\cellcolor{red!50}\xmark/\xmark \\

Impulse Response & \cellcolor{green!20}\cmark/\cmark &\cellcolor{red!50}\xmark/\xmark &\cellcolor{green!20}\cmark/\cmark &\cellcolor{red!50}\xmark/\xmark &\cellcolor{green!20}\cmark/\cmark &\cellcolor{red!50}\xmark/\xmark &\cellcolor{red!50}\xmark/\xmark &\cellcolor{red!50}\xmark/\xmark &\cellcolor{red!50}\xmark/\xmark \\

\midrule
\multicolumn{10}{l}{\textbf{Physical Distortions}} \\
Different Distances & \cellcolor{red!20}\cmark/\xmark/\xmark &\cellcolor{red!50}\xmark/\xmark/\xmark  &\cellcolor{red!20}\cmark/\cmark/\xmark &\cellcolor{red!50}\xmark/\xmark/\xmark &\cellcolor{red!50}\xmark/\xmark/\xmark &\cellcolor{red!50}\xmark/\xmark/\xmark &\cellcolor{red!20}\cmark/\xmark/\xmark &\cellcolor{red!50}\xmark/\xmark/\xmark &\cellcolor{red!50}\xmark/\xmark/\xmark \\ 

Different Hardware & \cellcolor{green!20}\cmark/\cmark/\cmark &\cellcolor{red!50}\xmark/\xmark/\xmark  &\cellcolor{green!20}\cmark/\cmark/\cmark &\cellcolor{red!50}\xmark/\xmark/\xmark &\cellcolor{red!50}\xmark/\xmark/\xmark &\cellcolor{red!50}\xmark/\xmark/\xmark &\cellcolor{red!20}\cmark/\xmark/\xmark &\cellcolor{red!50}\xmark/\xmark/\xmark &\cellcolor{red!50}\xmark/\xmark/\xmark \\

\midrule
\multicolumn{10}{l}{\textbf{AI Distortions}} \\
Adaln-VC & \cellcolor{red!50}\xmark & \cellcolor{red!50}\xmark & \cellcolor{red!50}\xmark &- &\cellcolor{red!50}\xmark &\cellcolor{red!50}\xmark &\cellcolor{red!50}\xmark &\cellcolor{red!50}\xmark &\cellcolor{red!50}\xmark \\
Fragment-VC & \cellcolor{red!50}\xmark & - & \cellcolor{red!50}\xmark &- &\cellcolor{red!50}\xmark &\cellcolor{red!50}\xmark &\cellcolor{red!50}\xmark &\cellcolor{red!50}\xmark &\cellcolor{red!50}\xmark \\
Medium-VC & \cellcolor{red!50}\xmark & \cellcolor{red!50}\xmark & \cellcolor{red!50}\xmark &- &\cellcolor{red!50}\xmark &\cellcolor{red!50}\xmark &\cellcolor{red!50}\xmark &\cellcolor{red!50}\xmark &\cellcolor{red!50}\xmark \\
YourTTS-VC & \cellcolor{red!50}\xmark & \cellcolor{red!50}\xmark & \cellcolor{red!50}\xmark &- &\cellcolor{red!50}\xmark &\cellcolor{red!50}\xmark &\cellcolor{red!50}\xmark &\cellcolor{red!50}\xmark &\cellcolor{red!50}\xmark \\
RVC & \cellcolor{red!50}\xmark & - & \cellcolor{red!50}\xmark &- &\cellcolor{red!50}\xmark &\cellcolor{red!50}\xmark &\cellcolor{red!50}\xmark &\cellcolor{red!50}\xmark &\cellcolor{red!50}\xmark \\
Tacotron2+G-Lim & \cellcolor{red!50}\xmark & - & \cellcolor{green!20}\cmark &- &\cellcolor{red!50}\xmark &\cellcolor{red!50}\xmark &\cellcolor{red!50}\xmark &\cellcolor{red!50}\xmark &\cellcolor{red!50}\xmark \\
Tacotron2+H-GAN & \cellcolor{red!50}\xmark & - & \cellcolor{green!20}\cmark &- &\cellcolor{red!50}\xmark &\cellcolor{red!50}\xmark &\cellcolor{red!50}\xmark &\cellcolor{red!50}\xmark &\cellcolor{red!50}\xmark \\
Tacotron2+H-GAN* & \cellcolor{red!50}\xmark & - & \cellcolor{green!20}\cmark &- &\cellcolor{red!50}\xmark &\cellcolor{red!50}\xmark &\cellcolor{red!50}\xmark &\cellcolor{red!50}\xmark &\cellcolor{red!50}\xmark \\
Fastspeech2+G-Lim & \cellcolor{red!50}\xmark & - & \cellcolor{green!20}\cmark &- &\cellcolor{red!50}\xmark &\cellcolor{red!50}\xmark &\cellcolor{red!50}\xmark &\cellcolor{red!50}\xmark &\cellcolor{red!50}\xmark \\
Fastspeech2+H-GAN & \cellcolor{red!50}\xmark & - & \cellcolor{green!20}\cmark &- &\cellcolor{red!50}\xmark &\cellcolor{red!50}\xmark &\cellcolor{red!50}\xmark &\cellcolor{red!50}\xmark &\cellcolor{red!50}\xmark \\
Fastspeech2+H-GAN* & \cellcolor{red!50}\xmark & - & \cellcolor{green!20}\cmark &- &\cellcolor{red!50}\xmark &\cellcolor{red!50}\xmark &\cellcolor{red!50}\xmark &\cellcolor{red!50}\xmark &\cellcolor{red!50}\xmark \\
YourTTS & \cellcolor{red!50}\xmark &\cellcolor{red!50}\xmark & \cellcolor{red!50}\xmark &- &\cellcolor{red!50}\xmark &\cellcolor{red!50}\xmark &\cellcolor{red!50}\xmark &\cellcolor{red!50}\xmark &\cellcolor{red!50}\xmark \\

\bottomrule
\end{tabular}
}
\caption{Comparison of Watermark Removal Attacks across Different Watermarking Schemes}
\label{tab:watermark_removal_attacks}
\end{table*}
To summarize our evaluation, we create a table (Table~\ref{tab:watermark_removal_attacks}) to present our findings.
In this table, we assess the robustness of all 9 watermarks against 3 types of distortions. Detailed explanations and observations can be found in Appendix E.

\section{Discussion}\vspace{-10pt}
\subsection{Are watermarks a reliable solution?}
\vspace{-10pt}
Based on the extensive measurements conducted in this study, our findings suggest a \textbf{negative answer}. Audio watermarking, despite its potential as an IP safeguard, suffers from inherent limitations that render it unreliable in practice. These limitations can be broadly categorized into two main issues: \textbf{vulnerability} and \textbf{irrevocability}.

\noindent\textbf{Vulnerability of Audio Watermarks}  
The vulnerability of audio watermarks arises from their architecture, optimization goals, and the dynamic nature of adversarial threats. Audio watermarks are inherently susceptible to diverse attack methods, many of which they were not designed to counter. With the rise of generative AI, which can modify audio rapidly and adaptively, the robustness of watermarks is increasingly challenged. Each new AI model or technique has the potential to compromise a watermark’s integrity, highlighting the current solutions' inability to keep pace with evolving threats. Without fundamental improvements, audio watermarks are unlikely to provide reliable long-term protection for audio IP in the face of advancing generative models.

\noindent\textbf{Irrevocability of Audio Watermarks}  
A major limitation of audio watermarking is its unpatchable nature. Unlike traditional security systems that can be updated to address vulnerabilities, audio watermarks are static once embedded in content. If an attack exploits a vulnerability, the watermark cannot be updated or replaced, leaving it permanently compromised. This lack of adaptability means that as new attack strategies emerge, all watermarks are eventually at risk of breach.

\subsection{Strategies to Strengthen Watermarks}
\vspace{-10pt}
To enhance the robustness of audio watermarks against evolving attacks, particularly in the context of generative AI, several strategies can be considered based on insights from the current watermarking study.

\noindent\textbf{Expand Attack Scenarios During Optimization:}  
Building on the resilience demonstrated by the Timbre watermark, robustness can be improved by incorporating a wide range of attack scenarios during the training phase. By exposing the watermark to diverse distortions, including signal-level, physical, and AI-induced attacks, the model can learn to anticipate and resist these variations. 

\noindent\textbf{Develop Content-Adaptive Watermarks:}  
Designing watermarks that adapt to the unique properties of each audio sample can improve robustness. Unlike static patterns, content-adaptive watermarks dynamically adjust their embedding characteristics based on the audio’s specific features, making it harder for attackers to use a generalized removal strategy. This approach enhances both robustness and imperceptibility by tailoring the watermark to its host audio.


\noindent\textbf{Create Patchable Watermarks:}  
Introducing a framework for patchable watermarks could address their immutability. Such watermarks would allow periodic updates or reinforcement without re-encoding the original audio. By embedding a flexible “base” watermark that can evolve as vulnerabilities are discovered, this approach enables the watermark to adapt to new threats, extending its effectiveness as an IP protection mechanism.

\section{Conclusion}\vspace{-11pt}
In this paper, we review audio watermarking schemes. We analyze vulnerabilities through signal-level, physical distortion, and AI-induced distortion attacks, finding all watermarks susceptible to VC and TTS models. Our results reveal that increased distances and generative AI attacks effectively weaken watermark robustness.

\bibliographystyle{IEEEtran}
\bibliography{bibliography}

\begin{thebibliography}{10}
\providecommand{\url}[1]{#1}
\csname url@samestyle\endcsname
\providecommand{\newblock}{\relax}
\providecommand{\bibinfo}[2]{#2}
\providecommand{\BIBentrySTDinterwordspacing}{\spaceskip=0pt\relax}
\providecommand{\BIBentryALTinterwordstretchfactor}{4}
\providecommand{\BIBentryALTinterwordspacing}{\spaceskip=\fontdimen2\font plus
\BIBentryALTinterwordstretchfactor\fontdimen3\font minus \fontdimen4\font\relax}
\providecommand{\BIBforeignlanguage}[2]{{%
\expandafter\ifx\csname l@#1\endcsname\relax
\typeout{** WARNING: IEEEtran.bst: No hyphenation pattern has been}%
\typeout{** loaded for the language `#1'. Using the pattern for}%
\typeout{** the default language instead.}%
\else
\language=\csname l@#1\endcsname
\fi
#2}}
\providecommand{\BIBdecl}{\relax}
\BIBdecl

\bibitem{boney1996digital}
L.~Boney, A.~H. Tewfik, and K.~N. Hamdy, ``Digital watermarks for audio signals,'' in \emph{Proceedings of the Third IEEE International Conference on Multimedia Computing and Systems}.\hskip 1em plus 0.5em minus 0.4em\relax IEEE, 1996, pp. 473--480.

\bibitem{cox1997secure}
I.~J. Cox, J.~Kilian, F.~T. Leighton, and T.~Shamoon, ``Secure spread spectrum watermarking for multimedia,'' \emph{IEEE Transactions on Image Processing}, vol.~6, no.~12, pp. 1673--1687, 1997.

\bibitem{harvard2023ai}
H.~L. School, ``Ai created a song mimicking the work of drake and the weeknd. what does that mean for copyright law?'' \url{https://tinyurl.com/4vr86kuz}, 2023, accessed: 2024-10-15.

\bibitem{fortune2019deepfake}
Fortune, ``Scammers used ai to mimic a ceo’s voice in unusual cybercrime case,'' \url{https://fortune.com/2019/09/04/ai-deepfake-scammer-ceo-voice-fraud/}, 2019, accessed: 2023-10-15.

\bibitem{LI2021171}
\BIBentryALTinterwordspacing
Y.~Li, H.~Wang, and M.~Barni, ``A survey of deep neural network watermarking techniques,'' \emph{Neurocomputing}, vol. 461, pp. 171--193, 2021. [Online]. Available: \url{https://www.sciencedirect.com/science/article/pii/S092523122101095X}
\BIBentrySTDinterwordspacing

\bibitem{Lukas2021SoKHR}
\BIBentryALTinterwordspacing
N.~Lukas, E.~Jiang, X.~Li, and F.~Kerschbaum, ``Sok: How robust is image classification deep neural network watermarking?'' \emph{2022 IEEE Symposium on Security and Privacy (SP)}, pp. 787--804, 2021. [Online]. Available: \url{https://api.semanticscholar.org/CorpusID:236975869}
\BIBentrySTDinterwordspacing

\bibitem{hua2016twenty}
G.~Hua, J.~Huang, Y.~Q. Shi, J.~Goh, and V.~L. Thing, ``Twenty years of digital audio watermarking—a comprehensive review,'' \emph{Signal processing}, vol. 128, pp. 222--242, 2016.

\bibitem{liu2024audiomarkbench}
H.~Liu, M.~Guo, Z.~Jiang, L.~Wang, and N.~Gong, ``Audiomarkbench: Benchmarking robustness of audio watermarking,'' \emph{Advances in Neural Information Processing Systems}, vol.~37, pp. 52\,241--52\,265, 2024.

\bibitem{liu2023detecting}
C.~Liu, J.~Zhang, T.~Zhang, X.~Yang, W.~Zhang, and N.~Yu, ``Detecting voice cloning attacks via timbre watermarking,'' \emph{arXiv preprint arXiv:2312.03410}, 2023.

\bibitem{san2024proactive}
R.~San~Roman, P.~Fernandez, H.~Elsahar, A.~D{\'e}fossez, T.~Furon, and T.~Tran, ``Proactive detection of voice cloning with localized watermarking,'' in \emph{International Conference on Machine Learning}, vol. 235, 2024.

\bibitem{chen2023wavmark}
G.~Chen, Y.~Wu, S.~Liu, T.~Liu, X.~Du, and F.~Wei, ``Wavmark: Watermarking for audio generation,'' \emph{arXiv preprint arXiv:2308.12770}, 2023.

\bibitem{singh2024silentcipher}
M.~K. Singh, N.~Takahashi, W.~Liao, and Y.~Mitsufuji, ``Silentcipher: Deep audio watermarking,'' \emph{arXiv preprint arXiv:2406.03822}, 2024.

\bibitem{audiowmark}
S.~Westerfeld, ``{audiowmark}: Robust audio watermarking library,'' \url{https://github.com/swesterfeld/audiowmark}, accessed: 2024-10-30.

\bibitem{pavlovic2022robust}
K.~Pavlovi{\'c}, S.~Kova{\v{c}}evi{\'c}, I.~Djurovi{\'c}, and A.~Wojciechowski, ``Robust speech watermarking by a jointly trained embedder and detector using a dnn,'' \emph{Digital Signal Processing}, vol. 122, p. 103381, 2022.

\bibitem{natgunanathan2017patchwork}
I.~Natgunanathan, Y.~Xiang, G.~Hua, G.~Beliakov, and J.~Yearwood, ``Patchwork-based multilayer audio watermarking,'' \emph{IEEE/ACM Transactions on Audio, Speech, and Language Processing}, vol.~25, no.~11, pp. 2176--2187, 2017.

\bibitem{saadi2019novel}
S.~Saadi, A.~Merrad, and A.~Benziane, ``Novel secured scheme for blind audio/speech norm-space watermarking by arnold algorithm,'' \emph{Signal Processing}, vol. 154, pp. 74--86, 2019.

\bibitem{zhao2021desynchronization}
J.~Zhao, T.~Zong, Y.~Xiang, L.~Gao, W.~Zhou, and G.~Beliakov, ``Desynchronization attacks resilient watermarking method based on frequency singular value coefficient modification,'' \emph{IEEE/ACM Transactions on Audio, Speech, and Language Processing}, vol.~29, pp. 2282--2295, 2021.

\bibitem{kingma2022autoencodingvariationalbayes}
\BIBentryALTinterwordspacing
D.~P. Kingma and M.~Welling, ``Auto-encoding variational bayes,'' 2022. [Online]. Available: \url{https://arxiv.org/abs/1312.6114}
\BIBentrySTDinterwordspacing

\bibitem{goodfellow2014generativeadversarialnetworks}
\BIBentryALTinterwordspacing
I.~J. Goodfellow, J.~Pouget-Abadie, M.~Mirza, B.~Xu, D.~Warde-Farley, S.~Ozair, A.~Courville, and Y.~Bengio, ``Generative adversarial networks,'' 2014. [Online]. Available: \url{https://arxiv.org/abs/1406.2661}
\BIBentrySTDinterwordspacing

\bibitem{ho2020denoising}
\BIBentryALTinterwordspacing
J.~Ho, A.~Jain, and P.~Abbeel, ``Denoising diffusion probabilistic models,'' 2020. [Online]. Available: \url{https://arxiv.org/abs/2006.11239}
\BIBentrySTDinterwordspacing

\bibitem{rombach2022highresolutionimagesynthesislatent}
\BIBentryALTinterwordspacing
R.~Rombach, A.~Blattmann, D.~Lorenz, P.~Esser, and B.~Ommer, ``High-resolution image synthesis with latent diffusion models,'' 2022. [Online]. Available: \url{https://arxiv.org/abs/2112.10752}
\BIBentrySTDinterwordspacing

\bibitem{caillon2021ravevariationalautoencoderfast}
\BIBentryALTinterwordspacing
A.~Caillon and P.~Esling, ``Rave: A variational autoencoder for fast and high-quality neural audio synthesis,'' 2021. [Online]. Available: \url{https://arxiv.org/abs/2111.05011}
\BIBentrySTDinterwordspacing

\bibitem{kong2020hifigangenerativeadversarialnetworks}
\BIBentryALTinterwordspacing
J.~Kong, J.~Kim, and J.~Bae, ``Hifi-gan: Generative adversarial networks for efficient and high fidelity speech synthesis,'' 2020. [Online]. Available: \url{https://arxiv.org/abs/2010.05646}
\BIBentrySTDinterwordspacing

\bibitem{liu2023audioldmtexttoaudiogenerationlatent}
\BIBentryALTinterwordspacing
H.~Liu, Z.~Chen, Y.~Yuan, X.~Mei, X.~Liu, D.~Mandic, W.~Wang, and M.~D. Plumbley, ``Audioldm: Text-to-audio generation with latent diffusion models,'' 2023. [Online]. Available: \url{https://arxiv.org/abs/2301.12503}
\BIBentrySTDinterwordspacing

\bibitem{shen2018natural}
J.~Shen, R.~Pang, R.~J. Weiss, M.~Schuster, N.~Jaitly, Z.~Yang, Z.~Chen, Y.~Zhang, Y.~Wang, R.~Skerry-Ryan \emph{et~al.}, ``Natural tts synthesis by conditioning wavenet on mel spectrogram predictions,'' in \emph{International Conference on Acoustics, Speech and Signal Processing (ICASSP)}, 2018, pp. 4779--4783.

\bibitem{dhariwal2020jukebox}
P.~Dhariwal, H.~Jun, C.~Payne, J.~W. Kim, A.~Radford, and I.~Sutskever, ``Jukebox: A generative model for music,'' \emph{arXiv preprint arXiv:2005.00341}, 2020.

\bibitem{arik2018neural}
\BIBentryALTinterwordspacing
S.~Arik, J.~Chen, K.~Peng, W.~Ping, and Y.~Zhou, ``Neural voice cloning with a few samples,'' in \emph{Advances in Neural Information Processing Systems}, S.~Bengio, H.~Wallach, H.~Larochelle, K.~Grauman, N.~Cesa-Bianchi, and R.~Garnett, Eds., vol.~31.\hskip 1em plus 0.5em minus 0.4em\relax Curran Associates, Inc., 2018. [Online]. Available: \url{https://proceedings.neurips.cc/paper_files/paper/2018/file/4559912e7a94a9c32b09d894f2bc3c82-Paper.pdf}
\BIBentrySTDinterwordspacing

\bibitem{ruder2017overviewgradientdescentoptimization}
\BIBentryALTinterwordspacing
S.~Ruder, ``An overview of gradient descent optimization algorithms,'' 2017. [Online]. Available: \url{https://arxiv.org/abs/1609.04747}
\BIBentrySTDinterwordspacing

\bibitem{Ren2024CopyrightPI}
\BIBentryALTinterwordspacing
J.~Ren, H.~Xu, P.~He, Y.~Cui, S.~Zeng, J.~Zhang, H.~Wen, J.~Ding, H.~Liu, Y.~Chang, and J.~Tang, ``Copyright protection in generative ai: A technical perspective,'' \emph{ArXiv}, vol. abs/2402.02333, 2024. [Online]. Available: \url{https://api.semanticscholar.org/CorpusID:267412857}
\BIBentrySTDinterwordspacing

\bibitem{cox2002digital}
I.~J. Cox, M.~L. Miller, J.~A. Bloom, J.~Fridrich, and T.~Kalker, \emph{Digital Watermarking and Steganography}.\hskip 1em plus 0.5em minus 0.4em\relax Morgan Kaufmann, 2002.

\bibitem{ERFANI2009809}
\BIBentryALTinterwordspacing
Y.~Erfani and S.~Siahpoush, ``Robust audio watermarking using improved ts echo hiding,'' \emph{Digital Signal Processing}, vol.~19, no.~5, pp. 809--814, 2009. [Online]. Available: \url{https://www.sciencedirect.com/science/article/pii/S1051200409000426}
\BIBentrySTDinterwordspacing

\bibitem{Kanhe_2015}
A.~Kanhe, G.~Aghila, C.~Y.~S. Kiran, C.~H. Ramesh, G.~Jadav, and M.~G. Raj, ``Robust audio steganography based on advanced encryption standards in temporal domain,'' in \emph{2015 International Conference on Advances in Computing, Communications and Informatics (ICACCI)}, 2015, pp. 1449--1453.

\bibitem{Hua_2015}
G.~Hua, J.~Goh, and V.~L.~L. Thing, ``Cepstral analysis for the application of echo-based audio watermark detection,'' \emph{IEEE Transactions on Information Forensics and Security}, vol.~10, no.~9, pp. 1850--1861, 2015.

\bibitem{Chetan2021}
M.~Chetan, P.~P. Bhat, V.~Shet, S.~B. Husenbhai, and A.~Bhat, ``Audio watermarking using modified least significant bit technique,'' in \emph{2021 International Conference on Circuits, Controls and Communications (CCUBE)}, 2021, pp. 1--5.

\bibitem{Cvejic2004}
N.~Cvejic and T.~Seppanen, ``Increasing robustness of lsb audio steganography using a novel embedding method,'' in \emph{International Conference on Information Technology: Coding and Computing, 2004. Proceedings. ITCC 2004.}, vol.~2, 2004, pp. 533--537 Vol.2.

\bibitem{Wei2009}
F.~S. Wei and D.~Qi, ``Audio watermarking of stereo signals based on echo-hiding method,'' in \emph{Proceedings of the 7th International Conference on Information, Communications and Signal Processing}, ser. ICICS'09.\hskip 1em plus 0.5em minus 0.4em\relax IEEE Press, 2009, p. 485–488.

\bibitem{HUA2016222}
\BIBentryALTinterwordspacing
G.~Hua, J.~Huang, Y.~Q. Shi, J.~Goh, and V.~L. Thing, ``Twenty years of digital audio watermarking—a comprehensive review,'' \emph{Signal Processing}, vol. 128, pp. 222--242, 2016. [Online]. Available: \url{https://www.sciencedirect.com/science/article/pii/S0165168416300263}
\BIBentrySTDinterwordspacing

\bibitem{Karajeh_2019}
\BIBentryALTinterwordspacing
H.~Karajeh, T.~Khatib, L.~Rajab, and M.~Maqableh, ``A robust digital audio watermarking scheme based on dwt and schur decomposition,'' \emph{Multimedia Tools Appl.}, vol.~78, no.~13, p. 18395–18418, Jul. 2019. [Online]. Available: \url{https://doi.org/10.1007/s11042-019-7214-3}
\BIBentrySTDinterwordspacing

\bibitem{Singh2014}
\BIBentryALTinterwordspacing
J.~Singh, P.~Garg, and A.~De, ``Multiplicative watermarking of audio in dft magnitude,'' \emph{Multimedia Tools and Applications}, vol.~71, no.~3, pp. 1431--1453, 2014. [Online]. Available: \url{https://doi.org/10.1007/s11042-012-1282-y}
\BIBentrySTDinterwordspacing

\bibitem{CVEJIC2004207}
\BIBentryALTinterwordspacing
N.~Cvejic and T.~Seppänen, ``Spread spectrum audio watermarking using frequency hopping and attack characterization,'' \emph{Signal Processing}, vol.~84, no.~1, pp. 207--213, 2004. [Online]. Available: \url{https://www.sciencedirect.com/science/article/pii/S0165168403002810}
\BIBentrySTDinterwordspacing

\bibitem{QU2023}
\BIBentryALTinterwordspacing
X.~Qu, X.~Yin, P.~Wei, L.~Lu, and Z.~Ma, ``Audioqr: deep neural audio watermarks for qr code,'' in \emph{Proceedings of the Thirty-Second International Joint Conference on Artificial Intelligence}, ser. IJCAI '23, 2023. [Online]. Available: \url{https://doi.org/10.24963/ijcai.2023/687}
\BIBentrySTDinterwordspacing

\bibitem{Nishimura2014AudioWB}
\BIBentryALTinterwordspacing
A.~Nishimura, ``Audio watermarking based on amplitude modulation and modulation masking,'' in \emph{IWIHC '14}, 2014. [Online]. Available: \url{https://api.semanticscholar.org/CorpusID:5872930}
\BIBentrySTDinterwordspacing

\bibitem{Ko2005TimespreadEM}
\BIBentryALTinterwordspacing
B.-S. Ko, R.~Nishimura, and Y.~Suzuki, ``Time-spread echo method for digital audio watermarking,'' \emph{IEEE Transactions on Multimedia}, vol.~7, pp. 212--221, 2005. [Online]. Available: \url{https://api.semanticscholar.org/CorpusID:12946882}
\BIBentrySTDinterwordspacing

\bibitem{Yong_dual-channel}
Y.~Xiang, I.~Natgunanathan, D.~Peng, W.~Zhou, and S.~Yu, ``A dual-channel time-spread echo method for audio watermarking,'' \emph{IEEE Transactions on Information Forensics and Security}, vol.~7, no.~2, pp. 383--392, 2012.

\bibitem{Natgunanathan2009ANP}
\BIBentryALTinterwordspacing
I.~Natgunanathan and Y.~Xiang, ``A novel pseudonoise sequence for time-spread echo based audio watermarking,'' \emph{GLOBECOM 2009 - 2009 IEEE Global Telecommunications Conference}, pp. 1--6, 2009. [Online]. Available: \url{https://api.semanticscholar.org/CorpusID:206685703}
\BIBentrySTDinterwordspacing

\bibitem{Guofu+M-sequency}
G.~Zhang, L.~Zheng, Z.~Su, Y.~Zeng, and G.~Wang, ``M-sequences and sliding window based audio watermarking robust against large-scale cropping attacks,'' \emph{IEEE Transactions on Information Forensics and Security}, vol.~18, pp. 1182--1195, 2023.

\bibitem{HU2014115}
\BIBentryALTinterwordspacing
H.-T. Hu, L.-Y. Hsu, and H.-H. Chou, ``Variable-dimensional vector modulation for perceptual-based dwt blind audio watermarking with adjustable payload capacity,'' \emph{Digital Signal Processing}, vol.~31, pp. 115--123, 2014. [Online]. Available: \url{https://www.sciencedirect.com/science/article/pii/S1051200414001365}
\BIBentrySTDinterwordspacing

\bibitem{Lalitha_DCT}
N.~V. Lalitha, P.~V. Prasad, and S.~U. Rao, ``Performance analysis of dct and dwt audio watermarking based on svd,'' in \emph{2016 International Conference on Circuit, Power and Computing Technologies (ICCPCT)}, 2016, pp. 1--5.

\bibitem{qiuling_adaptive}
\BIBentryALTinterwordspacing
Q.~Wu and M.~Wu, ``Adaptive and blind audio watermarking algorithm based on chaotic encryption in hybrid domain,'' \emph{Symmetry}, vol.~10, no.~7, 2018. [Online]. Available: \url{https://www.mdpi.com/2073-8994/10/7/284}
\BIBentrySTDinterwordspacing

\bibitem{Hu2015TheUO}
\BIBentryALTinterwordspacing
H.-T. Hu, L.-Y. Hsu, S.-Y. Lai, and Y.-J. Chang, ``The use of spectral shaping to extend the capacity for dwt-based blind audio watermarking,'' \emph{2015 5th International Conference on IT Convergence and Security (ICITCS)}, pp. 1--5, 2015. [Online]. Available: \url{https://api.semanticscholar.org/CorpusID:14202847}
\BIBentrySTDinterwordspacing

\bibitem{Ngo2015}
N.~M. Ngo and M.~Unoki, ``Robust and reliable audio watermarking based on phase coding,'' in \emph{2015 IEEE International Conference on Acoustics, Speech and Signal Processing (ICASSP)}, 2015, pp. 345--349.

\bibitem{Wang2021}
S.~Wang, W.~Yuan, Z.~Zhang, J.~Wang, and M.~Unoki, ``Synchronous multi-bit audio watermarking based on phase shifting,'' in \emph{ICASSP 2021 - 2021 IEEE International Conference on Acoustics, Speech and Signal Processing (ICASSP)}, 2021, pp. 2700--2704.

\bibitem{yang2024improvedphasecodingaudio}
\BIBentryALTinterwordspacing
G.~Yang, ``An improved phase coding audio steganography algorithm,'' 2024. [Online]. Available: \url{https://arxiv.org/abs/2408.13277}
\BIBentrySTDinterwordspacing

\bibitem{Esmaili_2003}
S.~Esmaili, S.~Krishnan, and K.~Raahemifar, ``A novel spread spectrum audio watermarking scheme based on time-frequency characteristics,'' in \emph{CCECE 2003 - Canadian Conference on Electrical and Computer Engineering. Toward a Caring and Humane Technology (Cat. No.03CH37436)}, vol.~3, 2003, pp. 1963--1966 vol.3.

\bibitem{zhang2020timefrequencyperspectiveaudiowatermarking}
\BIBentryALTinterwordspacing
H.~Zhang, ``A time-frequency perspective on audio watermarking,'' 2020. [Online]. Available: \url{https://arxiv.org/abs/2002.03156}
\BIBentrySTDinterwordspacing

\bibitem{Hamza_2005}
\BIBentryALTinterwordspacing
H.~\"{O}zer, B.~Sankur, and N.~Memon, ``An svd-based audio watermarking technique,'' in \emph{Proceedings of the 7th Workshop on Multimedia and Security}.\hskip 1em plus 0.5em minus 0.4em\relax New York, NY, USA: Association for Computing Machinery, 2005, p. 51–56. [Online]. Available: \url{https://doi.org/10.1145/1073170.1073180}
\BIBentrySTDinterwordspacing

\bibitem{liu2023dear}
C.~Liu, J.~Zhang, H.~Fang, Z.~Ma, W.~Zhang, and N.~Yu, ``Dear: A deep-learning-based audio re-recording resilient watermarking,'' in \emph{Proceedings of the AAAI Conference on Artificial Intelligence}, vol.~37, no.~11, 2023, pp. 13\,201--13\,209.

\bibitem{Moritz_2024}
\BIBentryALTinterwordspacing
M.~Moritz, T.~Olán, and T.~Virtanen, ``Noise-to-mask ratio loss for deep neural network based audio watermarking,'' in \emph{2024 IEEE 5th International Symposium on the Internet of Sounds (IS2)}.\hskip 1em plus 0.5em minus 0.4em\relax IEEE, Sep. 2024, p. 1–6. [Online]. Available: \url{http://dx.doi.org/10.1109/IS262782.2024.10704132}
\BIBentrySTDinterwordspacing

\bibitem{Chuxuan_enhance}
C.~Tong, I.~Natgunanathan, Y.~Xiang, J.~Li, T.~Zong, X.~Zheng, and L.~Gao, ``Enhancing robustness of speech watermarking using a transformer-based framework exploiting acoustic features,'' \emph{IEEE/ACM Transactions on Audio, Speech, and Language Processing}, vol.~32, pp. 4822--4837, 2024.

\bibitem{Maha_non-security}
M.~Charfeddine, E.~Mezghani, S.~Masmoudi, C.~B. Amar, and H.~Alhumyani, ``Audio watermarking for security and non-security applications,'' \emph{IEEE Access}, vol.~10, pp. 12\,654--12\,677, 2022.

\bibitem{li2024ideaw}
\BIBentryALTinterwordspacing
P.~Li, X.~Zhang, J.~Xiao, and J.~Wang, ``{IDEAW}: Robust neural audio watermarking with invertible dual-embedding,'' in \emph{Proceedings of the 2024 Conference on Empirical Methods in Natural Language Processing}.\hskip 1em plus 0.5em minus 0.4em\relax Miami, Florida, USA: Association for Computational Linguistics, Nov. 2024, pp. 4500--4511. [Online]. Available: \url{https://aclanthology.org/2024.emnlp-main.258}
\BIBentrySTDinterwordspacing

\bibitem{chou2019oneshotvoiceconversionseparating}
\BIBentryALTinterwordspacing
J.~chieh Chou, C.~chieh Yeh, and H.~yi~Lee, ``One-shot voice conversion by separating speaker and content representations with instance normalization,'' 2019. [Online]. Available: \url{https://arxiv.org/abs/1904.05742}
\BIBentrySTDinterwordspacing

\bibitem{lin2021fragmentvcanytoanyvoiceconversion}
\BIBentryALTinterwordspacing
Y.~Y. Lin, C.-M. Chien, J.-H. Lin, H.~yi~Lee, and L.~shan Lee, ``Fragmentvc: Any-to-any voice conversion by end-to-end extracting and fusing fine-grained voice fragments with attention,'' 2021. [Online]. Available: \url{https://arxiv.org/abs/2010.14150}
\BIBentrySTDinterwordspacing

\bibitem{gu2021mediumvcanytoanyvoiceconversion}
\BIBentryALTinterwordspacing
Y.~Gu, Z.~Zhang, X.~Yi, and X.~Zhao, ``Mediumvc: Any-to-any voice conversion using synthetic specific-speaker speeches as intermedium features,'' 2021. [Online]. Available: \url{https://arxiv.org/abs/2110.02500}
\BIBentrySTDinterwordspacing

\bibitem{casanova2023yourttszeroshotmultispeakertts}
\BIBentryALTinterwordspacing
E.~Casanova, J.~Weber, C.~Shulby, A.~C. Junior, E.~Gölge, and M.~A. Ponti, ``Yourtts: Towards zero-shot multi-speaker tts and zero-shot voice conversion for everyone,'' 2023. [Online]. Available: \url{https://arxiv.org/abs/2112.02418}
\BIBentrySTDinterwordspacing

\bibitem{rvc_project}
RVC-Project, ``Retrieval-based voice conversion webui,'' \url{https://github.com/RVC-Project/Retrieval-based-Voice-Conversion-WebUI}, 2023.

\bibitem{ren2022fastspeech2fasthighquality}
\BIBentryALTinterwordspacing
Y.~Ren, C.~Hu, X.~Tan, T.~Qin, S.~Zhao, Z.~Zhao, and T.-Y. Liu, ``Fastspeech 2: Fast and high-quality end-to-end text to speech,'' 2022. [Online]. Available: \url{https://arxiv.org/abs/2006.04558}
\BIBentrySTDinterwordspacing

\bibitem{Liu_2019}
Z.~Liu, Y.~Huang, and J.~Huang, ``Patchwork-based audio watermarking robust against de-synchronization and recapturing attacks,'' \emph{IEEE Transactions on Information Forensics and Security}, vol.~14, no.~5, pp. 1171--1180, 2019.

\bibitem{ljspeech17}
K.~Ito and L.~Johnson, ``The lj speech dataset,'' \url{https://keithito.com/LJ-Speech-Dataset/}, 2017.

\bibitem{panayotov2015librispeech}
V.~Panayotov, G.~Chen, D.~Povey, and S.~Khudanpur, ``Librispeech: an asr corpus based on public domain audio books,'' in \emph{2015 IEEE international conference on acoustics, speech and signal processing (ICASSP)}.\hskip 1em plus 0.5em minus 0.4em\relax IEEE, 2015, pp. 5206--5210.

\bibitem{zhang2022msinger}
L.~Zhang, R.~Li, S.~Wang, L.~Deng, J.~Liu, Y.~Ren, J.~He, R.~Huang, J.~Zhu, X.~Chen, and Z.~Zhao, ``M4singer: A multi-style, multi-singer and musical score provided mandarin singing corpus,'' in \emph{Thirty-sixth Conference on Neural Information Processing Systems Datasets and Benchmarks Track}, 2022.

\bibitem{Griffin_1984}
D.~Griffin and J.~Lim, ``Signal estimation from modified short-time fourier transform,'' \emph{IEEE Transactions on Acoustics, Speech, and Signal Processing}, vol.~32, no.~2, pp. 236--243, 1984.

\bibitem{oprea2021hganpowerganshands}
\BIBentryALTinterwordspacing
S.~Oprea, G.~Karvounas, P.~Martinez-Gonzalez, N.~Kyriazis, S.~Orts-Escolano, I.~Oikonomidis, A.~Garcia-Garcia, A.~Tsoli, J.~Garcia-Rodriguez, and A.~Argyros, ``H-gan: the power of gans in your hands,'' 2021. [Online]. Available: \url{https://arxiv.org/abs/2103.15017}
\BIBentrySTDinterwordspacing

\end{thebibliography}

\section*{Appendix}\vspace{-10pt}
\label{sec:appendix}

\subsection*{A. Attack Details}

\noindent\textbf{Pitch Shift:}
Changing the pitch of the audio without affecting its duration. This alters the frequency components of the signal, therefore making it hard for watermarking algorithms relying on certain features in frequency to detect this watermarked audio. Slight changes in pitch evade the detection of a watermark but may still keep the speech recognizable, hence it finds wide applications in attacks where intelligibility is to be preserved.

\noindent\textbf{Time Stretch:}
A process that modifies the audio duration without altering its pitch. Time-stretching destroys watermarks based on temporal patterns since it stretches or compresses the signal. Speeding up or slowing down the audio may render embedded numeric patterns unreadable to watermark detectors.

\noindent{\textbf{Gaussian Noise:}} 
Addition of Gaussian or white noise to mask low-level features in which watermarks are embedded. It works in the case of perceptual watermarking schemes, which depend on more quiet signal components. Gaussian noise addition does a much better job of subtly degrading the watermark without a noticeable effect on the whole listening experience.

\noindent{\textbf{Bitcrush:}} 
A simple reduction in bit depth within the audio reduces the resolution which can have a distorting effect on the watermark. Bitcrushing properly removes the detailed noise from the audio signal, where unfortunately subtle watermarking information may reside. Bitcrushing adds a "lo-fi" effect to the audio by reducing bit depth, which can destroy watermarks relying on fine-grained signal details.

\noindent{\textbf{MP3 Compression:}} 
Apply lossy compression to shrink the size of an audio file. MP3 compression would normally discard some information in hardly audible frequency ranges; very often, information corresponding to a watermark falls in such frequency ranges. MP3 compression is a standard procedure for audio distribution, and its capabilities for discarding non-vital frequency data render it very efficient at destroying some watermarks.

\noindent{\textbf{Background Noise:}} 
Introducing ambient sound into the audio, such as white noise, street noise, or crowd murmur, camouflages the watermark features that might be in low-amplitude regions or inside a narrow frequency band. Background noise can mask the watermark notably, especially for pieces of audio intended for use in public environments, which makes the extraction process difficult.

\noindent{\textbf{Audio Cropping:}} 
The removal of sections from the audio. This destroys temporal patterns relied on for watermark detection. Audio cropping is often applied in an attempt to cut out parts of the audio that may contain the watermark. In audio splicing, this is generally effective for most watermarks embedded in repeated patterns throughout the audio duration.

\noindent{\textbf{High-Pass and Low-Pass Filter:}} 
A filter removes either the low or high-frequency audio information. This kills the watermarks set in specific bands of frequencies, making them less detectable. High-pass filtering may remove the low-frequency embedded signals, and low-pass filtering may remove the high-frequency watermark data depending upon the embedding approach of a scheme.

\noindent{\textbf{Sample Suppression:}} 
Randomly suppress or remove samples within the audio to create dropouts. This attack disrupts continuous watermark signals without affecting overall intelligibility. Sample suppression is effective in watermarking schemes relying on continuous signal properties because it introduces gaps in the watermark pattern.

\noindent{\textbf{Resampling:}} 
The process of changing the sampling rate of the audio changes its frequency spectrum and can destroy or degrade the watermark embedded at the original sampling rate. The audio is resampled into a lower- or higher-rate sample, distorting the watermark data embedded at that rate through removal or aliasing.

\subsection*{B. Watermark Reproduction Result}

\begin{figure}[t]
\centering
\subfigure[PESQ Original vs WM]{\label{fig:pesq_accuracy_radar}\includegraphics[width=0.23\textwidth]{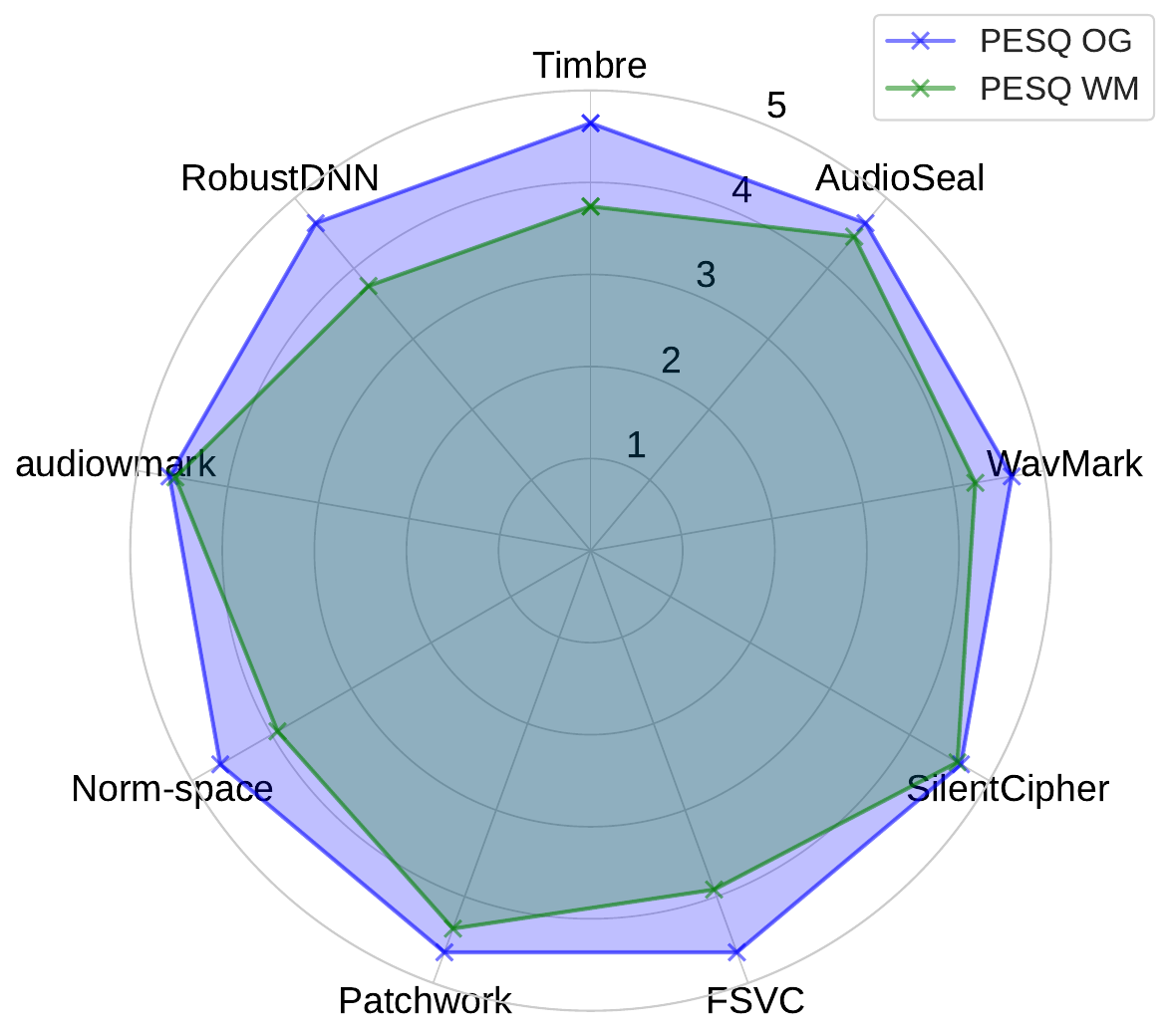}}
\subfigure[ViSQOL Original vs WM]{\label{fig:ViSQOL_accuracy_radar}\includegraphics[width=0.23\textwidth]{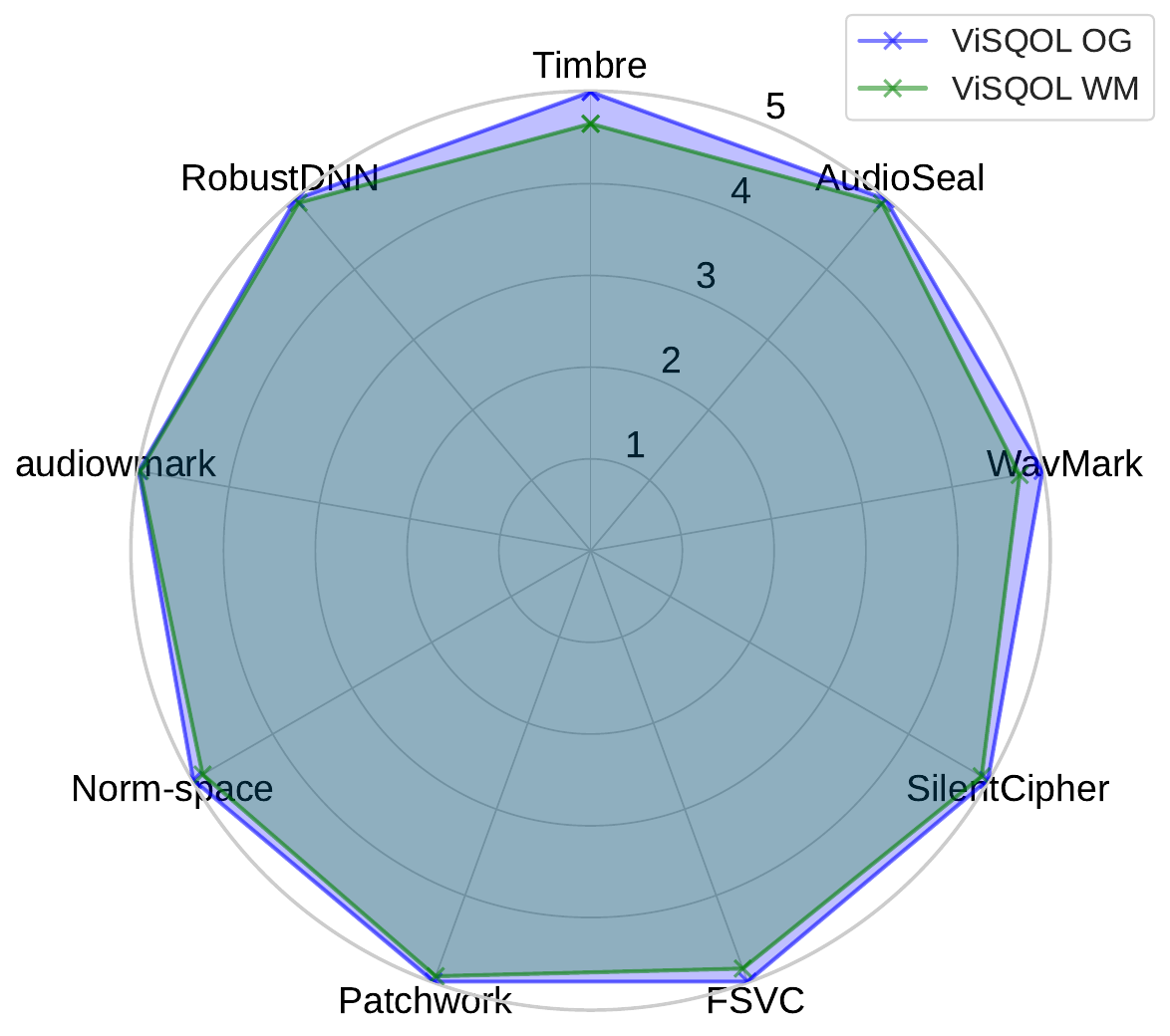}}
\vspace{-10pt}
\caption{Fidelity}
\label{fig:reproduce_fid}
\end{figure}
\noindent\textbf{Watermark Fidelity:}
use PESQ and ViSQOL as metrics. A smaller gap between the original and watermarked audio scores indicates a minimal impact of the watermark on audio quality. The results, presented in Fig.~\ref{fig:reproduce_fid}, show that the PESQ scores for the original audio are 4.6, while the watermarked audio scores vary between 3.8 and 4.6. Similarly, ViSQOL scores for the original audio are 5, with watermarked scores between 4.6 and 5. These results indicate that the watermarking process introduces only slight changes in audio quality, suggesting that all evaluated watermarks maintain high fidelity and are well-suited for applications requiring minimal perceptual degradation.

\begin{figure}[t]
\centering
\subfigure[LJspeech Acc: Original vs WM]{\label{fig:ljspeech_acc}\includegraphics[width=0.23\textwidth]{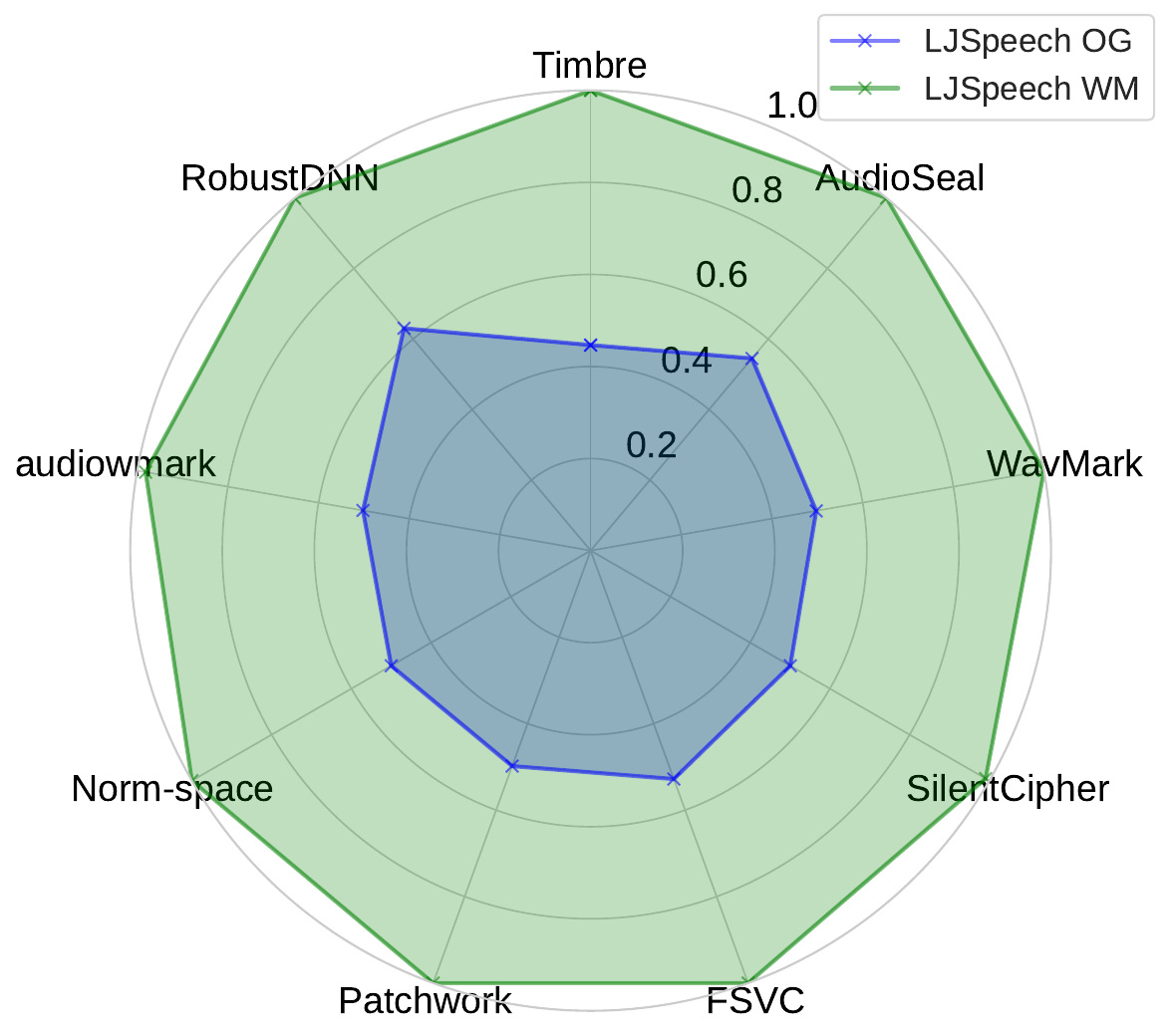}}
\subfigure[M4Singer Acc: Original vs WM]{\label{fig:music_acc}\includegraphics[width=0.23\textwidth]{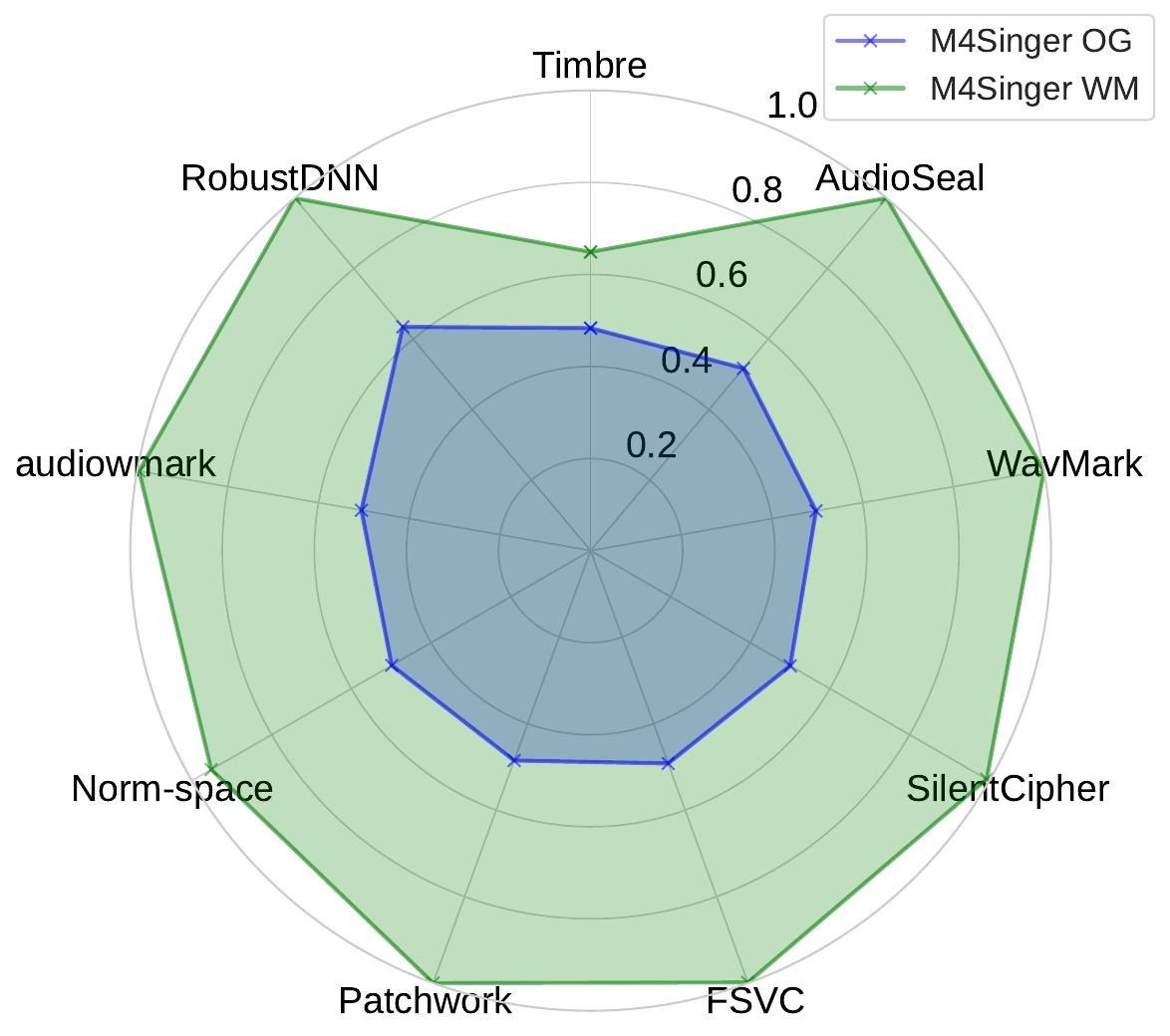}}
\vspace{-10pt}
\caption{Watermark Accuracy}
\end{figure}

\noindent\textbf{Bit Recovery Accuracy:}
In this evaluation, we assess whether watermarked audio can be accurately recovered by measuring the bit recovery accuracy. We test on two datasets: the LJspeech dataset for watermarking speech (Fig.~\ref{fig:ljspeech_acc}), and the M4Singer dataset for watermarking singing audio (Fig.~\ref{fig:music_acc}). For each dataset, we also check the original unwatermarked audio bit recovery accuracies (it is supposed to be 50\% as they do not contain any watermark). The results indicate that all watermarking methods perform well for watermarking speech, with only RobustDNN showing a slight tendency for false positives. However, the Timbre watermark does not perform as effectively using the M4Singer dataset, achieving only 65\% accuracy. This suggests limitations on the usage of watermarking music. Overall, these findings demonstrate that the evaluated watermarking methods maintain high accuracy for speech, with some variability in effectiveness when applied to singing.

\begin{figure}[h]
\centering
\subfigure[LJspeech Prob: Original vs WM]{\label{fig:ljspeech_prob}\includegraphics[width=0.23\textwidth]{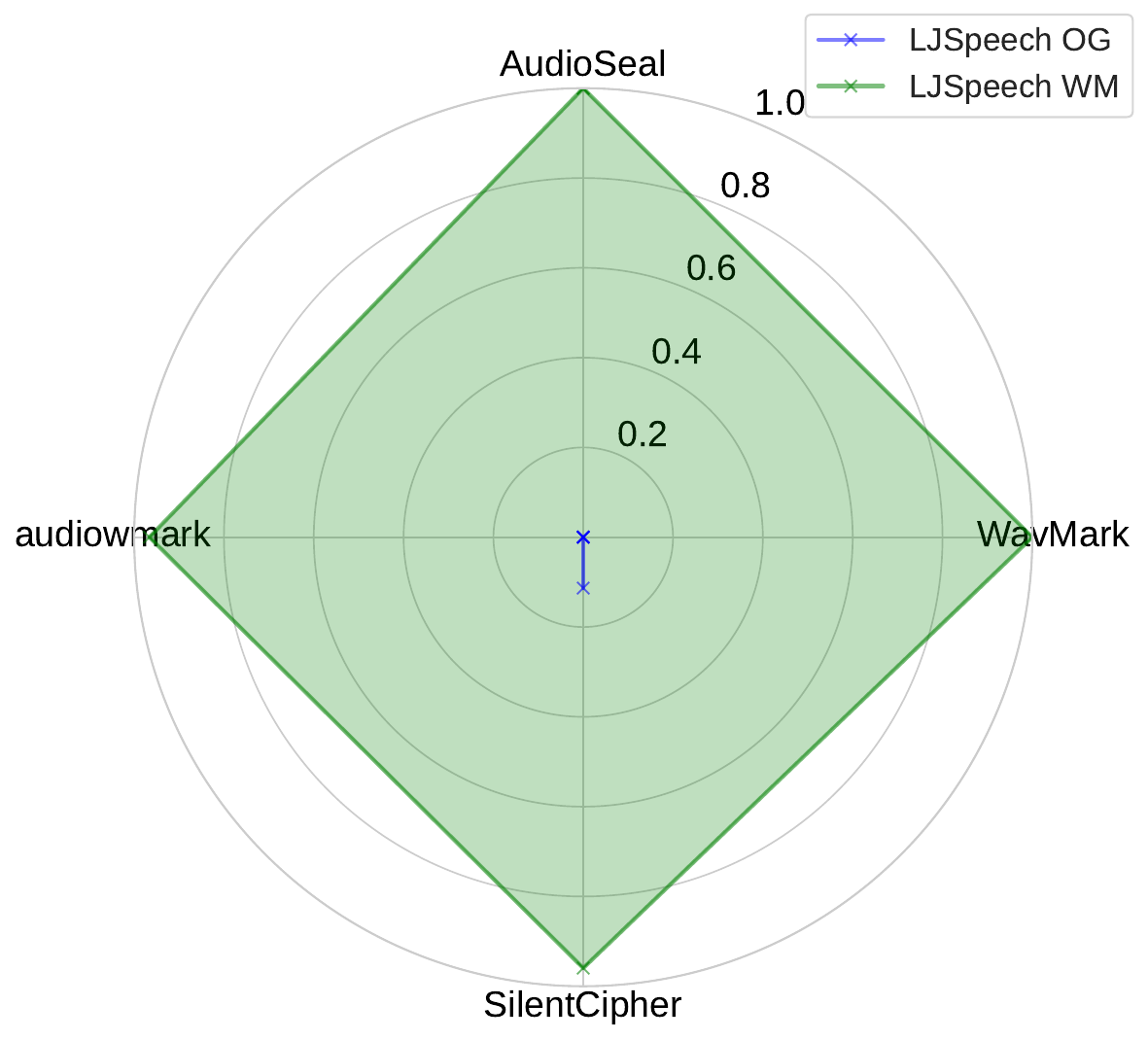}}
\subfigure[M4Singer Prob: Original vs WM]{\label{fig:music_prob}\includegraphics[width=0.23\textwidth]{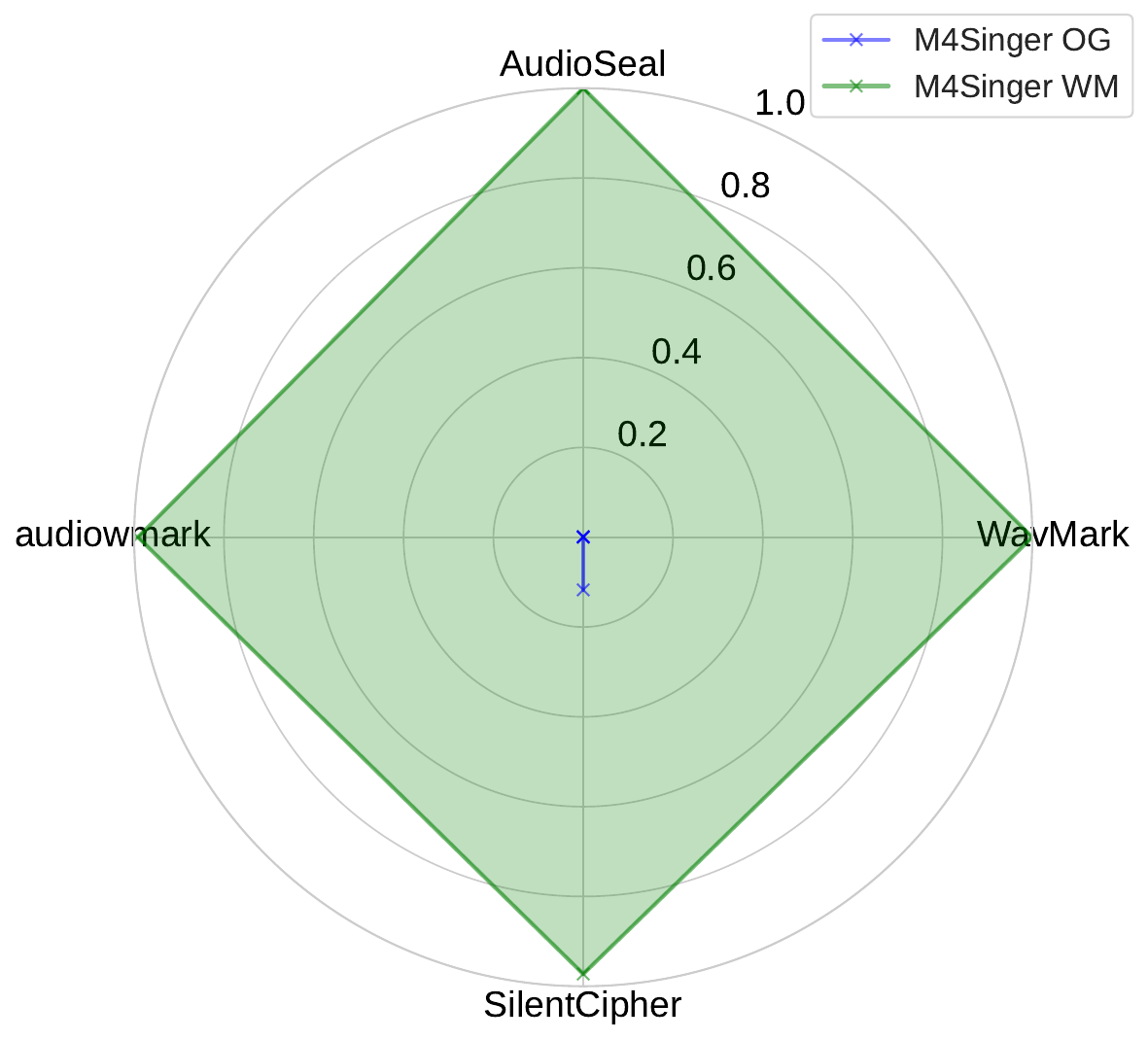}}
\vspace{-10pt}
\caption{Watermark Probability}
\label{fig:reproduce_proba}
\end{figure}
\noindent\textbf{Watermark Probability:}

Across the 9 evaluated watermarks, 4 of them offer a unique probability bit to indicate whether a given audio sample is likely to be watermarked. To assess the accuracy of these indicators, we feed both original unwatermarked and watermarked audio samples generated by different schemes into their respective watermark extractors. The results, presented in Fig.~\ref{fig:reproduce_proba}, illustrate that for both the LJspeech (speech) dataset and the M4Singer (singing) dataset, the probability bit consistently identifies watermarked samples, as the watermarked sample achieves watermark probability as 1, and the unwatermarked samples have watermark probability as near 0. This suggests that these four watermarking schemes reliably detect the presence of a watermark with minimal false positives, demonstrating robustness in identifying watermarked audio across different content types.

\subsection*{C. Detailed Effective Signal Distortion}
\begin{figure*}[t]
\centering     
\subfigure[Pitch Shift]{\label{fig:pitchshift_line}\includegraphics[width=0.23\textwidth]{{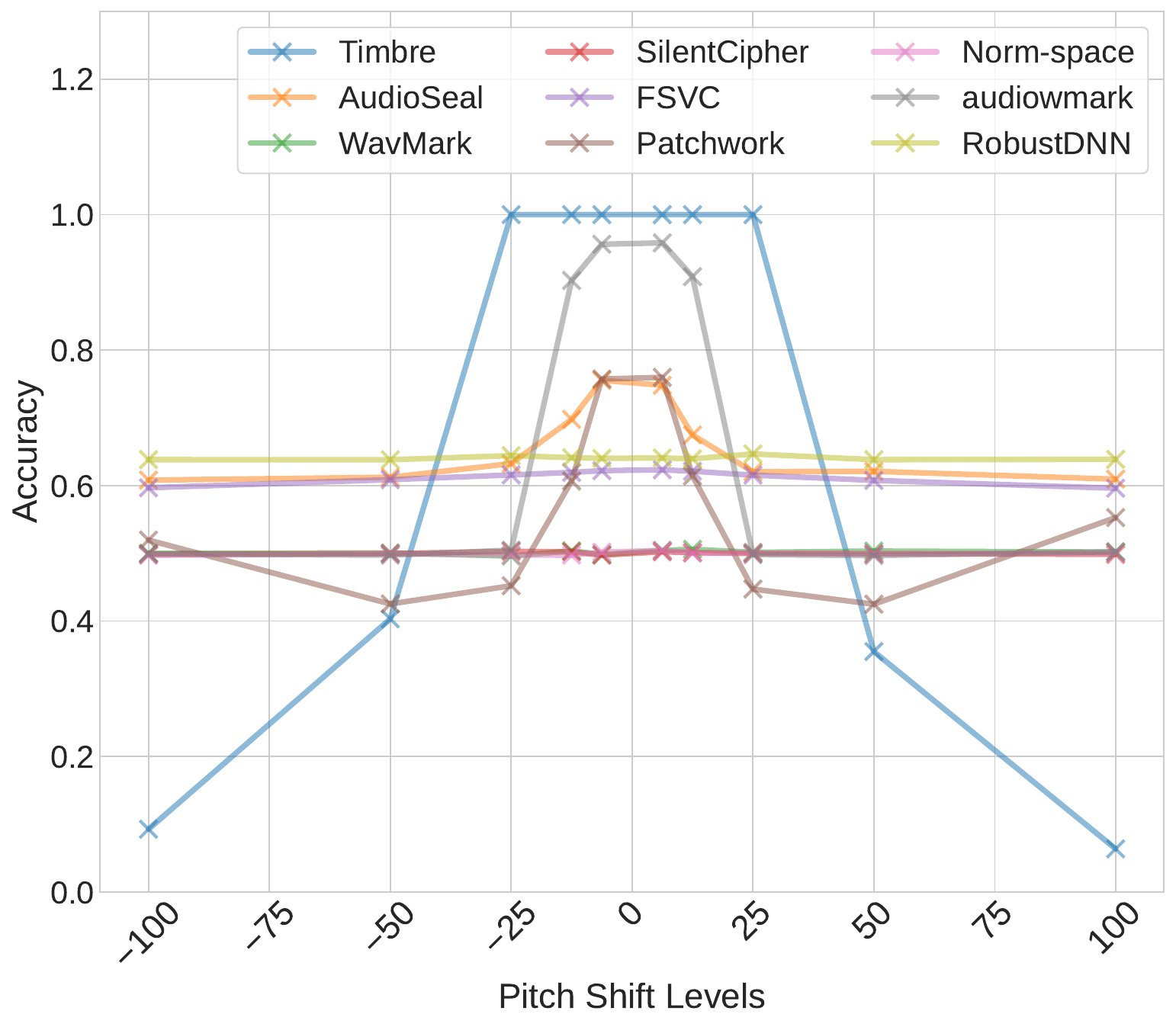}}}
\subfigure[Time Stretch]{\label{fig:time_stretch_line}\includegraphics[width=0.23\textwidth]{{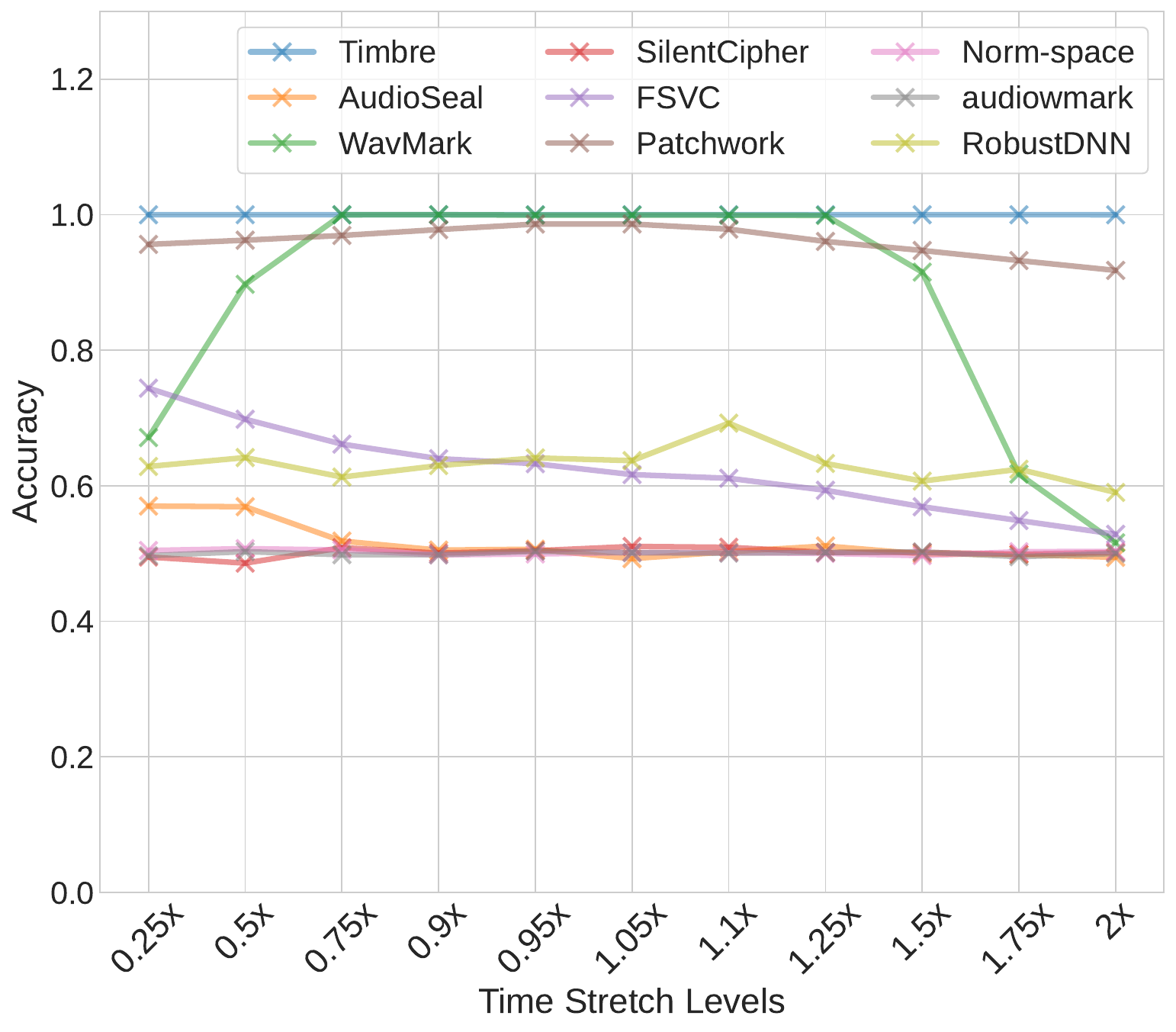}}}
\subfigure[Cutting Audio]{\label{fig:cuttingaudio_line}\includegraphics[width=0.23\textwidth]{{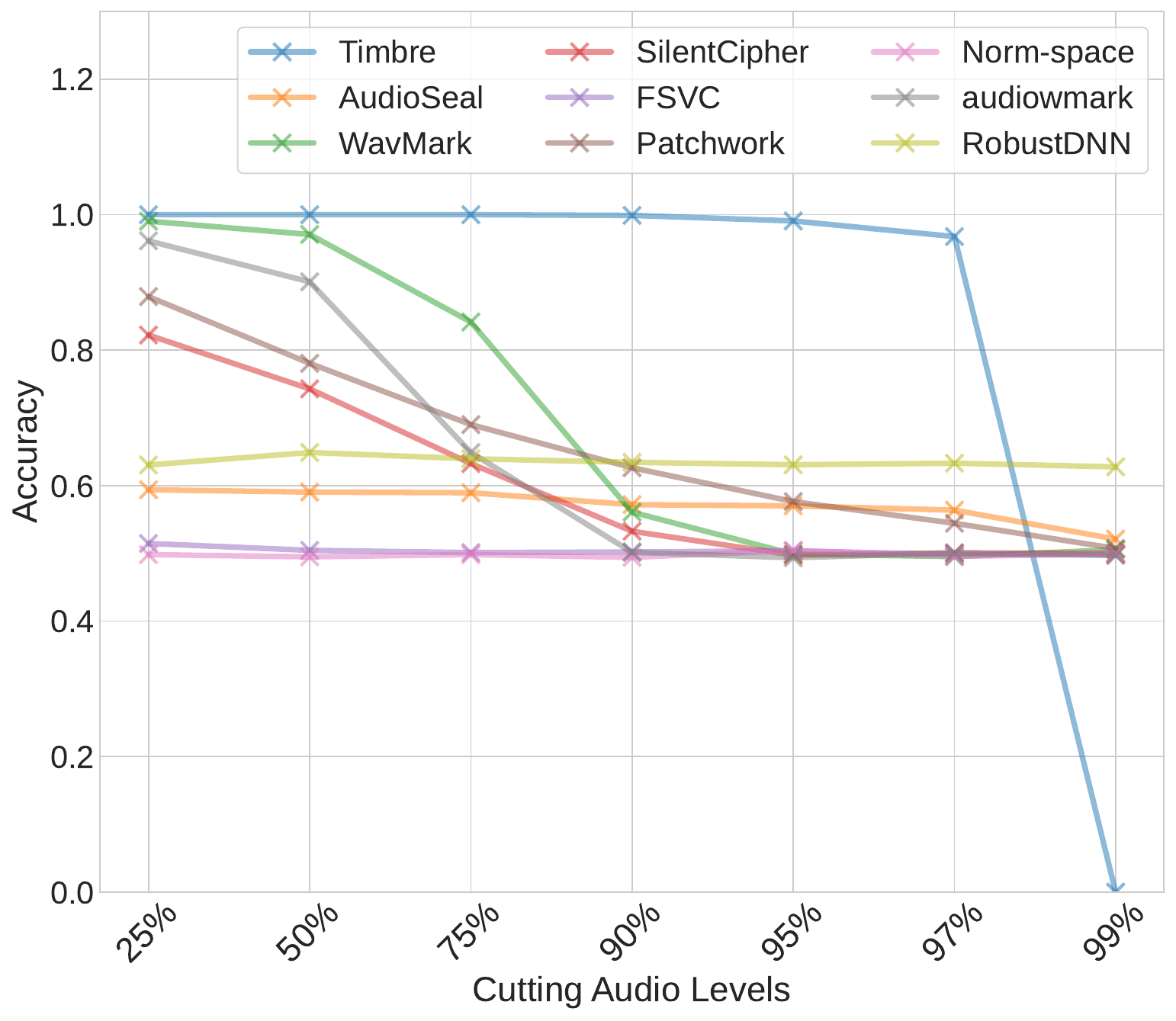}}}
\subfigure[Sample Suppression]{\label{fig:samplesuppression_line}\includegraphics[width=0.23\textwidth]{{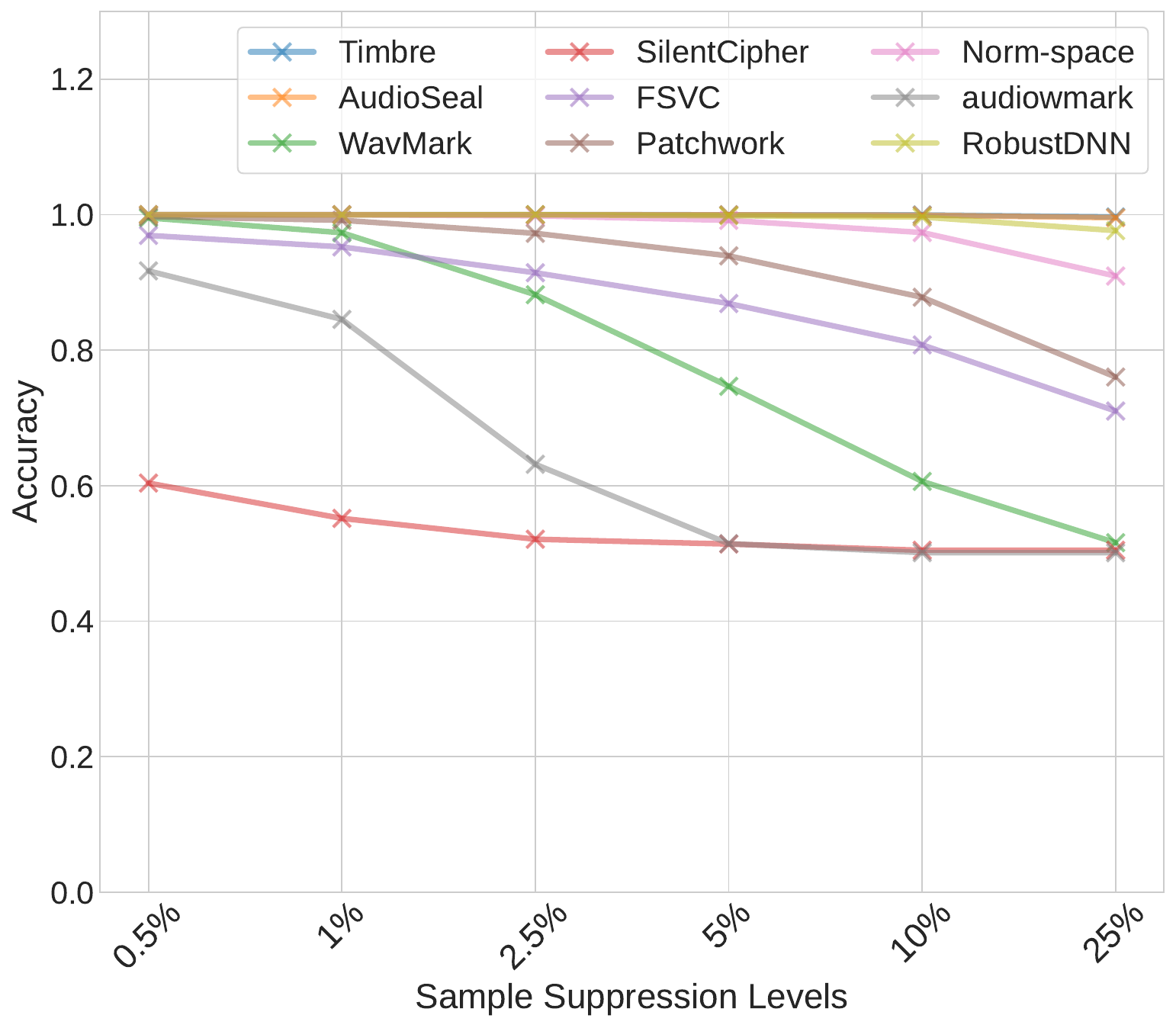}}}
\vspace{-10pt}
\caption{Most Effective Signal Distortion Attacks}
\label{fig:signal_detail}
\end{figure*}

We conducted detailed experiments with specific settings for each attack, as presented in Fig.~\ref{fig:signal_detail}. From these results, we observe the following key trends in watermark robustness across different distortion levels.

\noindent\textbf{Impact of Pitch Shift:} As shown in Fig.~\ref{fig:pitchshift_line}, most watermarks exhibit significant vulnerability. As pitch shift levels deviate from zero, all watermarks experience a sharp drop in accuracy. Timbre~\cite{liu2023detecting} initially maintains high accuracy close to the unshifted pitch (0 levels) but drops sharply beyond ±25 shift levels, eventually falling below 0.4 at higher shifts. This aligns with our earlier finding that pitch-based attacks can disrupt watermarks embedded through frequency patterns, as even moderate shifts alter the watermark signal. AudioSeal~\cite{san2024proactive} and audiowmark~\cite{audiowmark} also show some resilience within a narrow pitch shift range but decline quickly beyond that.

\noindent\textbf{Impact of Time Stretch:} In Fig.~\ref{fig:time_stretch_line}, Timbre, WavMark, and Patchwork demonstrate noticeable robustness, maintaining accuracy levels around 0.8-1.0 across various stretching factors from 0.5x to 1.5x. These watermarks appear more adaptable to time-domain modifications due to their repeated watermark embedding across periods, allowing them to withstand moderate to substantial changes in playback speed. In contrast, other schemes such as SilentCipher and FSVC display a steady decline in accuracy, often dropping below 0.6 as the time stretch factor increases, highlighting their sensitivity to time-domain shifts.

\noindent\textbf{Impact of Cutting Audio:} As demonstrated in Fig.~\ref{fig:cuttingaudio_line}, where segments of the audio are removed, Timbre once again shows strong robustness, maintaining accuracy near 1.0 until 90\% of the audio is cut, after which it declines sharply. WavMark and audiowmark retain moderate robustness up to 75\% cutting but begin to lose accuracy beyond this point. The remaining watermark schemes, including AudioSeal and FSVC, are significantly impacted, with accuracy levels dropping to 0.5 at higher levels of audio cutting, indicating their dependence on continuous audio segments for maintaining watermark integrity.

\noindent\textbf{Impact of Sample Suppression:} Fig.~\ref{fig:samplesuppression_line} demonstrates that when a portion of the audio samples are suppressed, Timbre and AudioSeal display remarkable resilience, with accuracy remaining near 1.0 even at 10\% suppression. FSVC and Patchwork also perform well under mild suppression but fall below 0.8 as suppression reaches 25\%. Other watermarks, such as SilentCipher and audiowmark, are much more vulnerable to sample suppression, with accuracy levels dropping steadily and reaching 0.5 at higher suppression percentages.

\subsection*{D. Physical Distortion on Watermark Probability}

We measure the robustness of watermark detection probability in physical settings, as shown in Table~\ref{tab:device_prob}.
At the close range with the default setup, WavMark achieves 100\% detection, while SilentCipher is at 8.98\% and other schemes near 0\%. With alternative devices, WavMark maintains 100\%, while SilentCipher improves to 29.9\% with the HyperX microphone but drops to 0
At the medium range (2.5 m), detection probabilities drop significantly, with WavMark and SilentCipher falling to 0\% and 8.57\%, respectively, while AudioSeal slightly improves to 1.25\%. At 5 m, most schemes show 0\% detection, except AudioSeal, which reaches 4.36\%.
Overall, WavMark demonstrates high robustness at close range but weakens with distance. AudioSeal shows low detection overall, with slight improvement at longer distances, while SilentCipher is inconsistent across devices and distances. These results highlight that close-range recording is most favorable for reliable detection.
\begin{table}[h]
\resizebox{\columnwidth}{!}{%
\begin{tabular}{|c|ccc|c|c|}
\hline
                                & \multicolumn{3}{c|}{\textbf{Close (0.5 m)}}                                                              & \textbf{Medium (2.5 m)} & \textbf{Far (5 m)} \\ \hline
\textbf{Watermark}              & \multicolumn{1}{c|}{\textbf{Default}} & \multicolumn{1}{c|}{\textbf{HyperX Mic}} & \textbf{Logitech Spk} & \textbf{Default}     & \textbf{Default}   \\ \hline
AudioSeal                       & \multicolumn{1}{c|}{0.07\%}           & \multicolumn{1}{c|}{0.24\%}              & 0.02\%                & 1.25\%               & 4.36\%             \\ \hline
\cellcolor[HTML]{FFFFFF}WavMark & \multicolumn{1}{c|}{100.00\%}         & \multicolumn{1}{c|}{100.00\%}            & 100.00\%              & 0.00\%               & 0.00\%             \\ \hline
SilentCipher                    & \multicolumn{1}{c|}{8.98\%}           & \multicolumn{1}{c|}{29.90\%}             & 0.00\%                & 8.57\%               & 0.00\%             \\ \hline
audiowmark                       & \multicolumn{1}{c|}{0.00\%}           & \multicolumn{1}{c|}{0.00\%}              & 0.00\%                & 0.00\%               & 0.00\%             \\ \hline
\end{tabular}%
}
\caption{Physical experiments demonstrate the probability of watermark detection under various settings.}
\label{tab:device_prob}
\end{table}

\subsection*{E. Attack Summary}

For \emph{Signal Distortions}, the first check mark indicates that the watermark withstands a high-audio-quality-guaranteed attack (a mild attack), while the second check mark shows resilience to a medium-audio-quality-guaranteed attack (a moderate attack). 
For \emph{Physical Distortions}, each cell's first symbol shows whether the watermark is robust under the default setting (0.5m distance with MacBook's built-in microphone and speaker). Additional check marks indicate robustness in alternative settings: for different distances, these are 2.5m and 5m; for different hardware, they represent the HyperX microphone and Logitech speaker.
For \emph{AI Distortions}, a check mark indicates that the watermark remains intact after passing through the generative model. \textbf{Across all distortion types, a check mark means the watermark achieves a bit recovery accuracy above 0.8.} We use color coding to convey robustness levels: green represents full robustness to all conditions, light red indicates partial robustness to mild attacks but vulnerability to stronger ones, and dark red signifies no robustness against the specified type of attack.

Based on Table~\ref{tab:watermark_removal_attacks}, Timbre Watermarking stands out as one of the more resilient schemes, showing substantial robustness across both Signal and Physical Distortions. In the Signal Distortions section, Timbre maintains resilience against a range of attacks, indicated by green cells across multiple distortions like MP3 Compression, Time Stretching, and Impulse Response augmentation, suggesting that it is well-equipped to handle various signal processing techniques without losing integrity. In contrast, watermarks like SilentCipher and audiowmark display significant vulnerability, with many dark red cells across Signal Distortions, indicating that they fail to withstand even mild attacks such as Gaussian Noise addition or Bitcrushing.
For Physical Distortions, the table reveals that most watermarks, including AudioSeal and Norm-Space, are less robust when exposed to varying recording distances and hardware, as shown by the predominant light and dark red cells in these columns. Timbre remains an exception here as well, demonstrating partial to full robustness (green and light red cells) across different physical settings. This suggests that Timbre has been optimized to perform well under real-world recording conditions where physical variations are common.

However, the AI Distortions section highlights a critical vulnerability across almost all watermarking schemes. Nearly all schemes, including Timbre, are marked in dark red, reflecting a lack of robustness against AI-based attacks such as VC and TTS regeneration. Only Timbre shows robustness to a few AI-based distortions such as TTS using Tacotron2 and Fastspeech2, where the adversary does not have any unwatermarked samples. This widespread weakness indicates that, while some watermarks can withstand traditional signal and physical distortions, generative AI remains a significant challenge, with current watermarks unable to consistently survive the transformations introduced by advanced AI models.

Overall, the table underscores the strengths and limitations of various watermarking schemes. Timbre watermarking demonstrates relatively higher resilience, especially against Signal and Physical Distortions, while most other schemes lack the robustness needed to handle either AI-based or high-intensity physical distortions, emphasizing the need for more adaptable watermarking solutions in the evolving landscape of audio manipulation.

\subsection*{F. Table Explanation}
Convolutional Neural Networks (CNNs), Frequency Singular Value Coefficient (FSVC), Frequency-Domain Coefficients Logarithmic Mean (FDLM), Fully Connected Layers (FC), Recovery Accuracy (BRA), Detection Rate (DR), Normalized Correlation Coefficient (NC), Bit Error Rate (BER), and Area Under the Curve (AUC). Fidelity and perceptual evaluation metrics include Document-to-Watermark Ratio (DWR), Objective Difference Grade (ODG), Spectral SNR (SSNR), Mean Opinion Score (MOS), Speech Embedding Consistency Score (SECS), Short-Time Objective Intelligibility (STOI), Segmental Objective Signal-to-Noise Ratio (SO-SNR), and Multiple Stimuli with Hidden Reference and Anchor (MUSHRA).



\end{document}